\documentclass[12pt]{article}

\usepackage{jheppub} 

\usepackage{braket}

\pdfoutput=1
\usepackage{graphicx}
\usepackage{dcolumn}
\usepackage{bm}
\usepackage{slashed}
\usepackage{color}
\usepackage{hyperref}

\usepackage[dvipsnames]{xcolor}

\usepackage{tikz}

\def\lsim{\mathrel{\rlap{\lower4pt\hbox{\hskip1pt$\sim$}}
     \raise1pt\hbox{$<$}}}         
\def\gsim{\mathrel{\rlap{\lower4pt\hbox{\hskip1pt$\sim$}}
     \raise1pt\hbox{$>$}}}         

\def\nn{\nonumber \\ }
\def\rd{{\rm d}}
\def\abs#1{ \left| #1 \right| }

\begin{document}

\title{Axion strings are superconducting}

\author[a,b]{Hajime Fukuda}
\emailAdd{hfukuda@lbl.gov}
\affiliation[a]{Theoretical Physics Group, Lawrence Berkeley National Laboratory,
Berkeley, CA 94720, USA}
\affiliation[b]{Berkeley Center for Theoretical Physics, Department of Physics,\\
University of California, Berkeley, CA 94720, USA}

\author[c]{Aneesh V.~Manohar}
\affiliation[c]{Department of Physics, University of California at San Diego, 9500 Gilman Drive,\\ La Jolla, CA 92093-0319, USA}

\author[a,b,d]{Hitoshi Murayama}
\affiliation[d]{Kavli IPMU (WPI), UTIAS, The University of Tokyo, Kashiwa, Chiba 277-8583, Japan}

\author[a,b]{Ofri Telem}

\date{\today}

\abstract{
We explore the cosmological consequences of the superconductivity of QCD axion strings. Axion strings can support a sizeable chiral electric current and charge density, which alters their early universe dynamics. We examine the possibility that shrinking axion string loops can become effectively stable remnants called vortons, supported by the repulsive electromagnetic force of the string current. We find that vortons in our scenario are generically unstable, and so do not pose a cosmological difficulty. Furthermore, if a primordial magnetic field (PMF) exists in the early universe, a large current is induced on axion strings, creating a significant drag force from interactions with the surrounding plasma. As a result, the strings are slowed down, which leads to an orders of magnitude enhancement in the number of strings per Hubble volume. Finally, we study potential implications for the QCD axion relic abundance. The QCD axion window is shifted by orders of magnitude in some parts of our parameter space.}

\maketitle

\section{Introduction}

One of the long standing puzzles in theoretical physics is the tiny value of $\overline \theta < 10^{-10} \ll 1$, usually referred to as the strong CP problem. The strong CP problem cannot be solved anthropically\,\cite{Ubaldi:2008nf}. A possible solution is by the Peccei-Quinn (PQ) mechanism\,\cite{Peccei:1977hh,Peccei:1977ur}, in which a broken anomalous chiral $\text{U}(1)$ symmetry dynamically sets the $\theta$ angle to zero\,\cite{Vafa:1984xg}. PQ symmetry breaking produces a light pseudo-Goldstone boson called the axion~\cite{Weinberg:1977ma,Wilczek:1977pj}. The axion is also a dark matter (DM) candidate.

If the PQ symmetry is broken during inflation, the corresponding axion field is frozen during the inflationary era. Eventually, when the Hubble expansion $H$ becomes smaller than the axion mass $H \ll m_a$, at which point the axion starts behaving like a massive particle, and the initial axion field value determines the relic axion density\,\cite{Preskill:1982cy,Abbott:1982af,Dine:1982ah}. Quantum fluctuations of the axion are generated, and the isocurvature fluctuation is generally large. To avoid observational constraints\,\cite{Aghanim:2018eyx}, the inflationary Hubble scale $H_\text{inf}$ must be as small as $H_\text{inf} \lesssim 10^7\,\text{GeV}$\,\cite{Hikage:2012tf,Kobayashi:2013nva}. In the case that PQ symmetry breaking occurs after inflation, the Kibble-Zurek mechanism\,\cite{Kibble:1980mv,Kibble:1976sj,Zurek:1985qw,Zurek:1996sj,Murayama:2009nj} leads to the formation of cosmic axion strings, topological defects associated with $\text{U}(1)_\text{PQ}$ symmetry breaking. In particular, the generated axion strings are vortices around which the phase of the scalar field rotates by $2\pi$. Axion strings are topologically stable because $\pi_1(U(1)_{\text{PQ}}) = \mathbb{Z}$, and the $U(1)_{\text{PQ}}$ symmetry is restored in their core. The dynamics of axion strings is crucial for understanding the relic density of  axion DM. In this paper we focus on post inflationary PQ breaking, and explore novel dynamics that governs the evolution of the resulting axion string network.

In standard cosmic string cosmology, an intially large number of axion strings appear at the PQ phase transition\,\cite{Zurek:1985qw,Murayama:2009nj}. As soon as they are created, they begin to reconnect with each other and the total length of string per Hubble horizon diminishes. Due to  friction caused by interactions with the thermal plasma, the number of strings in the Hubble horizon, $\xi$, may be larger than $\mathcal{O}(1)$ at first\,\cite{Nagasawa:1997zn}. As the universe cools, plasma friction weakens and the string abundance eventually enters a scaling regime, where $\xi\sim\mathcal{O}(1)$ of strings exist per Hubble volume. The scaling regime persists over many orders of magnitude in temperature until $T\gsim T_{\text{QCD}}$. The scaling of strings in the early universe has been extensively studied by numerical simulations\,\cite{Bennett:1989yp,Allen:1990tv,Vanchurin:2005yb,Olum:2006ix,Hiramatsu:2010yu,BlancoPillado:2011dq,Hiramatsu:2012gg,Kawasaki:2014sqa,Fleury:2015aca,Klaer:2017ond,Klaer:2017qhr,Gorghetto:2018myk,Vaquero:2018tib,Buschmann:2019icd,Hindmarsh:2019csc,Klaer:2019fxc,Saikawa_IPMU_slide,Gorghetto:2020qws}. The scaling behavior is the result of a balance between string length entering the horizon, and string length which is lost by the emission of unstable string loops due to efficient string reconnection.

At $T\gsim T_{\text{QCD}}$, the axion gets a mass from QCD non-perturbative effects and consequently domain walls are formed between the axion strings. At this point, each axion string becomes the boundary of $N_{\text{DW}}$ domain walls, where $N_{\text{DW}}$ is the coefficient of the PQ-QCD anomaly. Generically, when $N_{\text{DW}}>1$ the domain walls cannot decay\footnote{Unless other model building solutions are considered, for example some mild explicit breaking of PQ symmetry. See \cite{DiLuzio:2020wdo} and references within for the landscape of QCD axion models.}, and so they can overclose the universe. For this reason, most post-inflationary QCD axion scenarios are KSVZ-like \cite{Kim:1979if,Shifman:1979if} with $N_{\text{DW}}=1$. In this case, the string-domain wall systems are unstable, and they annihilate into cold axions which form the bulk of the DM relic density.

In this paper, we focus on the KSVZ scenario as an example with $N_{\text{DW}}=1$ and point out an important effect --- the axion string is (EM and color) superconducting\,\cite{Callan:1984sa}.\footnote{By superconducting, it is meant that the current on strings grows in proportion to electric fields imposed on the direction of strings. 
} While our discussion is completely generic and can be applied to {\it any} axion scenario, we use the KSVZ example for a quantitative discussion. The superconductivity of axion strings is widely known in the context of Chern-Simons theory, but has been overlooked in phenomenological cosmological studies. We find that axion string superconductivity changes the density of axion strings in the early universe, and hence the relic axion abundance. In particular, the presence of a primordial magnetic field (PMF), the axion relic abundance could be considerably larger than previous expectations. We also study if superconductivity allows for metastable remnants of the axion string, called vortons, as discussed previously in Ref.~\cite{Davis:1988jq,Carter:1993wu,Brandenberger:1996zp,Martins:1998gb,Martins:1998th,Carter:1999an} for local strings supporting vector-like currents\,\cite{Witten:1984eb}. We find that these are generically unstable in our scenario, as a result of zero mode decay induced by the curvature of the vortons.
\section{Fermion Zero-Modes and Axion String Superconductivity}

As a first step, we provide a quick review of the superconductivity of KSVZ axion strings. The argument is very general, and applies to other axion models as well. This section also serves to define our notation, and to introduce features of the model that will play a role in our analysis.

In addition to the Standard Model (SM) fields, the minimal KSVZ model~\cite{Kim:1979if,Shifman:1979if}  has two chiral fermions $\psi_L$ and $\psi_R$ which transform as $\mathbf{3}$ under color and are weak $SU(2)$ singlets, and a color-singlet complex scalar $\Phi$ with the interaction Lagrangian
\begin{align}
\label{eq:L_KSVZ}
\mathcal{L}_{\text{int}} =  - \left[ y_\Phi\, \Phi \bar{\psi}_L \psi_R + \text{h.c.} \right] - V(\Phi)\,.
\end{align}
The phase of $y_\Phi$ can be absorbed into $\Phi$ so that $y_\Phi$ is real.
The chiral transformation
\begin{align}
\label{eq:3.2}
\Phi &\to e^{i\alpha} \Phi, & \psi_L &\to e^{i\alpha} \psi_L, & \psi_R &\to  \psi_R,
\end{align}
is anomalous under QCD, $\theta \to \theta -  \alpha$, and is the PQ symmetry in this theory. This example has $N_{\text{DW}}=1$. We assume $\Phi$ acquires a vacuum expectation value (VEV), $\langle \Phi \rangle = f_a / \sqrt{2}$ which breaks the PQ symmetry and gives $\psi$ a mass $m_\psi=y_\Phi f_a/\sqrt 2$. The phase of $\Phi$ is the axion $a$, i.e.\ neglecting radial fluctuations
\begin{align}
\Phi &\to \frac{f_a}{\sqrt 2} \,e^{i a /f_a}\,,
\end{align}
$f_a$ is the axion decay constant, and $a/f_a$ has periodicity $2\pi$. As we are considering post inflationary PQ breaking, the KSVZ fermions are abundant in the early universe, and so they must decay to SM particles before big-bang nucleosynthesis (BBN) (see \cite{DiLuzio:2020wdo} and references within). To allow for decays, there have to be interactions between $\psi$ and SM particles. For example, KSVZ fermions with the quantum numbers of vertorlike quarks (VLQ) have Yukawa couplings via the Higgs field $H$ to the SM quark doublet $Q_L$, ($\widetilde H_i = \epsilon_{ij} H^{\dagger j}$)
\begin{align}
\label{eq:psiDecayOp}
\mathcal{L} &=  -y \,\overline Q^i_L\psi_R  H_i + \text{h.c.} & \text{or}&&
\mathcal{L} &=  -y \,\overline Q^i_L \psi_R  \widetilde H_i + \text{h.c.}
\end{align}
where $y$ is the Yukawa coupling of the KSVZ fermion to the SM Higgs. The Yukawa term can be of either form, depending on whether $\psi$ has weak hypercharge $Y=-1/3$ or $Y=2/3$, i.e.\ the quantum numbers of the SM fields $b_R$ or $t_R$. The interactions Eq.~\eqref{eq:psiDecayOp} preserve the PQ symmetry Eq.~\eqref{eq:3.2}. 
Note that because the gauge group of the standard model particles is actually limited to $G_\text{SM} \equiv \text{SU}(3) \times \text{SU(2)} \times \text{U}(1) \,/ \,{\mathbb Z}_6$, $\psi$ must have $Y=2/3\ {\rm mod}~1$, so that it can decay to the standard model particles.  The above two examples are the simplest choices that allow for direct Yukawa couplings without need for additional particles.

The axion string is a topological defect around which the axion (the phase of $\Phi$) winds by $2\pi$. An infinitely straight axion string in the $z$-direction has the form
\begin{align}
\label{eq:4}
\Phi(\mathbf{x},t) &= \frac1{\sqrt 2} \, f_a\, h(r_\perp) \,e^{i \theta}\,,
\end{align}
where $r_\perp=\sqrt{x^2+y^2}$ is the transverse distance from the string, and $\theta = \tan^{-1} (y/x)$ is the azimuthal angle. The radial solution $h(r_\perp)$ depends on the axion potential $V(\Phi)$, with $h(r_\perp) \to 1$ as $r_\perp \to \infty$, and $h(0)=0$ so that $\Phi$ is non-singular at the origin.

In the string background Eq.~\eqref{eq:4} for $\Phi$, the Dirac operator for the KSVZ fermions in the $xy$-plane has a zero eigenvalue, as guaranteed by an index theorem~\cite{Weinberg:1981eu}. As a result, the $3+1$ dimensional theory has a chiral massless fermion moving in the $z$-direction, which is a KSVZ fermion bound to the string~\cite{Jackiw:1981ee}. The transverse size of the massless fermion wavefunction is of order $1/m_\psi$. Because the KSVZ fermion interaction with $\Phi$ in Eq.~\eqref{eq:L_KSVZ} is chiral, the axion string has a trapped chiral fermion left (downward) moving fermion and its CPT conjugate anti-particle, originating from the KSVZ fields $\psi_L/\bar{\psi}_R$. Replacing $\Phi \to \Phi^*$ in Eq.~\eqref{eq:L_KSVZ} would result in a right (upward) moving trapped mode.
Since $\psi$ carries color and hypercharge, the zero-mode carries color and $U(1)$ current along the string. 

The $1+1$ dimensional theory of the chiral fermion has a $U(1)$ gauge anomaly. Below the electroweak phase transition, we simply consider the 1+1 dimensional gauge anomaly of $U(1)_{\text{EM}}$, 
\begin{align}
\label{eq:23}
\partial_\mu j^\mu_{\text{zero-mode}} &= - \frac{N_c e_\psi^2}{2\pi} E_z\,.
\end{align}
where $N_c = 3$, $e_\psi$ is the electric charge of $\psi$ multiplied by $e$, and $E_z$ is the electric field at the string core in the $z$ direction, leading to a violation of current conservation on the string.\footnote{Above the electroweak phase transition, one should instead use the analogous equation for hypercharge.} On the other hand the full $3+1$ dimensional theory is anomaly free, and has a conserved electromagnetic current. This ``paradox'' was resolved by Callan and Harvey\,\cite{Callan:1984sa} by noting that there is a Goldstone-Wilczek current~ \cite{Goldstone:1981kk}
\begin{align}\label{eq:london2}
J_{\text{GW}}^\mu\,=\,\frac{N_c e_\psi^2}{8\pi^2\,f_{a}}\epsilon^{\mu \nu\rho\sigma}\,\partial_\nu a\,F_{\rho\sigma}
\end{align}
in the bulk, leading to charge inflow onto the string exactly equal to Eq.~\eqref{eq:23}. Here $F_{\mu\nu}$ is the $U(1)$ field strength tensor, and we use the sign convention $\epsilon_{0123}=+1$.\footnote{So that $\epsilon^{0123}=-1$, which is the opposite of the sign convention in Ref.~\cite{Kaplan:1987kh}.} The above current is deposited on the string in the form of massless fermion zero-modes. This mechanism is referred to as anomaly inflow.

We have started from an explicit UV model to see the chiral anomaly on the string. However, the anomaly is determined solely by the IR Lagrangian. Let us derive the anomaly inflow from the coupling between the axion $a$ and gauge fields. Away from the string core, the fermion $\psi$ is massive and can be integrated out to give a low-energy Lagrangian for the interaction of the axion with the electromagnetic field
\begin{align}
L_{a} &=\frac12 (\partial_\mu a)^2 - \frac14 F_{\mu \nu} F^{\mu \nu} + \frac{N_c e_\psi^2}{16 \pi^2} \frac{a}{f_{a}} F^{\mu \nu} \widetilde F_{\mu \nu} \,,
\label{eq:24}
\end{align}
where
\begin{align}
\widetilde  F_{\mu \nu} &= \frac12  \epsilon_{\mu \nu \alpha \beta} F^{\alpha \beta}\,.
\label{eq:25}
\end{align}
Here, we omit the color gauge field for simplicity. The effective Lagrangian preserves electromagnetic gauge invariance, since the original $3+1$ dimensional theory had no electromagnetic anomaly, and was gauge invariant.

From Eq.~\eqref{eq:24}, the electromagnetic current in the bulk is
\begin{align}
\label{eq:CSCurrent}
J^\mu_{a} &= -\frac{\delta S_{a}}{\delta A_\mu} = -\frac{N_c e_\psi^2}{4 \pi^2} \widetilde F^{\mu \nu}\  \partial_\nu \frac{a}{f_{a}}  \,,
\end{align}
which is the same as the Goldstone-Wilczek current Eq.~\eqref{eq:london2}.
The divergence of $J^\mu_{a}$ is
\begin{align}
\partial_\mu J^\mu_{a} &= -\frac{N_c e_\psi^2}{4 \pi^2} \widetilde F^{\mu \nu}\  \partial_\mu   \partial_\nu \frac{a}{f_{a}}  \,,
\end{align}
which naively vanishes since $F_{\mu \nu}$ is antisymmetric. However,
the axion field has a winding number around the origin, leading to\,\cite{Callan:1984sa,Naculich:1987ci,Fukuda_Yonekura}
\begin{align}
\label{eq:26}
\left(\partial_x \partial_y - \partial_y \partial_x\right)  \frac{a}{f_{a}}  = 2\pi \delta^{(2)}(\mathbf{x}_\perp) \,,
\end{align}
so that
\begin{align}
\label{eq:27}
\partial_\mu J^\mu_{a} &= - \frac{N_c e_\psi^2}{2 \pi} \widetilde F^{x y}  \delta^{(2)}(\mathbf{x}_\perp) =   \frac{N_c e_\psi^2}{2 \pi} E_z \delta^{(2)}(\mathbf{x}_\perp) \,,
\end{align}
and is exactly canceled by the zero-mode current divergence Eq.~\eqref{eq:23}. In other words, the theory on the string must produce the same anomaly with the opposite sign to the inflow current to cancel the divergence of this current \emph{regardless of the UV theory}, as long as the UV theory preserves electromagnetic gauge invariance.

In the KSVZ model, we showed that the current is carried by a chiral fermion on the string so the trapped current density is chiral,\footnote{See Ref.\,\cite{Naculich:1987ci,Fukuda_Yonekura} for a discussion on regularization.}
\begin{align}
\label{eq:ChiralCharge}
\rho = -I,
\end{align}
where $\rho$ is the charge per unit length and $I$ is the current. The relative minus sign is because the zero-mode is left-moving. Taking the divergence
\begin{eqnarray}
\partial_\mu J^\mu = \partial_t \rho + \partial_z I = -\partial_t I - \partial_z I.
\end{eqnarray}
If the system is placed in a background electric field that is $z$-independent, $I$ does not depend on $z$, and
\begin{eqnarray}
\label{eq:LondonEq}
\partial_\mu J^\mu = -\partial_t I =  - \frac{N_c e_\psi^2}{2\pi} E_z\,,
\end{eqnarray}
i.e.\ the current increases linearly with the applied electric field.
Thus, there is no resistivity and \emph{the axion string is superconducting}. A similar argument holds for the color current.\footnote{Since the KSVZ fermion $\psi$ is a color triple and the interaction with $\Phi$ does not depend on color, there is a zero-mode $\psi_{0,\alpha}$ for each color component $\alpha=1,2,3$ of $\psi$. $\psi$ zero-modes can be combined to form color-singlet ``baryons'' $\epsilon^{\alpha \beta \gamma} \psi_{0,\alpha} \psi_{0,\beta} \psi_{0,\gamma}$. A background electric field is color-singlet, and will populate the string with such color singlet states.} 

A microscopic description of the current is useful for phenomenological applications later in this paper. The current is proportional to the number density of zero mode fermions traveling along the string. If the zero mode fermion states are occupy energy all levels up to the Fermi energy $\varepsilon_F$, the current on the string is
\begin{align}
\label{eq:I}
I &= \frac{N_c e_\psi }{2\pi} \varepsilon_F.
\end{align}
since the $1$-dimensional density of states is ${\rm d}k/(2\pi)= {\rm d} E/(2\pi)$ for each color. 
The Fermi energy picture allows us to understand the evolution of the current, Eq.\,\eqref{eq:LondonEq} from another viewpoint. In this language, the current evolution is due to the Fermi energy changing as a result of the applied electric field, and so new fermions ``appear from the vacuum" as negative energy states become positive energy states\,\cite{Peskin:1995ev}. For Fermi energy $\varepsilon_F$, the energy per unit length in bound fermions $\varepsilon_I = (\varepsilon_F/2) (N_c \varepsilon_F) / (2\pi)$, so that
\begin{align}
    \label{eq:currentE}
    \varepsilon_I = \frac{\pi I^2}{N_c e_\psi^2},
\end{align}
neglecting fermion-fermion interactions.
In addition to the energy of the zero modes, axion strings also have an intrinsic tension $\mu \simeq \ln(L f_a) f_a^2$, where $L$ is a typical distance between strings.\footnote{The energy per unit length diverges for global strings in the infinite volume limit. We have used the spacing between strings to cutoff the integral.} For
\begin{align}
    \label{eq:appCond}
    \varepsilon_I \ll \mu \iff I \ll e_\psi \sqrt{\frac{N_c \mu}{\pi}},
\end{align}
the fermion energy is negligible compared with the intrinsic string tension. Unless otherwise stated, the magnitude of the current is assumed to satisfy this condition, so that previous studies on axion string interactions, which do not include Eq.~\eqref{eq:currentE}, can be used. Nevertheless, interactions between axion strings and charged particles are modified because of the trapped currents, as studied in later sections.

Superconducting strings with zero-mode currents were first discussed in detail in Ref.\,\cite{Witten:1984eb}. This paper considers a vector-like UV theory with zero-modes propagating in both directions along the string. An important difference from this earlier analysis is the chiral nature of axion strings. One consequence of chirality is that electromagnetic fields can induce currents around the strings via Goldstone-Wilczek current inflow, Eq.\,\eqref{eq:london2}. In a large enough box (such as the entire universe), the axion winding is trivial and the Goldstone-Wilczek current inflow from infinity vanishes. However, the Goldstone-Wilczek current can still transport charge between strings, so that string loops develop charge even though the whole system remains neutral. Charge inflow and electrodynamics in the presence of chiral axion strings were studied in Refs.~\cite{Callan:1984sa,Naculich:1987ci,Kaplan:1987kh,Manohar:1988gv}, and the rotation of the polarization of electromagnetic waves propagating in the string background was studied in Refs.~\cite{Naculich:1987ci,Manohar:1988gv,Harari:1992ea,Fedderke:2019ajk}.

\section{Current Leakage}\label{sec:leak}

Besides the generation of zero mode currents on the string by the ambient EM field, we also have to take into account current leakage processes, which play an important role in the dynamics. In this section, we study mechanisms responsible for current leakage. These have mostly been analyzed for non-anomalous superconducting strings and are irrelevant in our case. For example, ref.~\cite{Witten:1984eb,Barr:1987xm} considered a vector-like theory with zero-modes moving in both directions along the string. In this case, the zero-modes can collide and produce fermion pairs that escape to infinity if the Fermi energy of the zero-modes exceeds $m_\psi$, the mass of the KSVZ fermion at infinity. This pair-production process limits the Fermi energy on the string~\cite{Witten:1984eb} by $\varepsilon_F \le m_\psi$, which translates into a maximum current $I \le 2 e_\psi m_\psi/(2\pi )$, where $1/(2\pi)$ is the density of states in $1+1$ dimension. The overall factor of $2$ is because a vector-like theory can have positive charges moving in one direction, and negative charges moving in the opposite direction.

In contrast, the axion string has chiral zero-modes, so zero mode particles and anti-particles are massless and move in the same direction.  For this reason, they cannot annihilate among themselves and produce massive fermions which escape to infinity; the process is forbidden by energy-momentum conservation and, there is no maximum current from zero mode collisions. For straight strings this can also be seen by making a Lorentz transformation\,\cite{Sen:1992yt}. The collision between a zero mode particle and its anti-particle on the long string is boost-equivalent to a massless particle decaying into massive particles, and so is kinematically forbidden.

Another potential source of leakage is the Goldstone-Wilczek current itself. This current, induced by the ambient EM field, can change the number of the chiral zero-mode on strings. However, for the long strings we the GW current and the zero mode current on the strings are in equilibrium. In particular, in the case of string loops, the Goldstone-Wilczek ccan never change the net charge of the loop but only redistribute it. This is because the overall axion winding around the loop (as seen from far away) is trivial.

\subsection{Leakage from plasma scattering}
Though the current on axion strings cannot dissipate by internal collisions among zero modes, it \textit{can} be reduced by scattering processes involving a plasma particle incident on the string. 

As in the calculation of the photo-ionization of hydrogen, here the rate for the the interaction of plasma particles with the current depends on the geometric overlap of the zero mode wavefuntion and the plane wave of the incoming plasma particle. In the hydrogen case, the ground state photo-ionization rate depends on the Fourier transform of the ground-state wavefunction, and is suppressed by $(\omega a_0)^3$ at low frequencies, where $\omega$ is the incident photon frequency and $a_0$ is the Bohr radius~\cite[\S37]{Schiff:1955vw}. In our problem, the zero-modes is localized in two transverse dimensions, so the cross section for ionization is suppressed by $p_\perp^2/m_\psi^2$, the ratio of the fermion zero-mode size and the wavelength of the incident particle.

At temperatures $T\sim m_\psi\sim y_{\Phi} f_a$, thermal plasma particles which scatter off the string have enough kinetic energy to knock the zero mode $\psi$ into the bulk. The cross section for such scattering process is geometric, $\sigma\sim f^{-2}_a$, but the process is Boltzmann suppressed for $T< m_\psi$. Consequently, it decouples below $T\sim m_\psi/b, \,b \sim 20$ (for $f_a \sim m_\psi \sim 10^{8\mbox-12}\,\text{GeV}$), when its rate drops below the Hubble expansion rate.

At temperatures much smaller than the fermion mass, $T\ll m_\psi$, the current can no longer leak into $\psi$ particles, but it can still leak by producing SM particles via the interaction Eq.\,\eqref{eq:psiDecayOp}, the rate of which depends on the Yukawa coupling $y$.
Prototypical scattering processes leading to current decrease are
\begin{align}
\label{eq:5.2}
\bar{Q}\,+\,\psi^0_{\text{string}} &\rightarrow \,\,g + h \,, \nonumber\\
 g\,+\,\psi^0_{\text{string}} &\rightarrow \,Q + h \,, \nonumber\\
h\,+\,\psi^0_{\text{string}} & \rightarrow\, Q + g \,,
\end{align}
via $t$- and $s$-channel $Q$ exchange,
where $\psi^0_{\text{string}}$ is the string zero-mode. The cross-section for these processes, which are related by crossing, is of order
\begin{align}
\label{eq:5.3}
\sigma_{\text{scat}} &\sim \frac{1}{16\pi s} \abs{ \frac{T y}{m_\psi}\, g_{s} } ^2 = \frac{1}{16\pi s} \abs{ \frac{\sqrt 2 T y}{y_\Phi f_a}\, g_{s} } ^2 \sim  \frac{\alpha_s y^2}{2 y_\Phi^2 f^{2}_a} \frac{T}{\varepsilon_F}
=\frac{N_c \alpha_s e_\psi y^2}{4 \pi  y_\Phi^2 f^{2}_a} \frac{T}{I} \, ,
\end{align}
where $g_s$ is the QCD coupling constant. The effective coupling of the zero-mode is suppressed by the transverse wavelength of the incoming fermion relative to the size of the string core, $p_\perp/m_\psi $ with $p_\perp \sim T$, times the coupling $y$ from Eq.~\eqref{eq:psiDecayOp}. The $1/(16 \pi s)$ factor is a dimensional estimate of the phase space where $s\sim{T\varepsilon_F}$ is the center-of-mass energy squared,  $\varepsilon_F$ is the zero-mode Fermi energy and $m_\psi = y_{\Phi} f_a$, with $y_{\Phi}$ from Eq.~\eqref{eq:L_KSVZ},
and we have used Eq.\,\eqref{eq:I} to rewrite $\varepsilon_F$ in $I$.
The resulting rate to ionize a zero-mode is then 
\begin{align}\label{eq:dest.r}
\Gamma_{\text{scat}} &= n_{\text{SM}}\,\sigma\,v_{\perp\text{rel}}\, \sim\,\frac{h_\star \zeta(3)T^3}{\pi^2} \sigma_{\text{scat}}
\end{align}
where we take $v_{\perp}\sim1$ because the zero mode is relativistic, and $h_\star=g_{B}+(3/4)g_{F}$ is the effective number of degrees of freedom which can interact with the zero-mode. Comparing this rate to the Hubble expansion rate, we see that the rate Eq.~\eqref{eq:dest.r} is negligible for temperatures $T<T_{\text{scat}}$ with 
\begin{align}
\label{eq:5.5}
T_{\text{scat}} &= \left( 1.2 \times 10^7 \,\text{GeV} \right)\,\frac{1}{y}\, \left(\frac{f_a}{10^{10}\,\text{GeV}}\right) \left(\frac{I/(N_c e_\psi) }{10^{10}\,\text{GeV}}\right)^{1/2} {\left(\frac{y^2_\Phi g_\star^{1/2}}{ h_\star }\right)}^{1/2}\, .
\end{align}
where $g_\star=g_B+(7/8)g_F$ for all particle species. 
For temperatures lower than $T_{\text{scat}}$ we can neglect current dissipation by scattering off the thermal background. The last factor in Eq.~\eqref{eq:5.5} is order unity, and will be neglected.

\subsection{Leakage from string oscillations}
Another potential leakage mechanism is the interaction of fermion zero-modes with the oscillation modes of the string. We can compute the decay rate by quantizing the oscillations, which are then treated as real scalars (``Nambu-Goto'' bosons) on the string world-sheet.\footnote{See the seminal book by Vilenkin and Shellrad \cite{Vilenkin:1991zk} for more information about the Nambu-Goto description of cosmic string oscillations.} The scalar is an elementary degrees of freedom in the Nambu-Goto action, labeled as $X^\mu$. The interaction in the language of the world-sheet and bulk theories is written as
\begin{align}
\label{eq:5.6}
X^\mu + \psi^0_{\text{string}} &\rightarrow \psi^* \to Q H\,.
\end{align}
On the world sheet, this process can be described by an effective operator
\begin{align}
\mathcal{O} &=	\partial_{\alpha} X^{\mu} (\partial_{\mu}\bar{\psi}^{*}) \gamma^{\alpha} \psi^{0}_{\text{string}}\,,
\end{align}
which respects the translational invariance of $X^{\mu}$, where we have taken the interaction coefficient to be unity.  Here, $\alpha$ is a coordinate on the world sheet. Switching to the canonical normalization for $X^{\mu} = f_{a}^{-1} \varphi^{\mu}$, $\mathcal{O}$ is a dimension-three operator with an overall $1/f_{a}$ suppression.  Therefore, the coupling $X^\mu + \psi^0_{\text{string}} \rightarrow \psi^*$ is suppressed by $1/f_a$, while the off-shell $\psi^*$ propagator gives an additional $1/(y_{\Phi}f_a)$ suppression, so the overall cross section is
\begin{align}
\label{eq:5.7}
\sigma_X &= \frac{y^2 s^2}{16 \pi y_{\Phi}^{2} f_a^4} = \frac{y^2 T^2 \varepsilon_F^2}{16 \pi y_{\Phi}^{2} f_a^4} = \frac{\pi y^2 T^2 I^2}{4  N_c^2 e_\psi^2 y_{\Phi}^{2}  f_a^4}\, ,
\end{align}
where $s^2/(16\pi)$ is a dimensional estimate of the phase space and the squared amplitude, and we have used Eq.~\eqref{eq:I}.
$\sigma_X$ is a cross section in $1+1$ dimension for $X^\mu$, and hence is dimensionless.
Multiplying by $T/\pi$ for the average number of $X^\mu$ modes gives the dissipation rate
\begin{align}
\label{eq:5.8}
\Gamma_X &= \frac{y^2 T^3 I^2}{4  N_c^2 e_\psi^2  y_{\Phi}^{2} f_a^4}\,,
\end{align}
and falls below the Hubble rate $\Gamma_X < H$ for $T < T_X$,
\begin{align}
\label{eq:5.9}
T_X &= \left( 3.8 \times 10^{2}\,\text{GeV} \right) \,\frac{1}{y^2}\,\left(\frac{f_a}{10^{10}\,\text{GeV}}\right)^4 \left( \frac{10^{10}\,\text{GeV}} {I/(N_c e_\psi)} \right)^2 {\left(\frac{y_{\Phi}^{2} g_\star^{1/2}}{h_\star}\right)}\, .
\end{align}
The last factor in Eq.~\eqref{eq:5.9} is order unity, and will be neglected.

\subsection{Overall leakage rate}
The overall leakage rate is given by
\begin{align}
\label{eq:5.10}
\Gamma_{\text{leak}} &= \Gamma_{\text{scat}} + \Gamma_X\,,
\end{align}
which in practice means whichever rate is the larger one, and so $T_{\text{leak}}$, the temperature at which current destruction is relevant is given by
\begin{align}
\label{eq:TDest}
T_{\text{leak}} &\simeq \min \left(T_{\text{scat}} , T_X \right)\,.
\end{align}

\begin{figure}
\begin{center}
\includegraphics[width=0.5\textwidth]{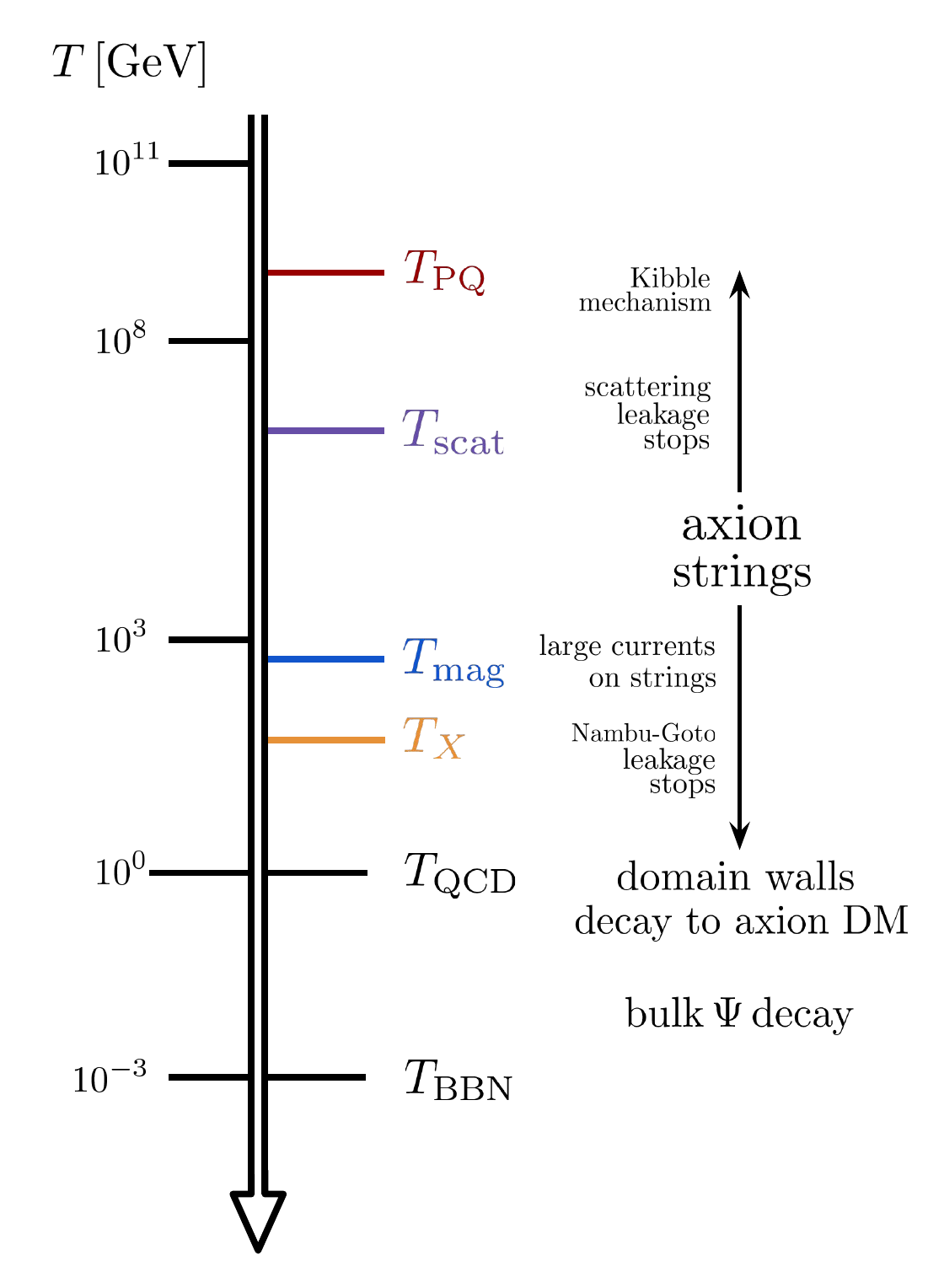}
\end{center}
\caption{The relevant scales in the evolution of superconducting axion strings. Besides the scales $T_{\text{scat}}$ and $T_X$ appearing in this section, we also include the scale $T_{\text{mag}}$ relevant for section~\ref{sec:PlasmaFriction}. The temperature when the bulk $\Psi$ decays is between $m_\Psi$ and $T_\text{BBN}$, depending on the value of $y$.}
\label{fig:scales}
\end{figure}

We summarize various scales appearing in this section in Fig.\,\ref{fig:scales} for typical values of couplings. The string network appears at $T \sim f_a$ and the Fermion zero-modes on the string escapes as the massive bulk Fermions by the collision between bulk particles until their temperature drops to $m_\psi/b$. The scattering between standard model particles and $X^\mu$ decouples at $T\text{scat}$ and $T_x$, respectively. The smaller one is the temperature when the leakage due to the scattering stops. The decay of current on strings with finite curvature is not shown here, but is discussed in following sections.

\subsection{Additional leakage processes for loops}\label{sec:curvmed}
For closed string loops, boost invariance is broken by $1/R$, where $R$ is the radius of curvature of the string, and the zero-mode can annihilate/decay into massive fermions. In particular, the trapped KSVZ fermions can decay directly into SM particles via processes such as $\psi \to Q + h$. The asymptotic expansion of the corresponding decay rate for large $R$ is computed in Appendix~\ref{app:a}. If the final state particles are massless, the decay rate is
\begin{align}
\Gamma &\sim \frac{\abs{y}^2 k}{16 \pi} \frac{1}{8 \pi^3} \frac{(k\Delta )^2}{(R k)^{2/3}}\label{eq:curvrate}
\end{align}
using the suppression factor in Eq.~\eqref{A.16}. Here $k$ is the momentum of the zero-mode $k \sim 2\pi I/(N_c e_\psi)$ 
from Eq.~\eqref{eq:I}. The decay rate vanishes as $R^{-2/3}$ as $R \to \infty$. 
In general, particles get a mass through the interaction and we need to take it into account.
It changes the asymptotic form of the decay rate to
\begin{align}
\Gamma &\sim \frac{\abs{y}^2 k}{16 \pi} \frac{\Delta^2 k^2}{\pi^2}\frac{1}{(k R)^{3/2}} e^{- \frac{2 R (m_1+m_2)^3}{3 k^2}}\label{eq:massiverate}
\end{align}
using the suppression factor in Eq.~\eqref{a.14}, where $m_1,m_2$ are the masses of the two particles. The decay rate of the current depends on the curvature and environment of the strings. We will discuss individual situations in later sections.

Another potential leakage mechanism on loops is the collision of zero-modes moving past each other in opposite directions. This interaction is exponentially small in the radius of the loop $\sim e^{-m_\psi R}$, from the asymptotic form of the zero-mode solution for large $r_\perp$.


\section{Vortons}
\label{sec:Vorton}

An interesting consequence of the charge and current on the string is that it impedes the shrinkage of string loops. Suppose we start with a string loop with length $L$ and current $I$. We assume, for the moment, that as long as Eq.\,\eqref{eq:appCond} is satisfied, the loop shrinks and loses its energy as axion radiation\,\cite{Martins:2000cs}. As discussed in Sec.~\ref{sec:leak}, when the temperature of the universe is lower than $T_\text{leak}$, current leakage by scattering with the thermal plasma is suppressed. Below we will address the curvature induced leakage mechanism for loops, but since it becomes active only for small loop radii, we first analyse loop stability in the absence of this leakage mechanism. In the absence of (curvature induced) leakage, the charge on a string loop,
\begin{align}
\label{eq:7.1}
Q &= -IL, \qquad (\text{since}\ \rho =-I)
\end{align}
is conserved and constant. As the loop shrinks, the current grows and eventually saturates. This is seen from the total energy of the loop, which is the sum of the tension and the current energy for $L \gg f_a^{-1}$ and $I \ll f_a$. The latter is given
\begin{align}
\label{eq:7.2}
E \simeq N_c^2\,\frac{\varepsilon_{F}^2 L}{2\pi} = \frac{2\pi Q^2}{{e_\psi}^2L}.
\end{align}
The total energy is
\begin{align}
\label{eq:7.3}
E_{\text{tot}}(L) \simeq \mu L + \frac{2\pi Q^2}{{e_\psi}^2 L}.
\end{align}
where $\mu \simeq \mathcal{C} f_a^2$ is the tension of the string and we define $\mathcal{C}$ as
\begin{equation}
\mathcal{C} \equiv \pi \log(f_a L)\,.
\end{equation}
As discussed above, this formula is valid only for $I \ll e_\psi \sqrt{\mathcal{C}/2\pi} f_a$. However, we assume that this is even valid for $I \sim e_\psi \sqrt{\mathcal{C}/2\pi} f_a$ for the purpose of estimating the stabilization radius of string loops, as was done in previous studies\,\cite{Carter:1993wu,Brandenberger:1996zp,Martins:1998gb,Martins:1998th,Carter:1999an}. This was shown for the case of vector-like superconducting strings in \cite{Babul:1987me}, and we expect it to hold in our case as well.\footnote{The axion coupling is another gauge coupling and we expect it does not change the qualitative picture either.} Setting aside the logarithmic $L$ dependence of $\mathcal{C}$, the minimum of the energy of the loop is reached when
\begin{align}
\label{eq:vlen}
L \sim L_v \equiv \sqrt{\frac{2\pi}{\mathcal{C}}} \frac{Q}{{e_\psi} f_a}  
\end{align}
or equivalently 
\begin{eqnarray}
\label{eq:CurrentUpperBound}
I \sim I_v \equiv {e_\psi} \sqrt{\frac{\mathcal{C}}{2\pi}} f_a.
\end{eqnarray}
Consequently, the mass of the loop at the minimum is given as
\begin{eqnarray}
\label{eq:VortonMass}
M_v \simeq \sqrt{8\pi\mathcal{C}} \frac{Q f_a}{e_\psi}.
\end{eqnarray}
The loop stops shrinking once the current reaches this value. Such static loops are called  {\it vortons}, first predicted for the superconducting cosmic string in Ref.\,\cite{Davis:1988jq} and extensively studied in the '90s\,\cite{Carter:1993wu,Brandenberger:1996zp,Martins:1998gb,Martins:1998th,Carter:1999an}. Later, it was analytically shown that the static loops exist if the chiral current reaches a certain value\,\cite{BlancoPillado:2000ep,Davis:2000cx}. In addition to the trapped current, the vorton also has a cloud of charge 
carried by the Goldstone-Wilczek current of the gauge field around the vorton\,\cite{Kaplan:1987kh}. However, this contribution is suppressed by $N_c e_\psi^2 / 4\pi^2$, and we ignore it for simplicity.

To estimate whether our loops are indeed stable or decay as a result of current leakage mechanisms, we consider three different ways in which they can decay. First, we consider vorton decay by tunneling of the current across the vorton. The tunneling decay rate is of order $\sim e^{-m_\psi L_v}$ and is negligible for $L_v \gg m_\psi^{-1}$. Using Eq.\,\eqref{eq:vlen}, this condition for $L_v$ is equivalent to
\begin{eqnarray}
\label{eq:vortonCond}
Q \gg \sqrt{\frac{2\mathcal{C}}{2\pi}} \frac{e_\psi}{y_\Phi}.
\end{eqnarray}
Next, we consider the leakage from scattering against plasma particles /string oscillations, analysed in detail in section~\ref{sec:leak}. These lead to the dissipation rate
\begin{align}\label{eq:7.4}
\dot I &= - \Gamma_{\text{leak}}\, I
\end{align}
with $\Gamma_{\text{leak}}$ given in Eq.~\eqref{eq:5.10}. This is because the rate Eq.~\eqref{eq:5.10} is the rate to ionize a single zero-mode state. The total ionization rate is given by multiplying by $N_0$, the zero-mode occupation number, and the current loss rate is thus given by multiplying by $e_\psi N_0$. On the other hand, $I=e_\psi N_0$, leading to Eq.~\eqref{eq:7.4}. Consequently, for a string loop with large enough charge, the loops that form below $T_\text{leak}$ can potentially be stabilized by their current and become vortons. In contrast, loops formed at higher temperatures lose their current to plasma/oscillation leakage processes, and subsequently decay to axions. 

Finally, we consider the effect of curvature-mediated leakage, discussed in section~\ref{sec:curvmed}. To estimate this rate, we have to first know the typical stabilization radius of our potential vortons. If the curvature-mediated leakage rate is too large at this radius, our vortons will decay. 

To estimate the typical stabilization radius in the absence of leakage, we can make use of Eq.~\ref{eq:vlen}, provided that we know the value of the overall charge $Q$. We now show how to estimate it. The charge $Q$ is induced on the long strings/loops by thermal fluctuations in the early universe, which lead to an average electric field of $E_{z} \approx T^{2}$, with a coherence time of $1/T$. Eq.~\eqref{eq:LondonEq} then gives
\begin{eqnarray}
\label{eq:ThermalCurrent}
|I| \sim N_c\, e_\psi \,\frac{T}{2\pi}\,,
\end{eqnarray}
at a temperature $T$. $I$ fluctuates over a thermal coherence scale $T^{-1}$, and can have either sign. While fluctuations result in non-zero charges on the strings, global charge conservation is maintained via the Goldstone-Wilczek current. The local charge density is $\rho = -I$.
Thus the total charge on a string loop of initial length $L$ is
\begin{eqnarray}
Q(L) = \frac{N_c\, e_\psi \,\sqrt{L\,T}}{2\pi},
\end{eqnarray}
taking the r.m.s.\ average over $L\,T$ independent segments.
Thus, for $L \gg T^{-1}$, Eq.\,\eqref{eq:vortonCond} is satisfied and all loops can potentially end up as vortons below $T_\text{leak}$.
We estimate the initial loop length at temperature $T$ as the string correlation length $L = L_0 \equiv \xi^{-1/2} H^{-1}$, where $\xi$ denotes the number of the strings per Hubble volume. The vorton charge is then
\begin{align}
Q(L_0)\simeq 1.2\times 10^3 \,e_\psi\, \left(\frac{g_\star}{106.75}\right)^{-{\frac14}} \left(\frac{T}{10^8\,\text{GeV}}\right)^{-{\frac12}} \left(\frac{\xi}{10^6}\right)^{-{\frac14}}.
\end{align}

Here, since the charge on the string moves in the same direction, the Goldstone-Wilczek current is assumed to average the charge to the r.m.s.\ value. By Eq.~\ref{eq:vlen}-\ref{eq:VortonMass}, the vorton size and mass are then given by
\begin{align}~\label{eq:stablen}
L_v &= N_c\sqrt{\frac{\mathcal{C}}{2\pi}} \, \frac{\sqrt{TL}}{{e_\psi} f_a} \simeq 1.7 \times 10^3 f_a^{-1} \left(\frac{g_\star}{106.75}\right)^{-{\frac14}} \left(\frac{T}{10^8\,\text{GeV}}\right)^{-{\frac12}} \left(\frac{\xi}{10^6}\right)^{-{\frac14}} \sqrt{\frac{\pi }{ \mathcal{C}}}\\
\label{eq:vmass}
M_v &= \sqrt{8\pi\mathcal{C}}\,\frac{Q(L_\text{loop}) f_a}{e_\psi} \simeq 2.3 f_a \left(\frac{106.75}{g_\star}\right)^{\frac14} \sqrt{\frac{\mathcal{C}}{\pi}}\, \sqrt{\frac{M_p}{T}}\, \xi^{-\frac{1}{4}}.
\end{align}

The stabilization length Eq.~\eqref{eq:stablen} was derived while neglecting the curvature mediated leakage process of section~\ref{sec:curvmed}. To estimate the rate for this process, we can use the stabilization length as a proxy for the radius of curvature in Eq.~\eqref{eq:massiverate}.  

If we assume the daughter particles, the quark and Higgs, are massless, the decay rate of the current, Eq.~\eqref{eq:curvrate}, is to be compared with the Hubble expansion rate for $k \sim \frac{2\pi}{N_c e_\psi} I_v$ and $R \sim L_v$. If we take the KSVZ Yukawa to be $y \gg 10^{-15}$, so that the $\Psi$ decay well before $T_\text{BBN} \sim \text{MeV}$, Eq.~\eqref{eq:curvrate} then tells us that the current on the would-be vortons quickly dissipates, and they subsequently decay. If, on the other hand $y\gsim 10^{-15}$ so that the bulk $\psi$ decays at relatively late times before BBN, $T \lesssim \,10 \, T_\text{BBN}$, a dominant fraction of the vortons can still remain post BBN, which is in serious tension with observation\,\cite{Kawasaki:2004qu}. In appendix~\ref{sec:vortonapp}, we further estimate the resulting vorton density for this scenario. Our estimate is also relevant for other UV models for the KSVZ fermion which could potentially lead to the decay of the bulk $\Psi$ before BBN, while suppressing the leakage Eq.~\eqref{eq:curvrate}. 

We stress that our estimate for the curvature-mediated leakage is sensitive to the effective mass of the decay products. For example, we should take into account the plasma mass for the outgoing quark and Higgs. According to Eq.~\eqref{eq:massiverate}, the decay rate depends on these masses exponentially, i.e. $\Gamma\sim \exp\left[- L_v (m_q + m_H)^3 / (I_v/e_\psi)^2\right]$ with $m_{q(H)}$ the quark (Higgs) effective mass. (Un)fortunately, this dependence is not large enough to suppress the decay rate. The situation could be different if for some reason the decay products get an even bigger effective mass in the vicinity of the string - for example by filling the Landau levels from its induced magnetic field. We leave the careful estimation of effective masses for future work, bearing in mind that they could have important implications for the dynamics. In particular, if the effective masses of the decay products end up suppressing the curvature-induced leakage, the resulting metastable ``vortons'' could be in tension with BBN bounds. 

\section{Axion String Evolution in a Primordial Magnetic Field}\label{sec:PMF}
It is widely believed that the intergalactic magnetic field played a role in the formation of the $\sim 10\,\mu\text{G}$ galactic magnetic fields observed today, via the galactic dynamo effect \cite{Parker:1955zz}. The intergalactic magnetic field (IGMF) is characterized by two parameters, the magnitude of the magnetic field $B$ and the typical coherence length $\lambda$. An observational lower bound on the IGMF today is set by the non-observation of secondary photons from the emission of highly collimated gamma rays by blazars. Blazar-produced gamma rays collide with extra-galacting beckground (EBL) photons, and produce highly boosted electron-positron pairs. These, in turn, undergo inverse Compton scattering on the CMB, producing secondary photons which could in principle be detected, if they are collinear enough with the original gamma rays. However, in the presence of a large enough IGMF, the electrons and positrons bend and no longer produce collinear secondary photons.
Recent non-observations of secondary photon put a lower limit of $B \gtrsim 3 \times 10^{-16}\, \text{G}$ \cite{Neronov:1900zz} on the IGMF today for $\lambda \gtrsim 0.1\,\text{Mpc}$. For shorter coherence lengths, the lower bound becomes even stronger by the factor $\sqrt{0.1\,\text{Mpc} / \lambda}$. Though the exact origin of today's IGMF isn't known, one very attractive possibility is that it originated from a primordial magnetic field (PMF). There is a vast literature exploring the different ways a PMF could have been generated in the early universe, a process referred to as magnetogenesis\,\cite{Durrer:2013pga,Subramanian:2015lua}. The strongest upper bound on the PMF scenario is $B \lesssim 10^{-11}\,\text{G}$, obtained from CMB perturbations\,\cite{Jedamzik:2018itu}.

In this section, we point out that if a PMF is generated in the early universe, it induces very large currents on axion strings, significantly altering their evolution. As a result, the relic abundance of axions may change by many orders of magnitude. Our estimates are based on a simplified analytic model \cite{Martins:1996jp}, and so should be taken with a grain of salt until a full simulation is conducted. However, we believe that they capture the right physics, and should be explored further using numerical tools. 

For later convenience, we define the comoving magnitude of the magnetic field $\tilde{B} \equiv B R^2$ and the comoving coherent scale $\tilde{\lambda} \equiv \lambda R^{-1}$. The dynamics of the magnetic field in the early universe is governed by its interaction with the highly conducting primordial plasma, and is accurately described by magnetohydrodynamics (MHD). The cosmological evolution of plasma-coupled PMF is a complicated problem that was studied by many authors \cite{Banerjee:2004df,Durrer:2013pga}, and we will not describe it in detail. Here we only include a simplified discussion, and refer the readers to exhaustive reviews \cite{Durrer:2013pga,Subramanian:2015lua} for more details.

The key quantity in our simplified analysis is the Alfv\'en velocity $v_A \sim \sqrt{\rho_B / \rho_\text{tot}} \propto B$\,\cite{Banerjee:2004df}, where $\rho_B$ is the energy density of the magnetic field and $\rho_\text{tot}$ is the total energy density of the plasma in the early universe. As the universe evolves, the primordial plasma becomes turbulent at scales smaller than the sound horizon $v_A \tau$, where $\tau$ is conformal time. Consequently, $B$ fluctuations with a correlation length smaller than the sound horizon get damped (processed), while those with a longer correlation length remain unprocessed. Assuming the typical coherence scale is initially small enough, the coherence scale at conformal time $\tau$ is given by $\lambda = v_A \tau \propto B\tau$. This simple linear relation between $B$ and $\lambda$ is maintained as the universe expands.

The evolution of $B$ is determined by the the unprocessed component of the magnetic field and the MHD plasma. Let $n$ denote the spectral index of the unprocessed part of the magnetic field and the plasma; $\tilde{P}(k) \sim {\tilde{k}}^{-n}$ for the wavenumber smaller than $\tilde{\lambda}^{-1}$, where $\tilde{P}$ is the comoving power spectrum of the magnetic field and of the MHD plasma and is related to the comoving energy density of the MHD flow $\tilde{\rho}$ by $\tilde{\rho} \sim \tilde{P}(\tilde{\lambda}^{-1}) \tilde{\lambda}^{-3}$. Since $\tilde{\rho} \sim \tilde{B}^2$ at $\tilde{\lambda}$, their conformal time dependence is given as\,\cite{Durrer:2013pga}
\begin{eqnarray}\label{eq:eqlab}
\tau^{-1}\,\tilde{\lambda}~\sim~\tilde{B}~\sim~\tau^{-\frac{3+n}{5+n}}\equiv \tau^{-\alpha}\, .
\end{eqnarray}
The spectral index $n$ can take several different values depending on the dynamics of magnetogenesis. For simplicity, we focus on a PMF coupled to a compressible plasma, where $n = 0$, and the helical PMF, whose evolution is equivalent to the case $n = -2$\,\cite{Durrer:2013pga}:
\begin{itemize}
\item{\makebox[7cm]{\textit{Compressible plasma}:\hfill$n=0$} {\makebox[1cm]{}} $\alpha=\frac{3}{5}$}
\item{\makebox[7cm]{\textit{Helical PMF}:\hfill$n=-2$} {\makebox[1cm]{}} $\alpha=\frac{1}{3}$}
\end{itemize}
Another case we consider, is a PMF generated during inflation - and so is acausal today. The spectrum for this PMF is expected to be scale invariant,\footnote{Note, however, that building a consistent inflationary PMF model is known to be difficult\,\cite{Demozzi:2009fu}.} and so the typical coherence scale cannot be defined. However, by a slight abuse notation we define $\tilde{\lambda} = v_a \tau$. The evolution of the magnetic field and $\tilde{\lambda}$ is then described by
\begin{itemize}
\item{\makebox[7cm]{\textit{Scale Invariant}:\hfill$n=-3$} {\makebox[1cm]{}} $\alpha=0$}.
\end{itemize}
We stress that in reality the magnetic field in the inflationary scenario can be correlated on super horizon scales. In the following, whenever we mention the scale invariant magnetic field, we simply take it to be comovingly constant ($\tilde{B}$=const.) and coherent over the entire Hubble horizon.

When analyzing the effect of the PMF on axion strings in the early universe, we will make use of the average comoving magnetic field $\tilde{B}(T)$ as a function of temperature. It is useful to relate this quantity to the intergalactic magnetic field $B_0$ observed today. Naively, the magnetic field decouples from matter at recombination, and the comoving amplitude of the magnetic field remains roughly constant afterwards. From Eq.\,\eqref{eq:eqlab}, we have
\begin{eqnarray*}
\tilde{B}(T) = \tilde{B}_0 \left(\frac{\tau_\text{rec}}{\tau}\right)^{\alpha} \sim \tilde{B}_0 \left(\frac{T}{\sqrt{T_\text{eq}T_\text{rec}}}\right)^{\alpha},~~~(\text{naive})
\end{eqnarray*}
where $T_0$ is the current temperature of the photon and $T_\text{rec(eq)}$ is the temperature recombination (the matter-radiation equality).
Indeed, a detailed study\,\cite{Banerjee:2004df}, including viscosity effects and MHD dynamics post matter-radiation equality results in a similar conclusion for $T \gtrsim 10\,\text{MeV}$;
\begin{eqnarray}
\label{eq:PMFearly}
\tilde{B}(T) \simeq \tilde{B}_0 \left(\frac1{2.3 \times 10^{-12}} \cdot\frac{T}{100\,\text{GeV}}\right)^{\alpha}.
\end{eqnarray}
In this paper, we use the above formula for the PMF as a function of temperature. Note that the coherence length at each time is automatically determined from the Alfv\'en velocity once $\tilde{B}$ is fixed. For $T > T_\text{rec}$\,\cite{Banerjee:2004df},
\begin{eqnarray}
\label{eq:Bearly}
\tilde{\lambda}^{\text{mag}}_B~=~1.55\cdot10^{-4}\,\,\text{pc}~\sqrt{\frac{r(T)}{0.01}}~{\left(\frac{T}{100\,\text{GeV}}\right)}^{-1},
\end{eqnarray}
where $r(T)$ is the ratio between the magnetic energy and a power of the entropy density\,\cite{Durrer:2003ja}, 
\begin{align}
r \equiv \frac{R^{-4} \tilde{B}^2}{10} \frac{1}{s^{4/3}}   .
\end{align}
The ratio $r$ would have been a constant if not for the coupling of the PMF to the primordial plasma.\footnote{In Ref\,\cite{Banerjee:2004df}, the magnetic constant is defined as $\mu_0 \equiv 4\pi$ to convert from SI units to  natural units. The value of $\mu_0$ is always arbitrary and we define $\mu_0 \equiv 1$ in this paper.} After recombination, the coherence length is frozen as well. In terms of $\tilde{B}_0 = B_0$, the coherent length of the PMF is\,\cite{Neronov:1900zz,Banerjee:2004df,Kamada:2018kyi}
\begin{align}
    \lambda_0 \simeq 11\,\text{pc} \left(\frac{B_0}{10^{-13}\,\,\text{G}}\right).
\end{align}
Combining this with the lower bound on the IGMF introduced in the beginning of this section, we obtain
\begin{align}
    B_0 \gtrsim 0.4 \times 10^{-13}\,\text{G}.
\end{align}
Thus, we take $10^{-13}\,\text{G}$ as the reference value of the causal ($n\ne -3$) PMF today. For acausal (inflationary) PMF, we take the direct lower bound $3 \times 10^{-16}\,\text{G}$, as the PMF is assumed to be coherent on a $0.1\,\text{Mpc}$ scale.

Having described in some detail the dynamics of a PMF in the early universe, we turn to consider its effect on the cosmic evolution of QCD axion strings. For each one of the PMF scenarios we consider, we vary the initial magnetic field\footnote{We use the subscript ``mag'' to denote values at the time of magnetogenesis.} $B_\text{mag}$ so that its value today is consistent with observations. Since axion strings are superconducting, their interaction with the PMF yields sizeable electric currents, which, as we shall see, back-react on their evolution. 

By Lorentz symmetry, a boosted magnetic field becomes an electric field. When the axion string moving with the velocity $v_s$ is embedded in the PMF, an electric current develops on the string generated by the effective electric field $E \sim B v_s$. The evolution of the current is governed by the electric field, Hubble expansion and  current dissipation, we obtain the current evolution equation\,\cite{Thompson:1988jn}, with the addition of a leakage term
\begin{equation}\label{eq:Bgrowth}
\dot{I} + I' = \frac{e_\psi^2}{2\pi}v_s B- \beta H I-\Gamma_{\text{leak}}I\,
\end{equation}
where $\dot{I}$ denotes the derivative with respect to the physical time and $I'$ does the derivative with respect to the spacial co\"ordinate on the string.
The first term on the right hand side is a direct consequence of Eq.~\eqref{eq:london2}, while the second term accounts for the redshift of the current due to Hubble expansion. The factor $\beta$ encodes the effect of the small scale structure of the string; if the string is infinitely long, the current redshift is as usual, $\beta = 1$ whereas if the string is very wiggly, Hubble expansion straightens the wiggles first so that $\beta \sim 0$. We assume $\beta \sim 1$ in this paper.  The current destruction rate $\Gamma_{\text{leak}}$ is given by Eq.~\eqref{eq:5.8} and Eq.~\eqref{eq:dest.r}, and we conservatively assume $y = y_\Phi = 1$ in the following calculations, corresponding to an unpressed leakage rate. Here, we may ignore the curvature-induced current decay, Eq.\,\eqref{eq:massiverate}, since the radius of curvature of the long strings is sufficiently small. We elaborate on this point below.
For simplicity, before solving Eq.~\eqref{eq:Bgrowth} numerically, we first take its root-mean-square average over a Hubble time. Since the string velocity is coherent over the typical inter-string distance, $L\equiv t/\sqrt{\xi}$, the root-mean-square current $I_{\text{RMS}}(t)$ includes a suppression factor of $\sqrt{\ell_\text{coh} \tau_\text{current}}$ with $\ell_\text{coh} = \min(\lambda, L)$ and $\tau_\text{current}\equiv \min(H^{-1}, \Gamma_{\text{leak}}^{-1})$.

In writing Eq.~\eqref{eq:Bgrowth}, we implicitly assumed that the main mode of energy loss by the strings is string reconnections, which results is a loss of both string length and charge. However, if some of the energy of the strings is directly radiated into axions without the loss of charge, this could induce an additional blue-shift factor, which enhances the current. This effect has been previously considered in \cite{Mijic:1988ag,Carter:2000fv}. However, we use the above formula as a conservative estimate.

According to Eq.\,\eqref{eq:Bgrowth}, the current at a given Hubble time is roughly
\begin{eqnarray}
I(T) \simeq \frac{e_\psi^2}{2\pi}\,v_s\, B_\text{eff} \,\tau_\text{current} \sim \frac{e_\psi^2}{2\pi}\,v_s \sqrt{\ell \tau_\text{current}^{-1} r(T)}\, T^2\, \tau_\text{current},
\end{eqnarray}
where $B_\text{eff} \equiv B \sqrt{\ell_\text{coh} \tau_\text{current}}$.
If it were not for friction and $\ell_\text{coh} \sim \tau_\text{current}$, $v_s$ would be $\mathcal{O}(1)$ and $I \sim \frac{e_\psi^2}{2\pi} \sqrt{r(T)} M_p$, which would be much larger than the thermal fluctuation given in Eq.\,\eqref{eq:ThermalCurrent}. Thus, potentially, the current on the axion string by the PMF may significantly affect the evolution of the string.
Of course, the current on the string induces friction, as discussed on Sec.\,\ref{sec:Vorton}. The string velocity is decelerated by the friction force as Eq.\,\eqref{eq:StrVelFric}. Thus, in order to estimate the current, we need to follow the evolution of the string including friction and evaluate the velocity.

\subsection{String Evolution with the PMF}\label{sec:PlasmaFriction}
The early universe evolution of cosmic string networks has been the subject of a vast number of analytic and numerical analyses. These have mostly been performed for non-superconducting local and global strings, as well as for local superconducting strings. We currently do not know of any dedicated simulations of the properties of anomalous superconducting global strings, and so we encourage the study of their reconnection and emission properties. In the meantime, we extrapolate some of these properties from those of global/local-superconducting strings. In particular, we assume that:
\begin{itemize}
\item Their reconnection properties are similar to those of local superconducting strings and global strings up to $\mathcal{O}(1)$ corrections.
\item Their axion emission rate is similar to that of non-superconducting global strings - since it is orthogonal to the dynamics of the zero mode currents on the strings.
\item The plasma friction on the superconducting strings is similar to that of non-anomalous global superconducting strings \cite{Dimopoulos:1997xa}. This should not be different for axion strings, since it is not linked to the PQ symmetry being anomalous. 
\end{itemize}

Under these assumptions, we can analyse the early universe evolution of superconducting axion strings using the analytic velocity dependent one-scale model (VOS)\,\cite{Martins:1996jp,Martins:2000cs}, which averages the time evolution of the string energy and velocity using the Nambu-Goto action. In terms of the average inter-string distance $L$ and the average string velocity $v_s$, the equations are
 \begin{align}\label{eq:VOS}
2\frac{dL}{dt} &= 2LH(1+v^2_s) + c_L v_s + v^2_s L \tau^{-1}_\text{fric}, \\ \nonumber
\frac{dv_s}{dt} &= (1-v^2_s)\left[\frac{k_s}{L} - v_s\left(2H + \tau^{-1}_\text{fric} \right)\right].
\end{align}
In the first equation, the first term on the right hand side is from Hubble expansion. The second term encodes the effect of string loop emission by string reconnection. $c_L$ is a $\mathcal{O}(1)$ factor of the loop-chopping efficiency, taken as $c_L\sim 0.66$\,\cite{Martins:2018dqg}, and we assume it remains constant. The third term is from friction due to the current. The friction relaxation timescale, $\tau_\text{fric}$, is defined in Eq.\,\eqref{eq:FricTime} for the thermal current, but the formula also holds for any current, regardless of its source. As for the second equation, here the first term is from acceleration of the string by its curvature, which depends on the small scale structure of the string, encoded in the factor $k_s = 0.25$ \cite{Martins:2018dqg}. The second term is deceleration by Hubble expansion and friction.

Finally, we note that the VOS model for global strings has been discussed in \cite{Martins:2018dqg}, and it involves two modifications to Eq.~\ref{eq:VOS}. The first is the logarithmic dependence of the string tension on the correlation length $L$. We have explicitly incorporated this effect into all of our equations. The second is an extra term $sv^6/\mathcal{C}$ on the right hand side of the $L$ equation, representing the energy loss axion radiation as in \cite{Martins:2018dqg}. We checked that it leads to a negligible modification of our dynamics (it is subleading with respect to $c_L v$), and so we omit it for simplicity. Lastly, we checked that $\mathcal{O}(1)$ changes in the values of $c_L$ and $k_s$ lead to $\mathcal{O}(1)$ changes in the resulting number of strings. Below we sill see that the current leads to at least a 4 order of magnitude effect in the number of strings, and so we simply fix our values for $k_s$ and $c_L$ to those of \cite{Martins:2018dqg}. 

Since $\tau_\text{fric}$ is determined by the current on the string as calculated in Eq.\,\eqref{eq:FricTime}, we combine the Hubble-time average of Eq.~\eqref{eq:Bgrowth} and Eq.~\eqref{eq:VOS} to determine the evolution of our current carrying string network embedded in a PMF. For the PMF time evolution, we simply assume that the magnetic field instantaneously appears at $T = T_\text{init}$ and evolves according to Eq.\,\eqref{eq:PMFearly} for the causal PMF, $n \ne -3$. In the following, we vary $T_\text{init}$ from around $10\,\text{GeV}$ to $10^3\,\text{GeV}$ as an example. In this setup, the string evolution 
experiences four distinct stages: (1) the evolution before the PMF appears (2) a transient stage in which the current grows and the string decelerates (3) the stretching phase (4) the friction dominated regime. We find the string never enters the scaling regime after a PMF appears. Below we introduce these stages, and the key processes that characterize them.

In Eq.~\eqref{eq:Bgrowth}, we ignored the curvature-induced current decay. Typical curvature of strings are of order the string correlation length, $L$. Since the string network is in the thermal bath, the standard model quark and Higgs have the thermal masses of order $g_s^2T$ and $g_WT$, respectively, where $g_{s(W)}$ is the strong (weak) gauge coupling. Thus, we may ignore Eq.\,\eqref{eq:massiverate} if the momentum of the current $k$ satisfies $k \ll \sqrt{L T^3}$. In the calculation presented below, we confirmed that this condition is satisfied.

\subsubsection{Evolution before the PMF appears}
Before the PMF appears, the evolution of the axion string follows the standard evolution\,\cite{Kibble:1980mv,Martins:1996jp,Martins:2000cs,Vilenkin:2000jqa,Zurek:1985qw,Murayama:2009nj}, with the extra effect of the thermally induced current on the string. According to the standard picture, the string evolution consists of two phases, the friction dominated regime and the scaling regime.
\begin{itemize}
\item \textit{Friction dominated regime.}\\
Just after the phase transition, numerous strings appear\,\cite{Zurek:1985qw,Murayama:2009nj}. They strongly interact with each other and quickly decrease forming loops. The number of strings immediately stops decreasing after the phase transition , as friction due to the thermal current becomes effective, and so the strings stop moving. As a result, the number of the string per Hubble horizon, $\xi$, obeys Eq.\,\eqref{eq:xiThermFric}.
As the temperature of the universe drops, friction becomes weaker and eventually negligible. The Kibble temperature, the temperature when friction becomes negligible, is when $\tau_\text{fric} \sim H^{-1}$. To be explicit, it is 
\begin{eqnarray}
\label{eq:KibbleTemp}
T_\text{Kibble} = \sqrt{\frac{\pi^2g_\star}{90}} \frac{\mu}{M_p} \simeq 140\, \mathcal{C} \left(\frac{g_\star}{106.75}\right)^{1/2}  \left(\frac{f_a}{10^{10}\,\text{GeV}}\right)^2 \,\text{GeV}\,.
\end{eqnarray}
\item \textit{Scaling regime.}\\
Below $T_{\text{Kibble}}$, the universe is cold enough and friction is negligible. The string velocity becomes $\mathcal{O}(1)$ constant and only $\xi = \mathcal{O}(1) = \text{constant}$ strings exist in one Hubble volume due to efficient reconnection. As a result, the ratio between the string and radiation energy is constant. This is the scaling regime of the string. 
Using the VOS model, one may calculate the typical distance and velocity of the string assuming $L \propto t, v = \text{constant}$ as
\begin{align}\label{eq:Scaling}
L_{\text{scaling}}(t)~&=~\sqrt{k_s\,(k_s+c_L)}\,t \,, &
v_{\text{scaling}}(t)~&=~\sqrt{\frac{k_s}{k_s+c_L}}\,.
\end{align} 
Note that in recent years there has been a growing interest in a potential logarithmic violation of this scaling law, $\xi > 1$, backed by dedicated simulations \cite{Gorghetto:2018myk,Vaquero:2018tib,Buschmann:2019icd,Hindmarsh:2019csc,Klaer:2019fxc,Saikawa_IPMU_slide,Gorghetto:2020qws,Yamaguchi:2005gp,Martins:2018dqg} though we do not discuss this further.
\end{itemize}

\subsubsection{Transient stage: current growth and string deceleration} 

\begin{figure*}[t]
\begin{center}
\includegraphics[width=0.485\linewidth]{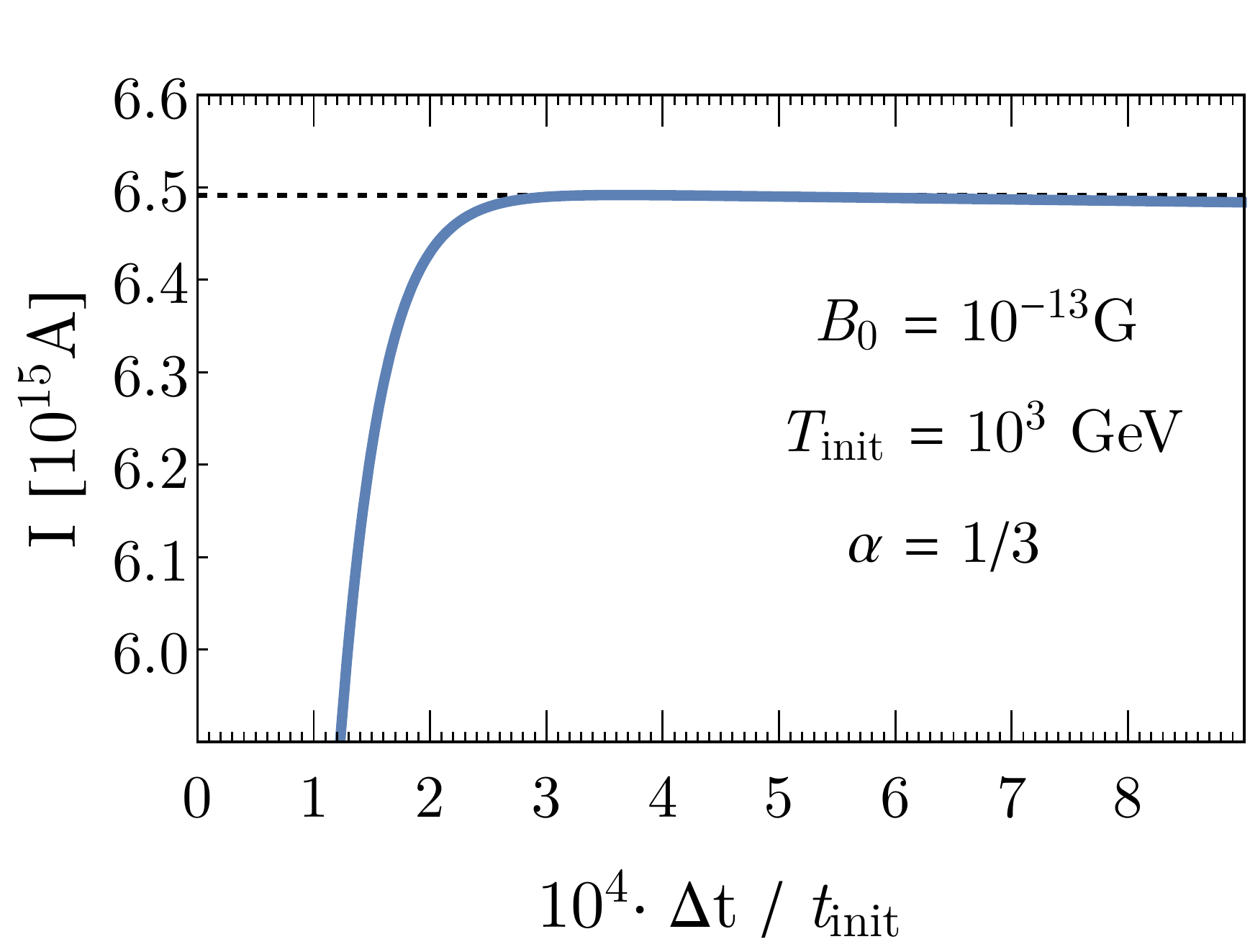}~
\includegraphics[width=0.5\linewidth]{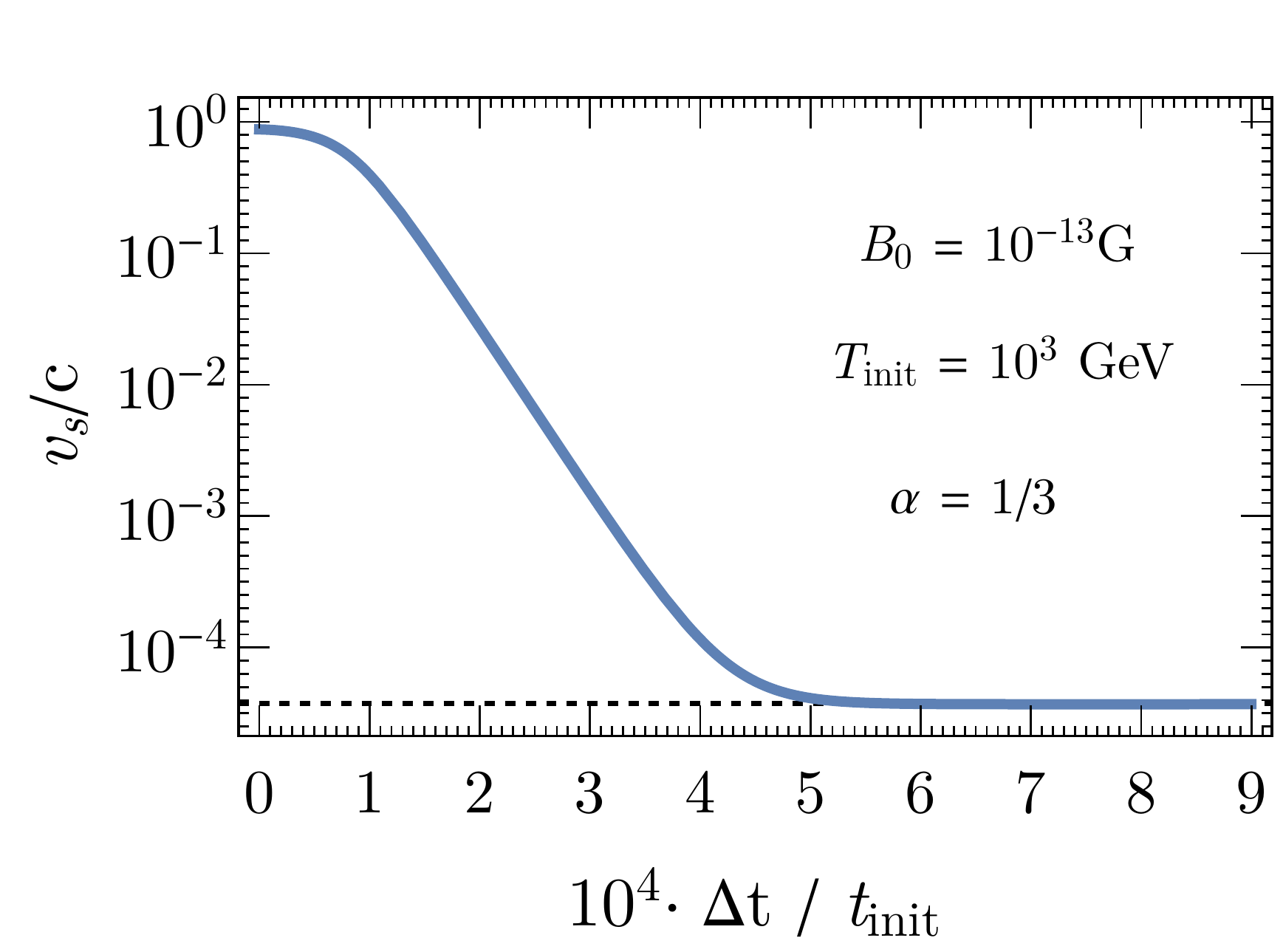}
\caption{Evolution of $I$ and $v$ in the transient stage after $t_{\text{init}}$. \textit{Left}: $I(t)$. \textit{Right}: $v_s(t)$. The dashed lines indicate the plateau values from Eq.~\ref{eq:plat}. $f_a$ is taken to be $10^{10}\,\text{GeV}$.}\label{fig:stage1}
\end{center}
\end{figure*}

At time $T_\text{init}$, the current growth is ``turned on'' according to Eq.~\eqref{eq:Bgrowth}. Then the current grows rapidly, and there is large plasma friction on the string. The friction relaxation time $\tau_\text{fric}$ is much shorter than the Hubble time leading to string deceleration, which in turn feeds back on the current growth equation via its velocity dependence. We call this the transient stage. The whole process ends in a much shorter period than the Hubble time. Since the VOS equation, Eq.\,\eqref{eq:VOS} assumes time-averaging over scales $\sim H^{-1}$, such shorter time scale physics cannot be discussed using the VOS equation. For simplicity, we assume $L$ does not change during the transient stage and solve Eq.~\eqref{eq:Bgrowth} taking spatial averages for $I$ and $v_s$. In Fig.~\ref{fig:stage1} we present the numerical solution of this initial stage. As we can see in the plot, the current $I$ grows rapidly until it reaches a plateau, and the velocity decreases accordingly. Indeed, the transient stage ends very quickly, which justifies our assumption. We can estimate the plateau values of $I$ and $v$ by taking $R,\,H,\,B,\,\lambda$ and $L$ as constants equal to their values at $t_{\text{init}}$. The plateau values are then obtained by demanding  
 \begin{align}\label{eq:plat}
0&=\frac{dv_s}{dt} = (1-v^2_s)\left[\frac{k_s}{L}\,-\,v_s\left(2H\,+\,\tau_\text{fric}^{-1}\right)\right]\,, \nonumber\\[10pt]
0&=\frac{d I}{dt} = \frac{e_\psi^2}{2\pi} v_s B-\Gamma_\text{leak}I - \beta H I\, \, .
\end{align}
We also show these plateau values for $v_s$ and $I$ in Fig.~\ref{fig:stage1}. Numerically, we use these values as initial conditions of the next phase.

\subsubsection{Stretching phase} 

The initial transient phase terminates when $v_s$ reaches a temporary plateau at $v_s\ll1$. At this point reconnections are suppressed, because the mean inter-string distance is larger than $v_s H^{-1}$, and so the number of the string per {\it comoving} Hubble volume is constant. This phase is called the stretching regime, since the string is stretched by the Hubble expansion. Let us see how this dynamics emerges from in VOS model. Neglecting the velocity dependent terms in the $L(t)$ Eq.~\ref{eq:VOS}, we obtain
\begin{align}\label{eq:VOSstretch}
\frac{dL}{dt} &= LH\,.
\end{align}
Using $H = 1 / (2t)$, the solution of the above equation is simple; $L(t)\sim \sqrt{t}$. Indeed, the number of strings per comoving Hubble volume, $(aH)^{-2} / L(t)^2 \sim t^0 = \text{constant}$.

\subsubsection{Plasma friction domination} 

The stretching regime ends when friction becomes weaker, $v_s$ increases, and $v_s H^{-1}$ becomes comparable to the typical string distance. At this point reconnections commence, but the plasma friction is still strong and the whole network evolution again follows the friction dominated regime. The typical string distance and velocity asymptotically follow  Eq.\,\eqref{eq:xiThermFric} with $L = t / \sqrt{\xi}$. Let us make a rough estimate of $\xi$. From Eq.\,\eqref{eq:xiThermFric}, we obtain
\begin{align}
\xi_0 &= \sqrt{\dfrac{90}{\pi^2 g_\star}} \dfrac{I M_p}{\mu}, &
v_{s0} &= \left(\sqrt{\xi}\right)^{-1} = \left( \dfrac{\pi^2 g_\star}{90} \right)^{1/4} \sqrt{\dfrac{\mu}{I M_p}},
\end{align}
where the subscript $0$ denotes the asymptotic value. On the other hand, the current is
\begin{align}
    I_0 \simeq \frac{e_\psi^2}{2\pi} B_\text{eff} v_{s0}\tau_\text{current}.
\end{align}
Thus, we expect
\begin{align}
\label{eq:xiAsymp}
    \xi_0 \simeq \left(\frac{e_\psi^2}{2\pi} B_\text{eff} \tau_\text{current}\right)^{2 / 3} \left(\sqrt{\dfrac{90}{\pi^2 g_\star}} \frac{M_p}{\mu}\right)^{2/3}
\end{align}
as the asymptotic value. For temperatures lower than $T_\text{leak}$, current dissipation is ineffective and we expect
\begin{align}
    \xi_0 \propto B^{2/3} \ell_\text{coh}^{1/3} f_a^{-4/3}.
\end{align}
Note that for all regions of interest, $\xi \gg 1$ and the axion string {\it never} enters the scaling regime if the PMF exists, as we numerically show later.

\subsubsection{Numerical result} 
In Fig.~\ref{fig:xires} we plot the number of strings per Hubble volume $\xi=(t/L)^2$ and the velocity $v_s$ as a function of temperature. We take $f_a = 10^{10}\,\text{GeV}$, $T_\text{init} = 10^{3}\,\text{GeV}$ and $B_0=3\cdot10^{-13}\,\text{G}$ as an example and show the result for $\alpha = 1/3$ and $3/5$. For comparison, we also show $\xi$ and $v_s$ without the PMF.

The figure demonstrates the dynamics discussed above. First, before $T_\text{init}$, the axion string is at the onset of the scaling regime. Without the PMF, the thermal current decays as the temperature drops and the string network enters the scaling regime, although there is an $\mathcal{O}(1)$ change in $\xi$, Eq.\,\eqref{eq:xiThermFric}. However, if the PMF appears, it produces a large current on the string and the string is immediately decelerated by plasma friction. For several orders of magnitude in temperature after $T_\text{init}$, the string velocity is too small, and string reconnections are suppressed. Thus, the number of strings per comoving Hubble horizon is conserved and $\xi$ grows as $t \sim T^2$. This regime of string evolution is called the ``stretching phase.'' As the universe gets cold, the PMF and friction force become weaker, allowing the string to accelerates because of its curvature. Eventually, the string network starts to reconnect again and the network goes over to another friction dominated regime. The temperature at which the stretching phase ends depends on the magnitude of the current on the strings, which in turn, depends on the PMF. The larger $\alpha$ is, the faster the PMF decays. Since we assume the present magnitude of the PMF is the same for both $\alpha = 1/3$ and $3/5$, the PMF at early times is stronger for larger values of $\alpha$. Thus, axion strings embedded in a PMF with $\alpha = 3/5$ spend more time in the stretching regime.

For causal magnetic fields, the temperature of magnetogenesis is bounded from above by the requirement that the PMF does not dominate the expansion of the universe. For that reason, we take $T_\text{init} = 10^{3}\,\text{GeV}$, which is the relevant upper bound for $\alpha=3/5$. For this value of $T_{\text{init}}$, the $\alpha = 1/3$ case enters the friction dominated regime at around $T\sim10\,\text{GeV}$, while in the $\alpha=3/5$ case, the network is approximately still in the stretching regime as late as $1\,\text{GeV}$. In comparison, if $T_\text{init}$ is lower than $10^3\,\text{GeV}$, the network is still in the stretching regime at the time of domain wall formation. Eventually, The tension of the domain wall wins over the friction force and the string-domain wall network shrinks and fragments.

\begin{figure}[t]
\begin{center}
\includegraphics[width=0.49\linewidth]{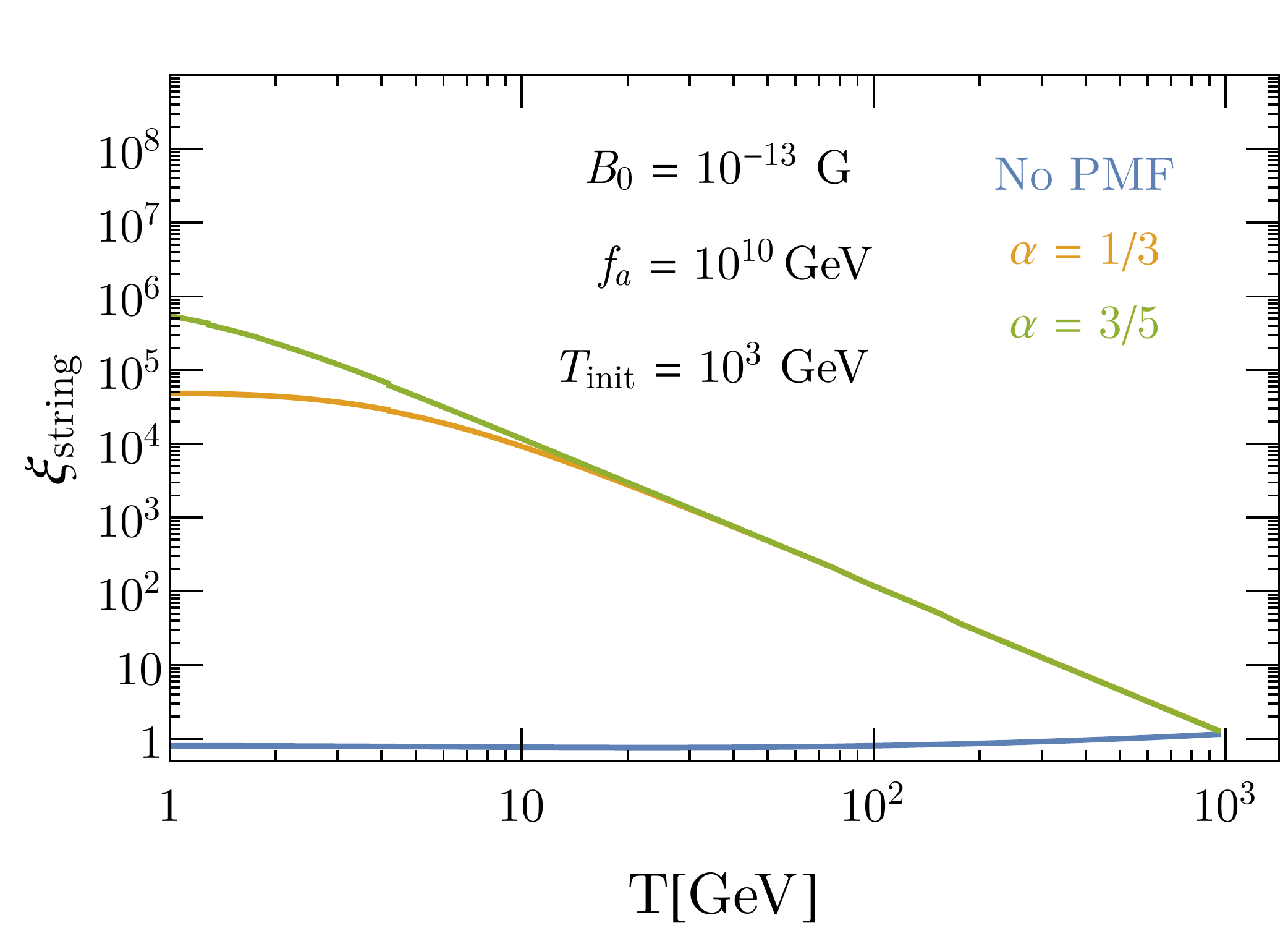}~~
\includegraphics[width=0.5\linewidth]{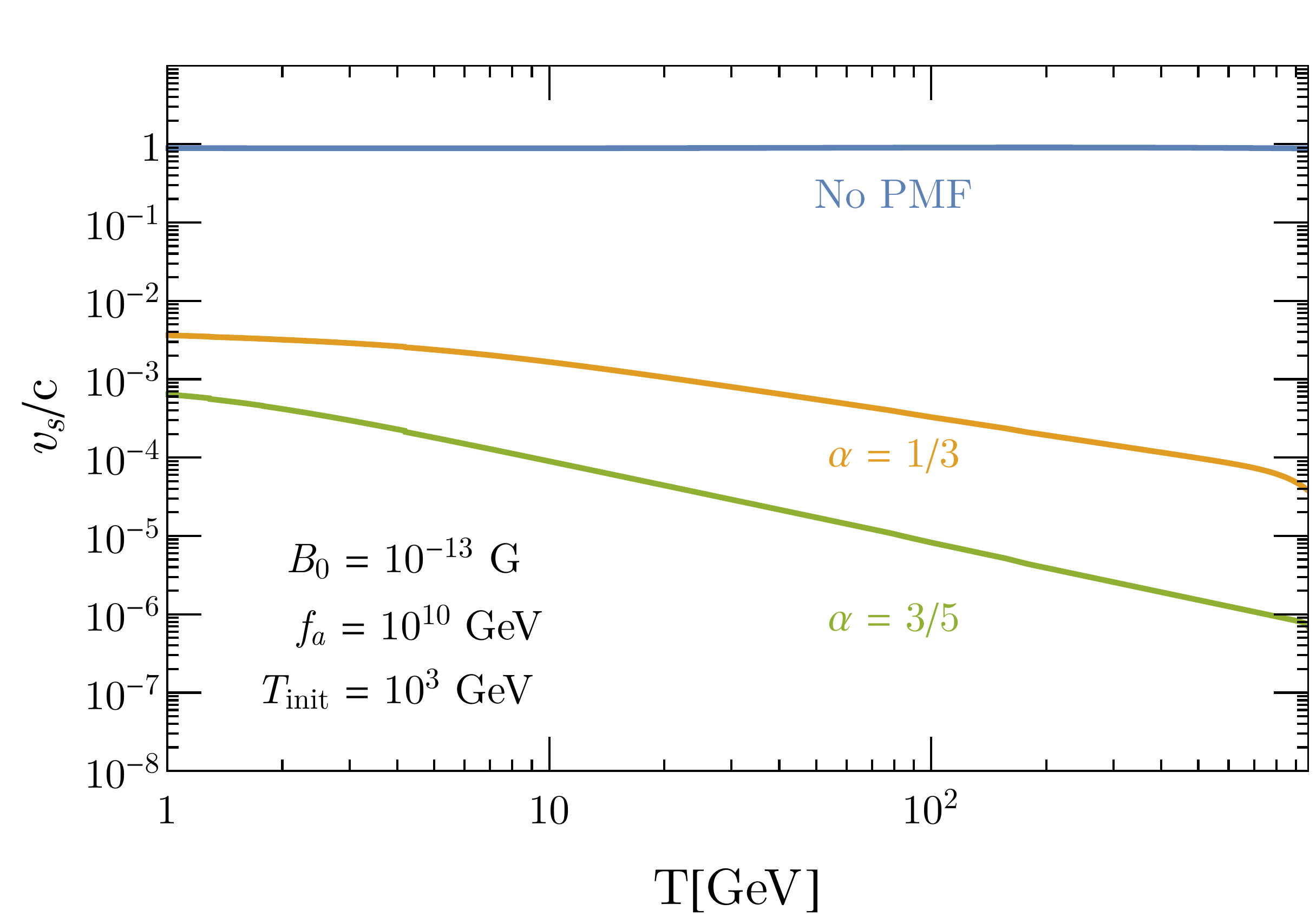}
\caption{Evolution of current carrying axion strings vs. temperature. \textit{Left}: Number of strings per Hubble volume $\xi_{\text{string}}$. \textit{Right}: Average string velocity $v_s$. The velocity drops immediately when the current is turned on, and then starts growing again due to string curvature.}
\end{center}
\end{figure}\label{fig:xires}

\subsection{Axion Relic Abundance}

Here, we estimate the relic density of axion radiated from axion strings embedded in a PMF. Naively, an increased number of strings per Hubble in leads to a proportional increase in the number density of radiated axions. However, as reference \cite{Gorghetto:2020qws} points out, the actual increase in the axion relic density is roughly proportional to $\sqrt{\mathcal{C}\,\xi}$. The reason for that is as follows:
The contribution to the relic abundance is biggest when the axion mass starts affecting the dynamics. Since the inter-string distance is $H^{-1}/\sqrt{\xi}$, each emitted axion carries a momentum of order $\sqrt{\xi} H$ and the number of emitted axions is then $\xi \mu H^{-1} / \sqrt{\xi} H \sim \sqrt{\xi}$. Assuming the number of axions is conserved, the final relic abundance is proportional to  $\sqrt{\xi}$. In reality, three corrections should be added to the above estimates. First, since the tension of the axion string is not constant but proportional to $\mathcal{C}$, we have to include this factor in our estimate. Secondly, we need to take into account how the axion mass varies with temperature. As a result, Ref.\,\cite{Gorghetto:2020qws} concludes
\begin{align}\label{eq:marcoest}
  \Omega_{\text{ax}}\propto(\mathcal{C}\, \xi)^{\frac{1}{2} + \frac{1}{4 + \gamma}},
\end{align}
where $\Omega_{\text{ax}}$ is the relic abundance of the axion today,  $\gamma$ is the temperature dependence of the axion mass at around $\lesssim 1\,\text{GeV}$, $m_a \propto T^{-\gamma / 2}$ and we take $\gamma\simeq 8$. The estimate Eq.~\eqref{eq:marcoest} has also been validated by numerical simulations in \cite{Gorghetto:2020qws}.

Finally, we have to account for the different emission properties of current-carrying axion strings with respect to current-less axion strings. We do this in two stages. First, we need to account for the different reconnection/loop emission rate in our scenario. This can be directly accounted for by comparing the energy loss rate of our long strings to that of standard axion strings. To account for this difference, we multiply Eq.~\ref{eq:marcoest} by the suppression factor
\begin{equation}
\mathcal{S}\equiv\Gamma_{a}/(\rho_s/t)\, , 
\end{equation}
which could be as small as $10^{-2}$ in the stretching regime.\footnote{The VOS equation is for scales around $L$. The suppression factor is, as we show, much smaller than $\mathcal O(1)$, which suggests smaller scale physics might be relevant.} 

\begin{figure*}[t]
\begin{center}
\includegraphics[width=0.7\linewidth]{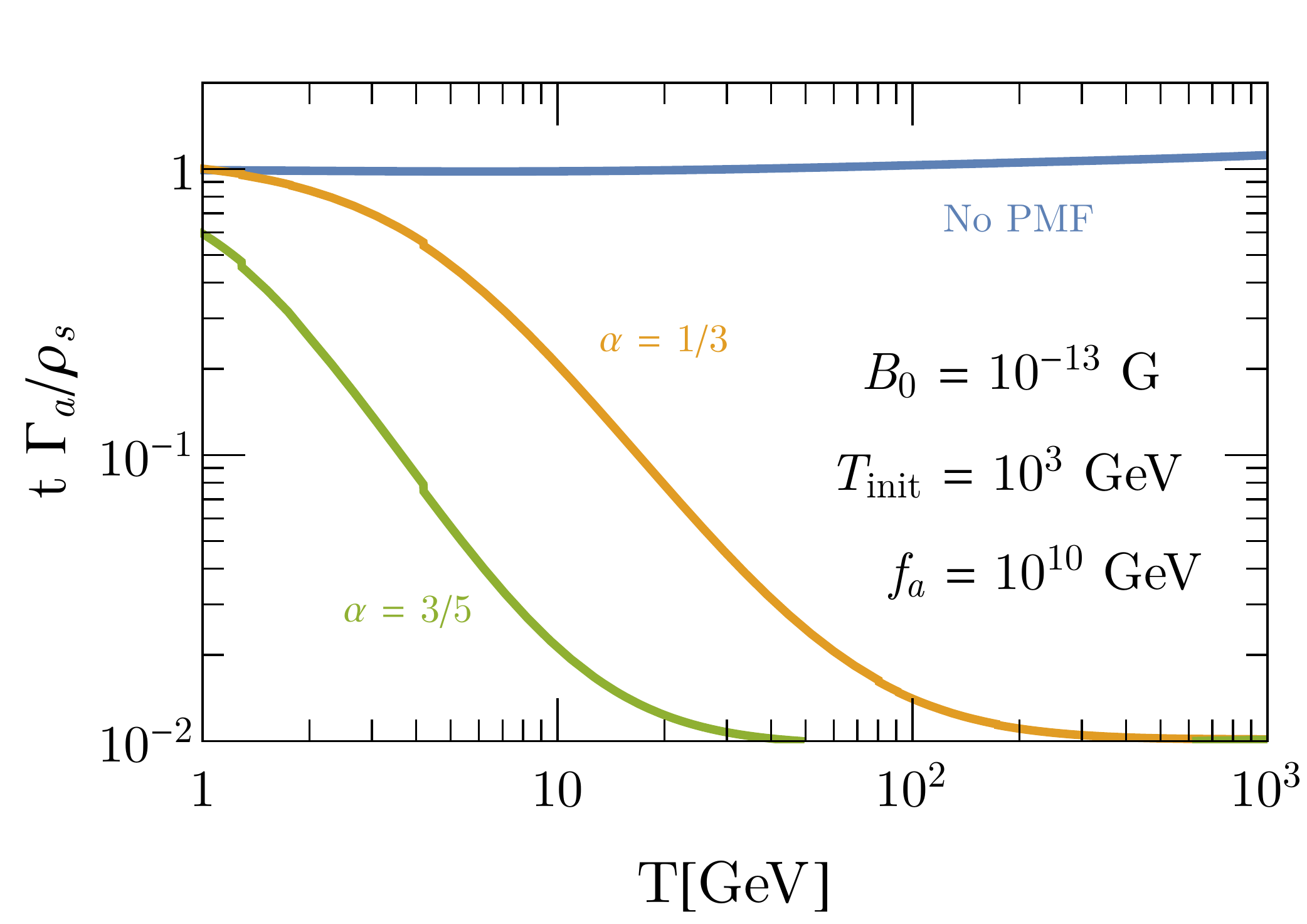}
\caption{Energy emission rate from axion strings into the axion field, normalized by the approximation $\rho_s/t$. In the friction dominated regime, this ratio becomes unity. In the stretching regime, the actual rate is suppressed by up to $10^{-2}$ due to the low reconnection rate.}\label{fig:gamma}
\end{center}
\end{figure*}

To calculate $\Gamma_{a}$ in our scenario,
we evaluate $\dot{\rho}_s$ for our network evolution, as compared to $\dot{\rho}^{\text{free}}_s$, an equivalent network of non-interacting strings with the same instantaneous string density at time $t$, $\rho_s|_{t}=\rho^\text{free}_s|_{t}$. The difference between the two terms is the instantaneous energy radiated into loops, and eventually into the axion field \cite{Gorghetto:2018myk}:
\begin{equation}
\Gamma_a~=~\dot{\rho}^{\text{free}}_s-\dot{\rho}_s\, .
\end{equation}
In Fig.~\ref{fig:gamma} we present the rate $\Gamma_a$ calculated in this manner, for the network evolution described in Fig.~\ref{fig:xires}. As can be seen in the plot, this rate is suppressed in the stretching regime, but grows to 1 in the friction dominated regime. Note that even in the presence of plasma friction, the dominant source of energy loss for our strings is \textit{reconnections}, which go like $c_L v_s$ in Eq.~\ref{eq:VOS}, and \textit{not} plasma friction, which enters as $v^2_s$ in the equation for $L$ (equivalently, the string energy). This is directly evident from Fig.~\ref{fig:gamma}.

Finally, we need to account for the different emission properties of \textit{loops} in our scenario with respect to the standard case, where all of the loop energy is effectively radiated to axions. In our case, however, the energy of the shrinking loops can also be transferred to the plasma by the friction. If this effect dominates, it leads to a large suppression in the density of radiated axion DM. This is an important point to explore in future work, but it is beyond the scope of the current paper. In particular, the resulting axion relic density strongly depends of loop emission properties at later time, $T\sim 1\,\text{GeV}$. In this case the effect of domain walls must be taken into account. Here we do not delve further into this analysis, but simply assume that axion emission prevails as the dominant energy loss for loops at $T\sim 1\,\text{GeV}$.

\begin{figure*}[htbp]
\begin{center}
\includegraphics[width=0.52\linewidth]{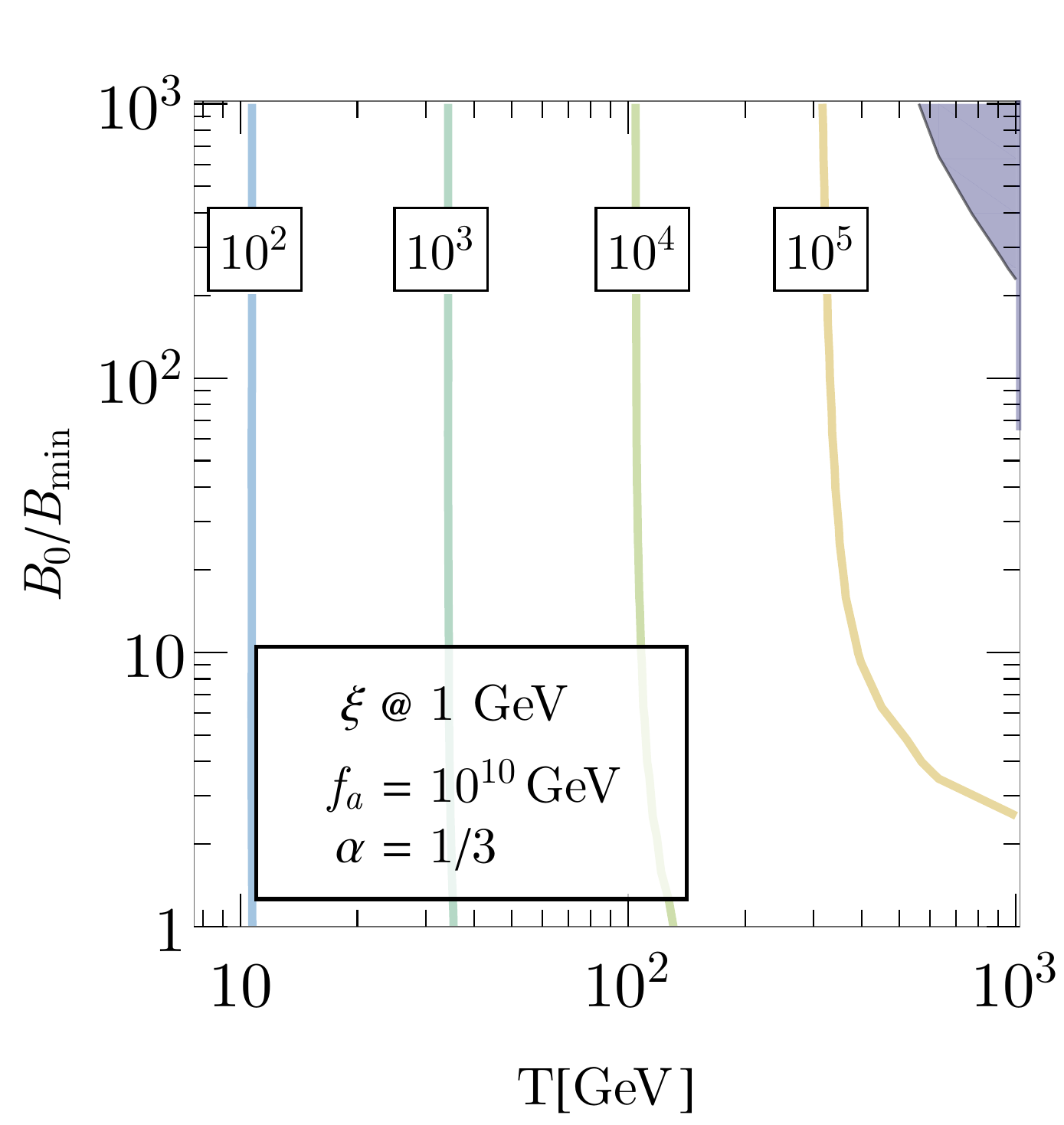}~~
\includegraphics[width=0.5\linewidth]{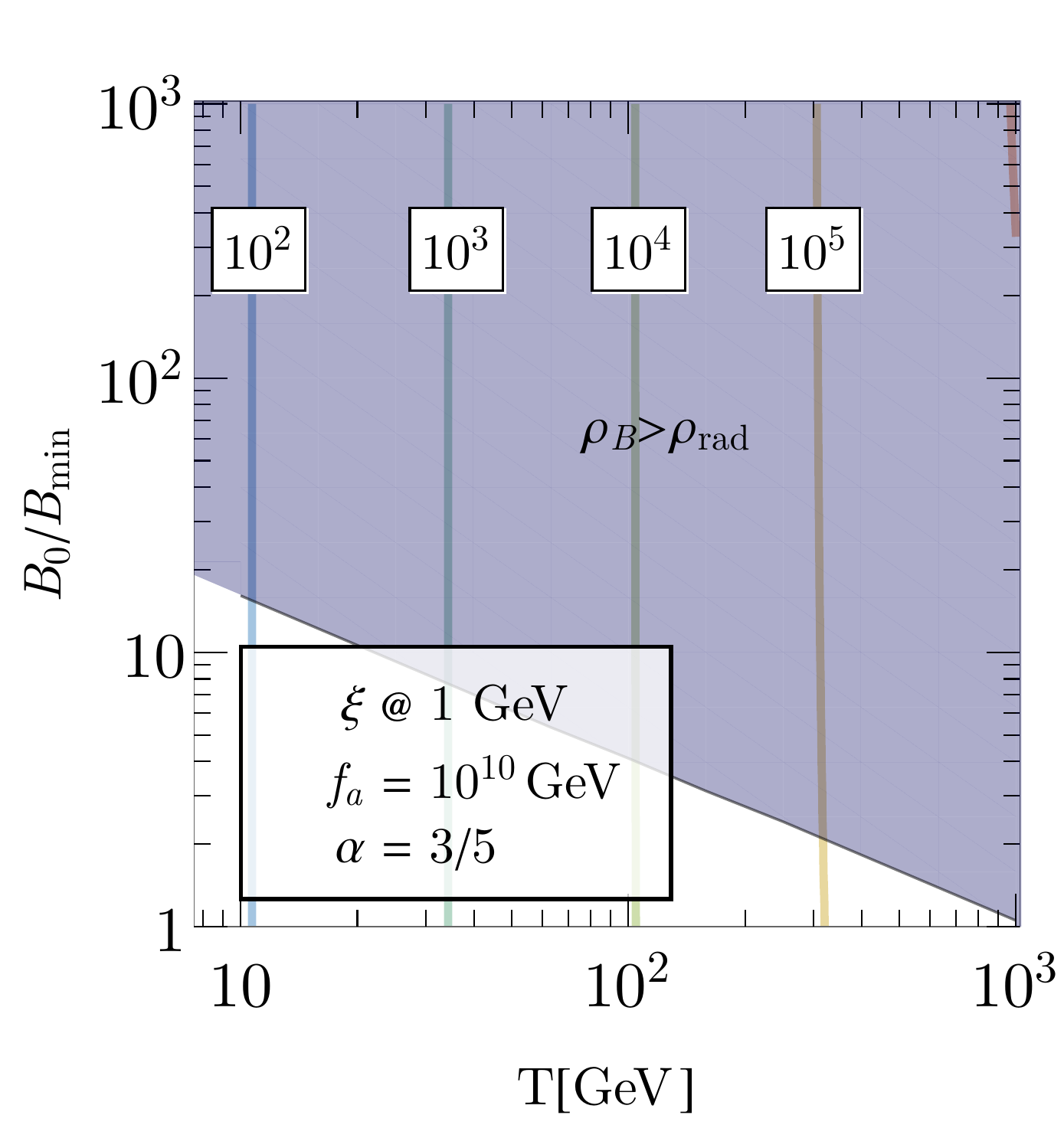}\\
\caption{$\xi$ at $T=1\,\,\text{GeV}$ as a function of $T_{\text{init}}$ and $B_0$ today, scaled by the observed lower bound $B_{\text{min}}\equiv1\cdot10^{-13}\,\text{G}$. The contours are labelled by the value of $\xi$ at $T=1\,\text{GeV}$. $f_a$ is taken to be $10^{10}\,\text{GeV}$. \textit{Left}: $\alpha=1/3$, \textit{Right}: $\alpha=3/5$.}\label{fig:ps}
\end{center}
\end{figure*}

In Fig.~\ref{fig:ps} we show contour plots of $\xi$ at $1\,\text{GeV}$ for the causal PMF, $n \ne -3$, as a function of the temperature $T_{\text{init}}$ at which current starts to accumulate, and the magnetic field $B_0$ today. $f_a$ is taken to be $10^{10}\,\text{GeV}$ here. Note that throughout this section, we assume current leakage is efficient enough, $y = 1$ in Eq.\,\eqref{eq:TDest}. The shaded region is excluded by the magnetic field dominating the universe at magetogenesis\,\cite{Durrer:2013pga}. We also confirm that the total energy of the current does not exceed the energy of the magnetic field so that its backreaction on the PMF is negligible.

We present two different contour plots for $\alpha\,=\,1/3$ and $\alpha\,=\,3/5$. For lower $T_{\text{init}}$, $\xi$ at $1\,\text{GeV}$ does not depend on the magnitude of the PMF. This is because the string network is still marginally in the stretching regime at $1\,\text{GeV}$. Indeed, $\xi$ for $\alpha\,=\,1/3$ and $\alpha\,=\,3/5$ are almost the same. On the other hand, in particular for $\alpha\,=\,1/3$, if $T_{\text{init}}$ is high enough, $\xi$ becomes independent of  $T_{\text{init}}$. There, the string network enters the friction dominated regime at $1\,\text{GeV}$ and $\xi$ gets closes to the asymptotic value, $\xi_0$.

\begin{figure*}[htbp]
\begin{center}
\includegraphics[width=0.6\linewidth]{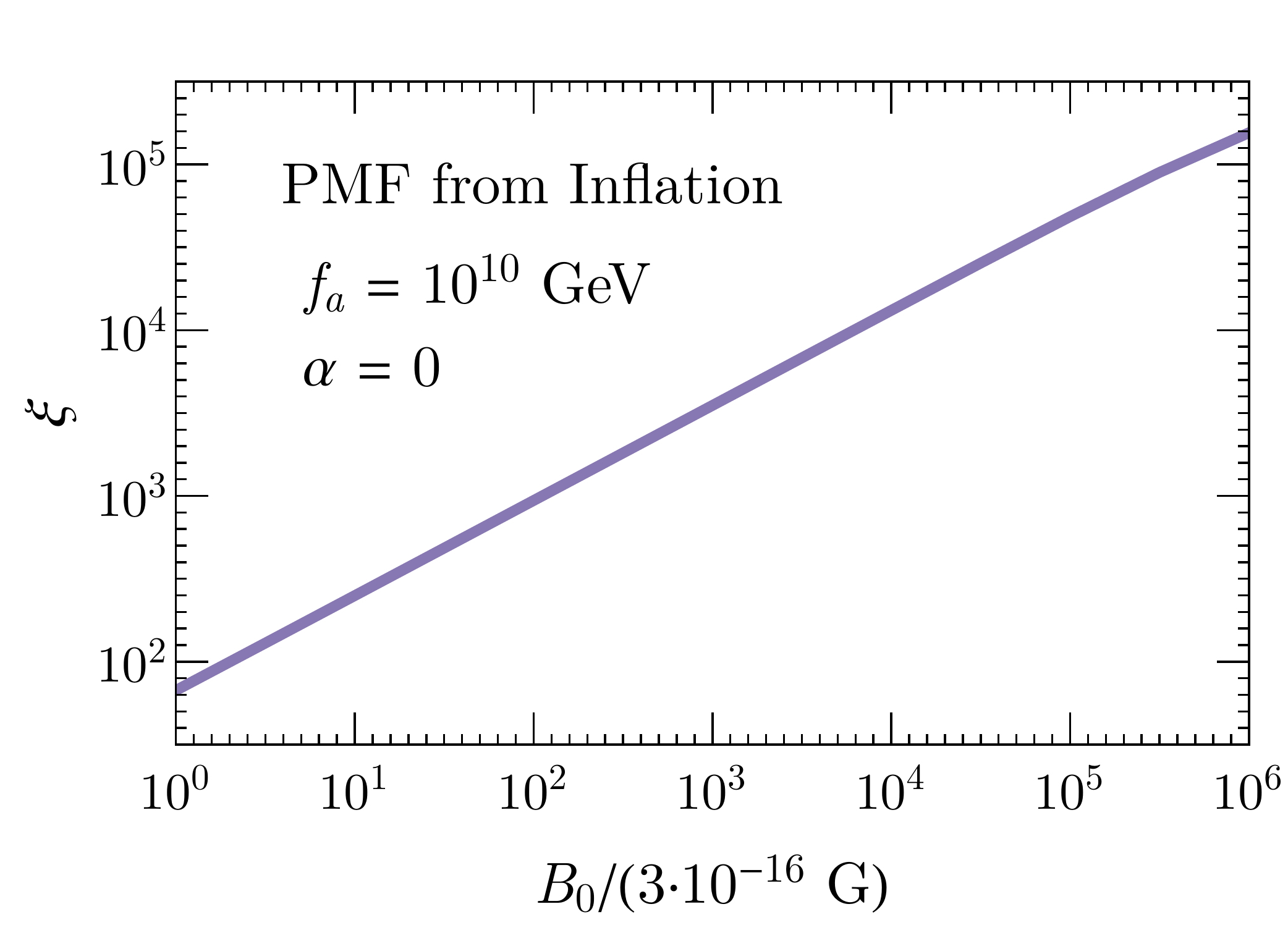}
\caption{$\xi$ at $T=1\,\,\text{GeV}$ as a function of $B_0$ today, scaled by the observed lower bound $B^\text{inf}_{\text{min}}\equiv3\cdot10^{-16}\,\text{G}$, for the inflationary PMF, whose coherent length is assumed to be larger than the current Hubble horizon. $f_a$ is taken to be $10^{10}\,\text{GeV}$.}
\label{fig:scan_inf}
\end{center}
\end{figure*}

In Fig.\,\ref{fig:scan_inf}, we show $\xi$ for the inflationary PMF in terms of the magnetic field today. Note that since inflation ends before the Peccei-Quinn phase transition, we may assume $T_\text{init} = T_\text{PQ} = f_a$. Since the inflationary PMF exists before the PQ symmetry breaking, the network enters the friction dominated regime well above $T=1\,\,\text{GeV}$ and $\xi$ takes its asymptotic value $\xi \propto B_0^{4/7}$ (see Eq.\,\eqref{eq:xiAsymp} with $\ell_\text{coh} \propto \xi^{-1/2}$).

\begin{figure*}[htbp]
\begin{center}
\includegraphics[width=0.5\linewidth]{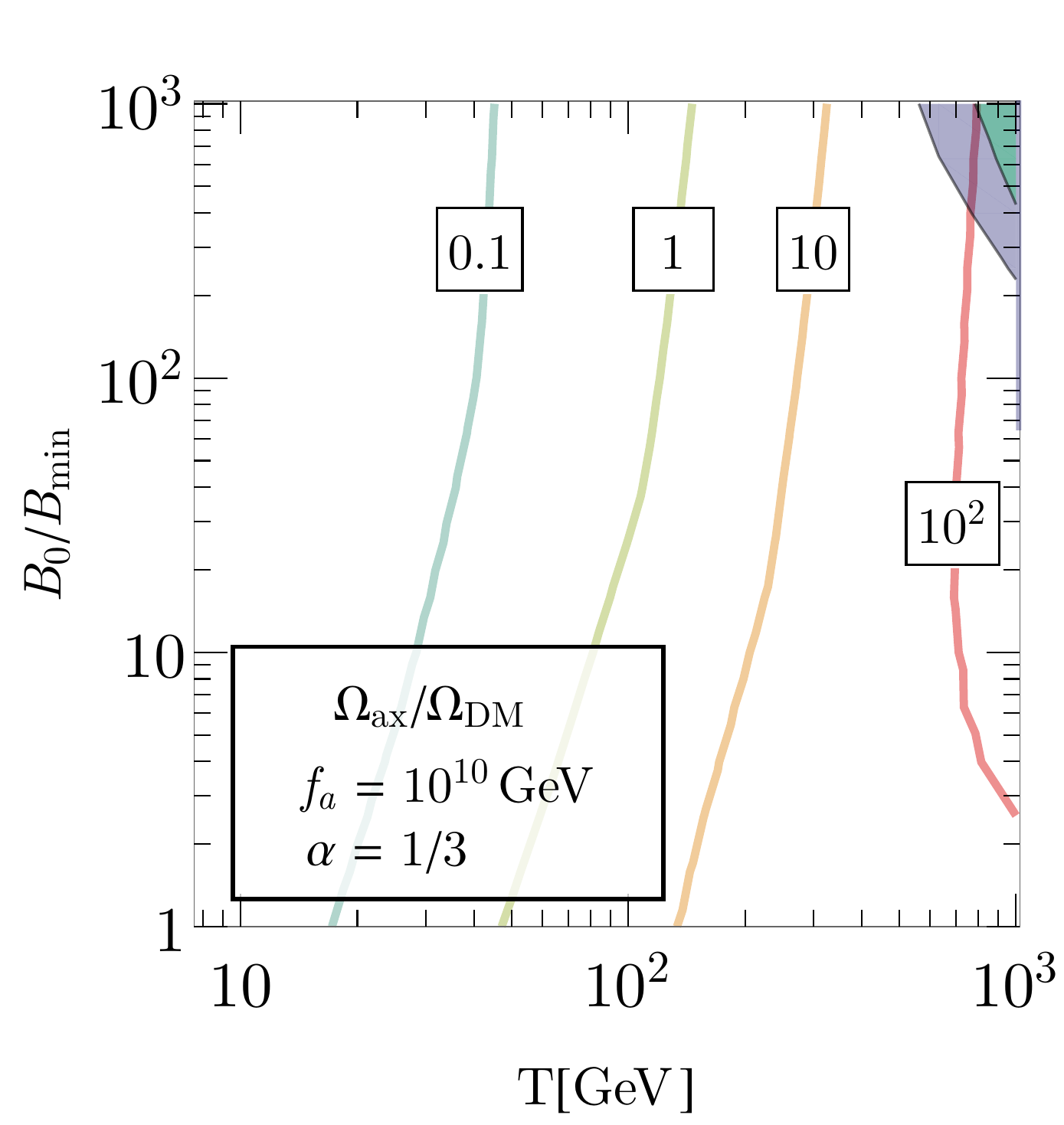}~~
\includegraphics[width=0.5\linewidth]{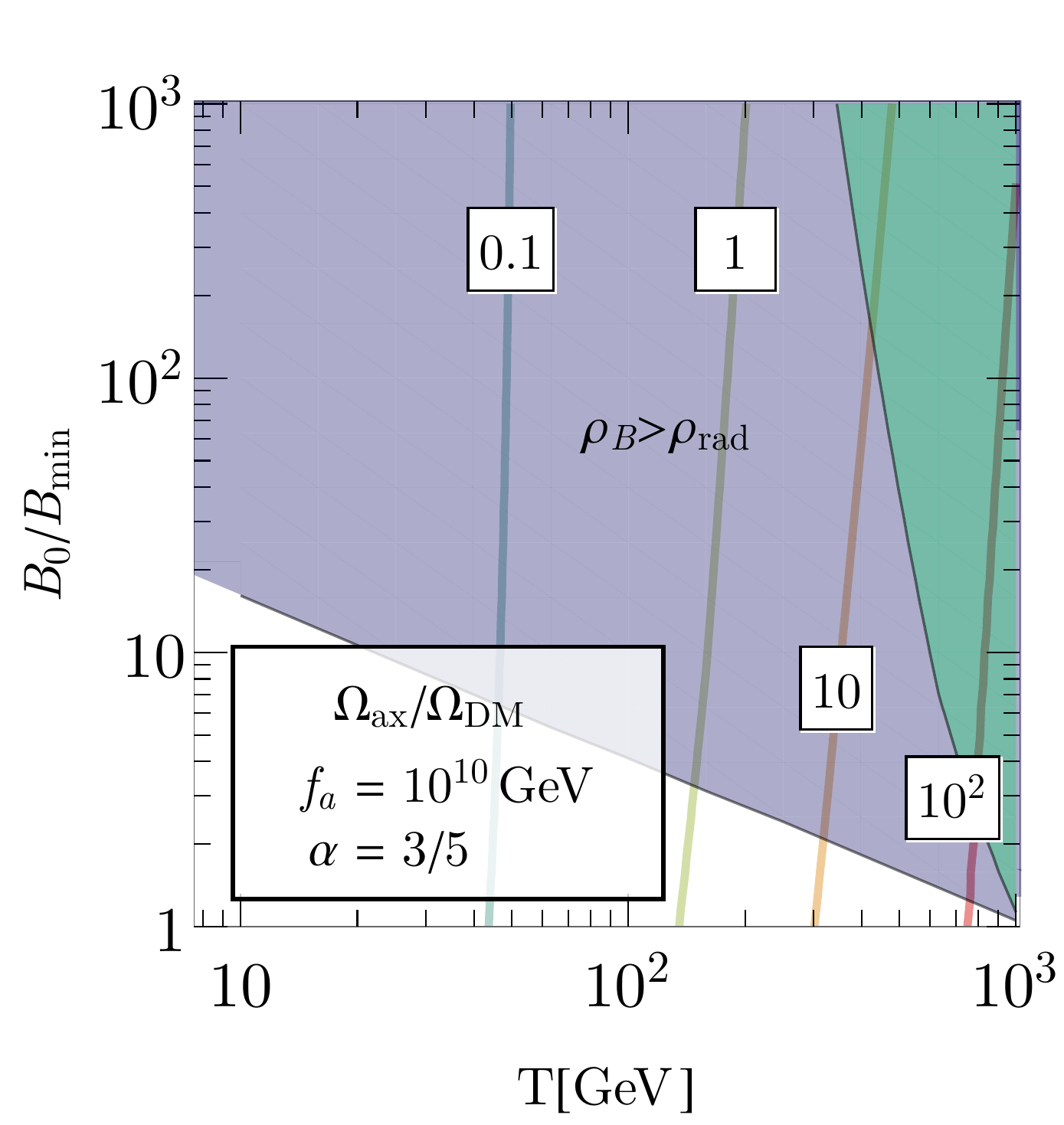}\\
\caption{$\Omega_\text{ax} / \Omega_\text{DM}$ as a function of $T_{\text{init}}$ and $B_0$ today, scaled by the observed lower bound $B_{\text{min}}\equiv1\cdot10^{-13}\,\text{G}$. The contours are labeled by the value of $\Omega_\text{ax} / \Omega_\text{DM}$. $f_a$ is taken to be $10^{10}\,\text{GeV}$. \textit{Left}: $\alpha=1/3$, \textit{Right}: $\alpha=3/5$. The purple shaded region in the left panel is excluded by the back reaction of the current on the PMF. The purple shaded region in the right panel is excluded by the PMF dominating the evolution of the universe. The green regions indicate $\Omega_{\text{vorton}}>\Omega_c$, but if $y\gg10^{-15}$, these vortons decay away before dominating the universe.}\label{fig:Xi}
\end{center}
\end{figure*}

\begin{figure*}[htbp]
\begin{center}
\includegraphics[width=0.6\linewidth]{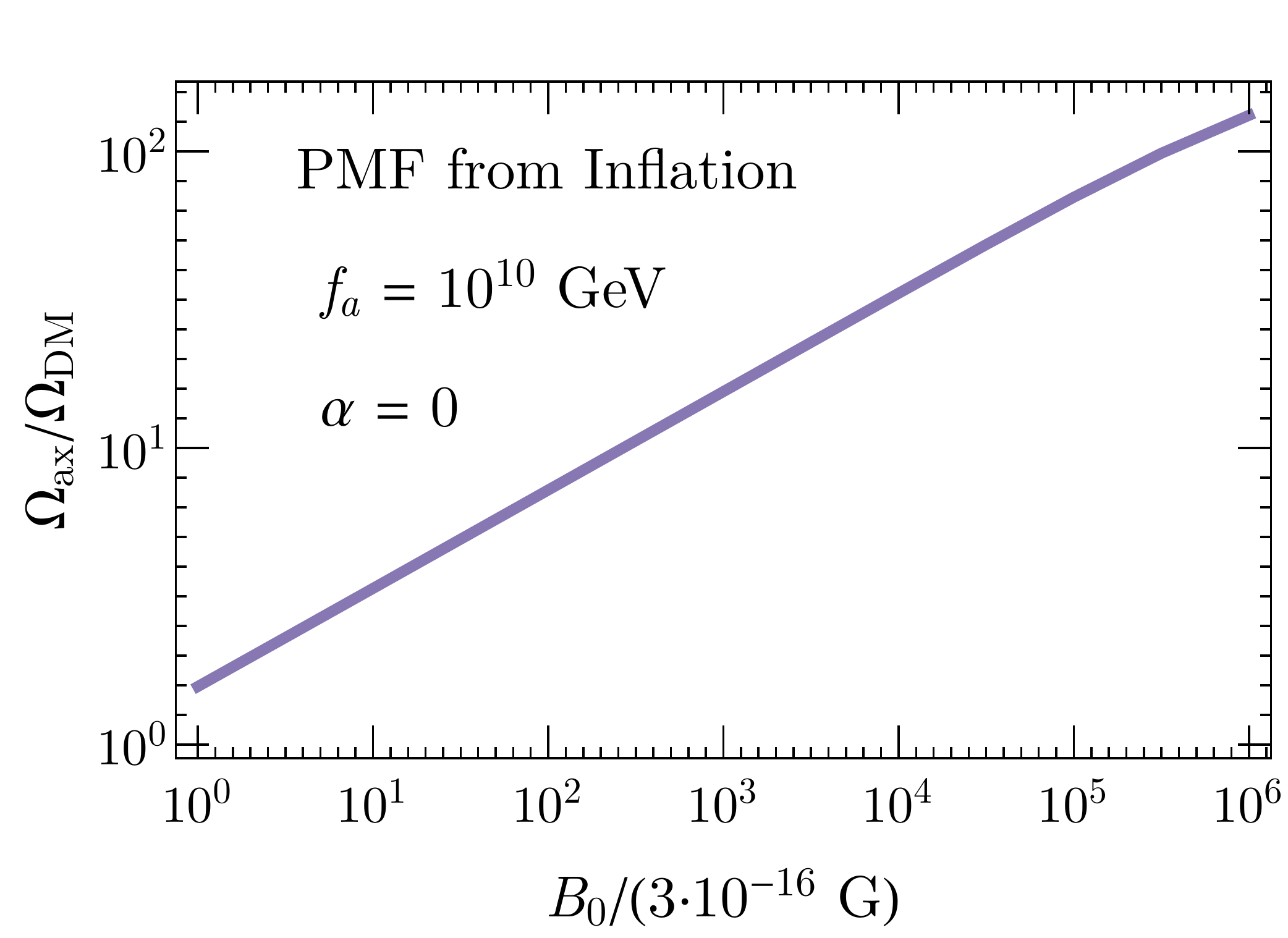}
\caption{For inflationary PMF, $\Omega_{\text{ax}} / \Omega_\text{DM}$ as a function of $B_0$ today, scaled by the observed lower bound $B_{\text{min}}\equiv3\cdot10^{-16}\,\text{G}$. $f_a$ is taken to be $10^{10}\,\text{GeV}$.}\label{fig:XiInf}
\end{center}
\end{figure*}

Finally, we show the axion abundance. In Fig.\,\ref{fig:Xi}, we show $\Omega_\text{ax}$ determined at $1\,\text{GeV}$ for the causal PMF, as a function of the temperature $T_{\text{init}}$ at which current starts to accumulate, and the magnetic field $B_0$ today. We refer to Ref.\,\cite{Gorghetto:2018myk} for the value of $\Omega_\text{ax}$ when $B_0 = 0$ and scale it by $\mathcal{S} (\mathcal{C} \xi)^{1 / 2 + 1 / (4 + \gamma)}$. The parameter region where vorton abundance could exceed the dark matter abundance is indicated by the green shaded region. Note that this part of the parameter space is likely still viable, as long as $y\gg10^{-15}$ and so the vortons are unstable. Compared with Fig.\,\ref{fig:ps}, $\Omega_\text{ax}$ is suppressed by the suppression factor. In Fig.\,\ref{fig:ps}, we find for lower $T_\text{init}$, $\xi$ is independent of $B_0$, since the string network is in the stretching phase and the ratio of the comoving Hubble horizon solely determines $\xi$. However, the suppression factor is smaller for larger $B_0$, since the friction becomes larger and the reconnection occurs less, resulting in smaller $\cal S$. As a result, for fixed $T_\text{init} \lesssim 10^2\,\text{GeV}$, where the network is in the stretching phase, the abundance is larger for the smaller $B_0 (> B_\text{min})$. Also, we see $\xi$ does not depend on $\alpha$ if the string network is in the stretching regime. However, the PMF is stronger for $\alpha = 3/5$ scenario and $\cal S$ becomes smaller, as we discuss. Consequently, for fixed $T_\text{init} \lesssim 10^2\,\text{GeV}$, the axion abundance for $\alpha = 1/3$ is larger.
In Fig.\,\ref{fig:XiInf}, we show $\Omega_\text{ax}$ determined at $1\,\text{GeV}$ and $\Omega_v$ for the inflationary PMF in terms of the magnetic field $B_0$ today (scaled by the observed lower bound $B_{\text{min}}\equiv3\cdot10^{-16}\,\text{G}$), for $f_a = 10^{10}\,\text{GeV}$.
As we have discussed, for the inflationary PMF, the string network enters the friction dominated regime at $1\,\text{GeV}$. Thus, the suppression factor is no longer effective and $\Omega_\text{ax}$ scales as $\sim \sqrt{\xi}$.

To derive the implications for the cosmological bound on $f_a$, we now present our results for the axion relic density, while scanning over $f_a$. For lower values of $f_a$, the string tension is lower and the effect of friction is larger. This makes $\xi$ larger for these small values of $f_a$. On the other hand, $\Omega_{\text{ax}}(\xi = 1)$ increases with $f_a$, as is well known. The axion abundance is determined by the balancing of these two factors.

In the case where the network is deep in the friction dominated regime by $T\sim\,1\,\text{GeV}$, we can use Eq.\,\eqref{eq:xiAsymp} to estimate $\xi \sim \mu^{-2/3}$ with $\xi$ dependence of $\ell_\text{coh}$ ignored, and so $\Omega_{\text{ax}} \sim f_a^{-2/3}$.  On the other hand, $\Omega_{\text{ax}}(\xi = 1) \sim f_a$. Thus, in total, we expect $\Omega_{\text{ax}} \sim f_a^{1/3}$ in this limit. In practice the network is either in the stretching regime or on the cusp between the two regimes, and so there is an extra $f_a$ dependent contribution from the suppression factor $\mathcal{S}$.

\begin{figure*}[htbp]
\begin{center}
\includegraphics[width=0.51\linewidth]{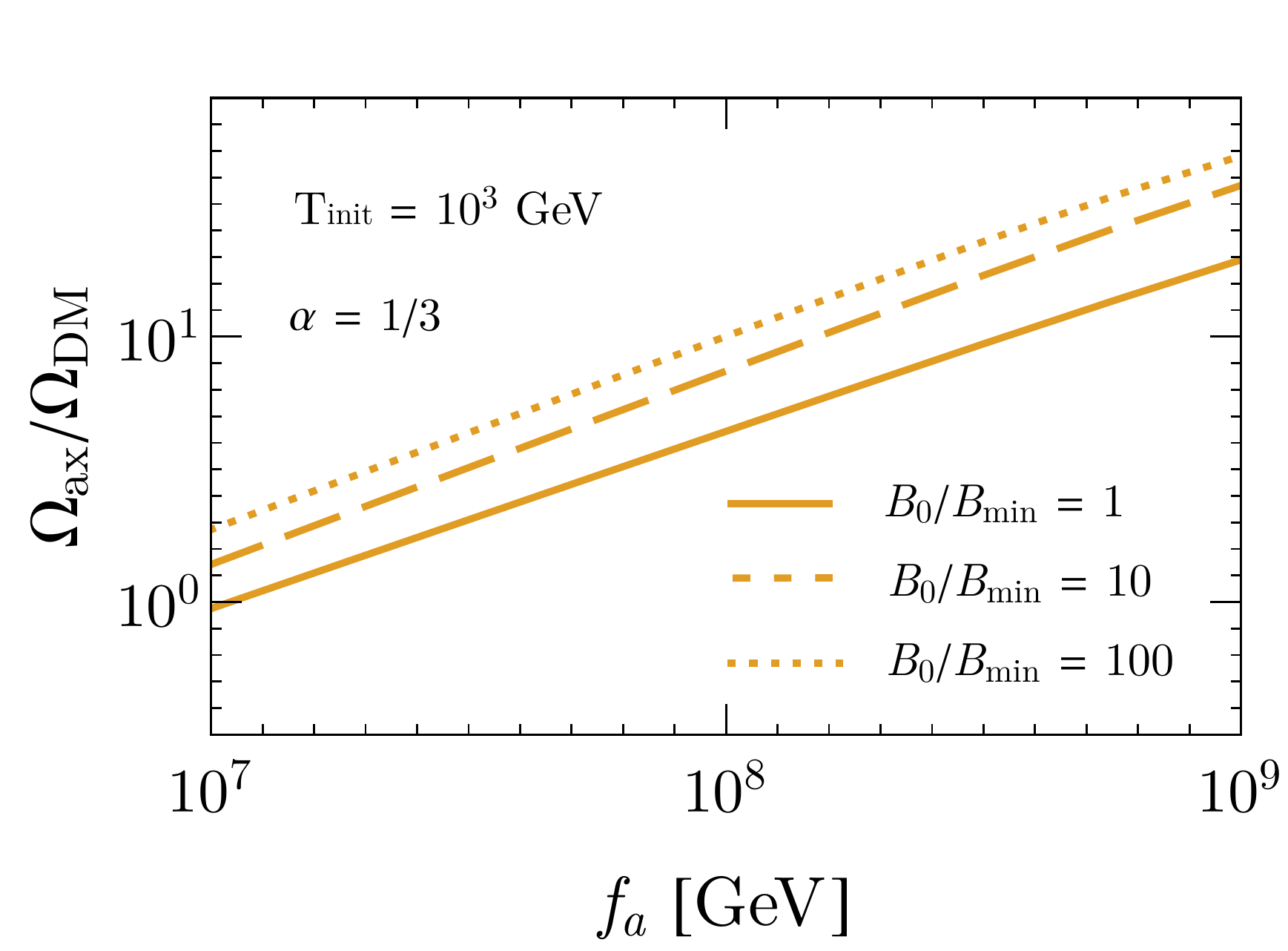}~~
\includegraphics[width=0.52\linewidth]{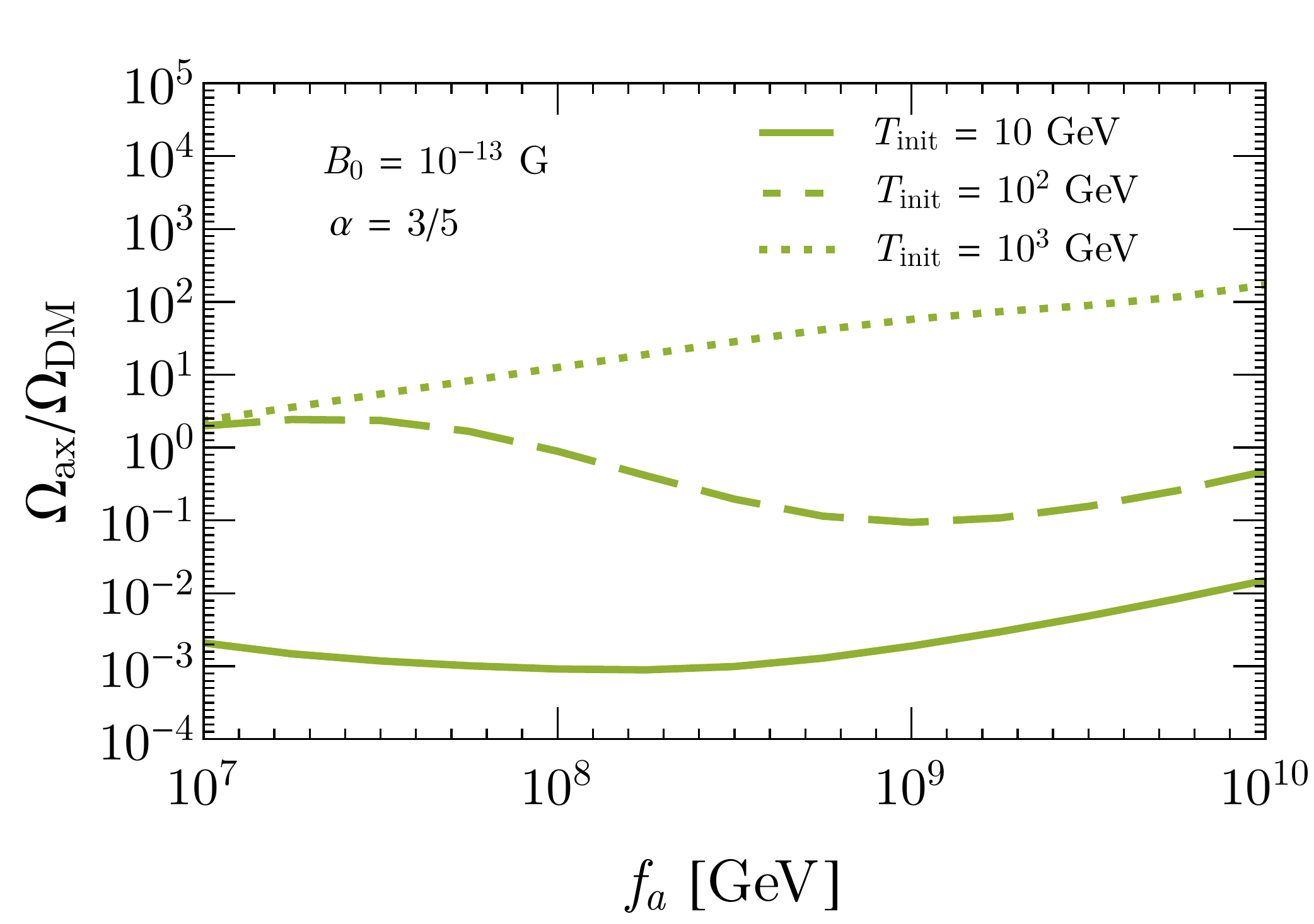}\\
\caption{$\Omega_{\text{ax}}/\Omega_\text{DM}$ as a function of $f_a$ for a fixed $T_\text{init}$ and $B_0$ today, scaled by the observed lower bound $B_{\text{min}}\equiv 10^{-13}\,\text{G}$, for the causal PMF. \textit{Left}: $\alpha=1/3$ and we change $B_0$ and see how $\Omega_{\text{ax}}$ behaves for $T_\text{init} = 10^3\,\text{GeV}$. \textit{Right}: $\alpha=3/5$ and we change $T_\text{init}$ and see how $\Omega_{\text{ax}}$ behaves for $B_0 = B_\text{min}$.}
\label{fig:OmegaInFa}
\end{center}
\end{figure*}

Our numerical results are shown in Fig.\,\ref{fig:OmegaInFa}.
For $\alpha = 1/3$, we show $\Omega_{\text{ax}} / \Omega_\text{DM}$ vs $f_a$ for 3 different values of $B_0$. The string network is not deep in the stretching regime and the suppression factor $\mathcal S|_{1\,\text{GeV}}$ is $\mathcal{O}(1)$ in the entire parameter space. Thus, as we expect, $\Omega_{\text{ax}}$ is a mildly increasing function of $f_a$, although the exponent is different from $1/3$, since the string network is only at the cusp of the friction dominated phase. We find, for example, if $B_0 = B_\text{min}$, the correct relic abundance is given for $f_a \sim 10^7\,\text{GeV}$. For $\alpha = 3/5$, we show $\Omega_{\text{ax}} / \Omega_\text{DM}$ in terms of $f_a$ with changing $T_\text{init}$. In this case, since the magnetic field in the early universe is larger, the whole network may be in the moderate stretching phase and the suppression factor, $\mathcal{S}$, may be small.

\begin{figure*}[htbp]
\begin{center}
\includegraphics[width=0.51\linewidth]{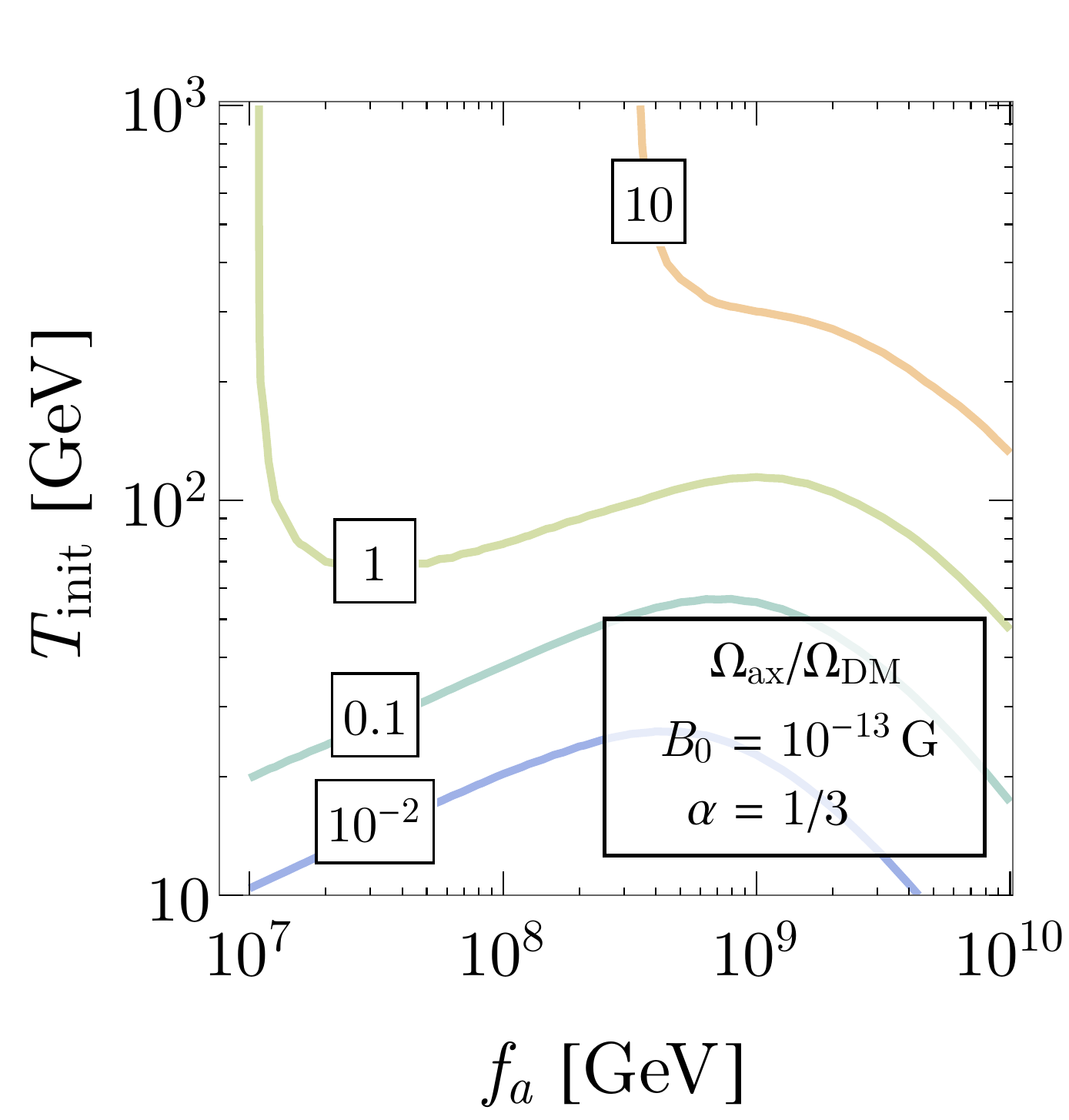}~~
\includegraphics[width=0.52\linewidth]{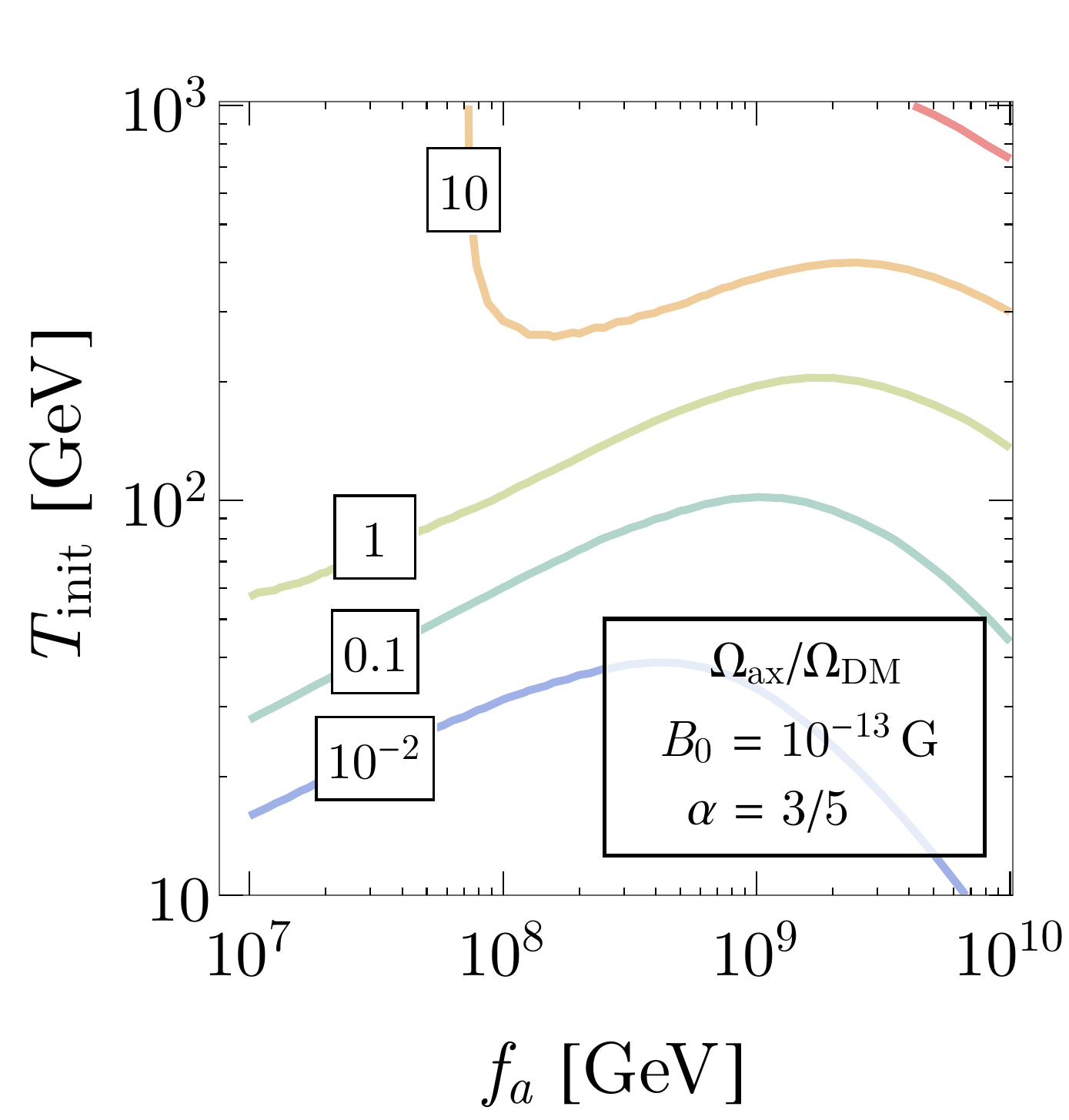}\\
\caption{Contour plots of $\Omega_{\text{ax}}/\Omega_\text{DM}$ as a function of $f_a$ and $T_\text{init}$ for $B_0 = 10^{-13}\,\text{G}$ today. The contours are labelled by the value of $\Omega_\text{ax} / \Omega_\text{DM}$. \textit{Left}: $\alpha=1/3$ \textit{Right}: $\alpha=3/5$}
\label{fig:OmegaInFaTinit}
\end{center}
\end{figure*}

For smaller $f_a$, the eventual axion relic density is affected not only by the PMF, but also by friction from the thermal current on the strings, prior to magnetogenesis. This sets a different initial condition with a higher value of $\xi$ at the onset of the stretching regime. As a result, the number of strings in the Hubble volume at $T_\text{init}$ is larger for smaller $f_a$ and the transition from the stretching to the friction dominated phase occurs earlier. Thus, $\Omega_{\text{ax}}/\Omega_\text{DM}$ is no longer monotonically decreasing in $f_a$, as can be seen in the right panel of Fig.~\ref{fig:OmegaInFa} and in Fig.\,\ref{fig:OmegaInFaTinit}.

Finally, in Fig.\,\ref{fig:OmegaInFaTinit} we present a 2D scan of $\Omega_{\text{ax}}/\Omega_\text{DM}$ as a function of $f_a$ and $T_\text{init}$ for the minimal value $B_0 = 10^{-13}\,\text{G}$ of the magnetic field today. As mentioned before, for $T_\text{init} \lesssim 10^2\,\text{GeV}$, the string network for both $\alpha=1/3$ and $\alpha=3/5$ is in the stretching regime, and so their value of $\xi$ is simlar. Even so, the value of $\Omega_{\text{ax}}$ for $\alpha = 3/5$ is smaller than the  one for $\alpha = 1/3$, because of its greater friction that leads to a slower string velocity and a smaller value of $\mathcal{S}$.

\begin{figure*}[htbp]
\begin{center}
\includegraphics[width=0.6\linewidth]{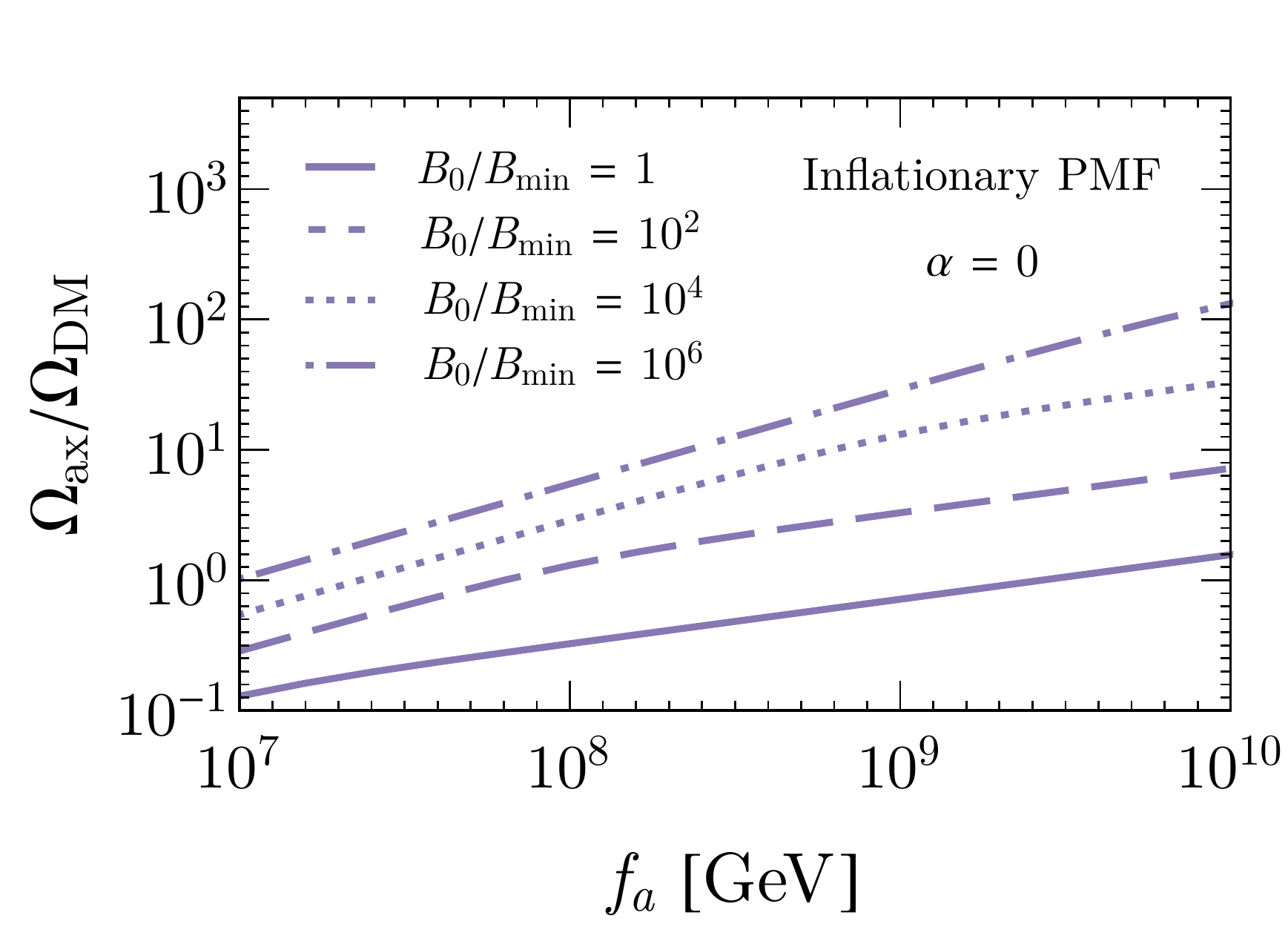}
\caption{For inflationary PMF, $\Omega_{\text{ax}} / \Omega_\text{DM}$ at $T=1\,\,\text{GeV}$ as a function of $f_a$ for four values of $B_0$ today, with  $B_{\text{min}}\equiv3\cdot10^{-16}\,\text{G}$.}\label{fig:OmegaInFaInf}
\end{center}
\end{figure*}

The axion abundance is also presented for the inflationary scenario in Fig.\,\ref{fig:OmegaInFaInf}. In this case, we may assume the asymptotic value of $\xi$. As a result, $\Omega_{\text{ax}}/\Omega_\text{DM}$ scales like $f_a \sqrt{\xi_0} \sim f_a^{3/7}B^{2/7}$ as one can check in the figure. As can be seen from the plot, the inflationary PMF scenario with axion strings is in tension with the astrophysical limit of $f_a \gtrsim 10^8\,\text{GeV}$ from supernova cooling \cite{Zyla:2020zbs}.

\section{Conclusion and Discussion}

In this paper, we point out the comsmological implications of the chiral superconductivity of axion strings. 
We examined the possibility that the axion string network leaves stable remnants, called the vortons, and found that they largely decay away as a result of curvature induced current leakage. If this leakage is suppressed by a small coupling $y\sim10^{-15}$, then vortons could potentially lead to tension with BBN bounds, as well as to overclosing the universe.

In addition, we analyzed the evolution of the axion string network in the presence of a PMF within the analytic VOS model of string evolution. We found that the induced current on the string is sizeable, and so it generates a dominant friction force on the string which suppresses string reconnections. This increases the number of strings per Hubble volume by \textit{many orders of magnitude}, and could greatly modify the relic abundance of axion DM. We stress that our results have been obtained within the simplified VOS model for cosmic string evolution, and so need to be validated by future numerical studies.

\acknowledgments

The authors would like to thank D. Arovas, J. Foster, M. Gorghetto, A. Hook,  K. Kamada, R. Lasenby, K. Nakayama,  T. Piran, B. Safdi, K. Saikawa, S. Shirai, and K. Yonekura, for useful discussions. We give special thanks to R. Lasenby for pointing out the possibility of curvature-mediated decay, which renders vortons unstable. HF and HM would like to thank J. Dror and J. Leedom for collaboration during the initial stage of this project.
HF, HM and OT were supported by the Director, Office of Science, Office of
High Energy Physics of the U.S. Department of Energy under the
Contract No. DE-AC02-05CH11231, and AM under DOE Grant No.\ DE-SC0009919.
The work of HM was also supported by the NSF grant
PHY-1915314, by the JSPS Grant-in-Aid for
Scientific Research JP20K03942, MEXT Grant-in-Aid for Transformative Research Areas (A)
JP20H05850, JP20A203, by WPI, MEXT, Japan, and Hamamatsu Photonics, K.K.

\begin{appendix}

\section{Energy-Momentum Conservation for a String Loop}
\label{app:a}

In this appendix, we study the consequences of energy-momentum conservation for a zero-mode trapped on a circular loop of radius $R$. We will model the string loop by a circle of radius $R$ in the $x-y$ plane, and study the decay of a zero-mode with energy $k$ to two particles, $k \to q_1 + q_2$. Such as decay is forbidden in for a straight string, and we want the asymptotic form of the decay rate as $R \to \infty$. See also Ref.\,\cite{Ibe:2021ctf}.

Assume the zero mode trapped on the string has wavefunction
\begin{align}
\psi_0(r,\phi,z,t) &= \sqrt{\frac{1}{ \Delta^2 L}} e^{-i \omega t} e^{i k R \phi }\ \theta(\abs{r-R} \le \Delta/2)\ \theta(\abs{z} \le \Delta/2)
\label{a.1}
\end{align}
where $(r,\phi,z)$ are cylindrical coordinates, and $L=2\pi R$ is the length of the loop. $kR \gg 1$ must be an integer.
$\psi_0$ is constant in the $\mathbf{\hat z}$ direction between $[-\Delta/2,\Delta/2]$ and for radii $r$ between $R-\Delta/2$ and $R+\Delta/2$. Eq.~\eqref{a.1} models the transverse form of the zero-mode wavefunction as a step function. The details of the functional form of $\psi_0$ in the transverse direction are not important; the consequences of energy-momentum conservation arise from the $t$ and $\phi$ dependence of $\psi_0$. We are interested in the decay of zero-modes with momentum $k \Delta \ll 1$, so that their momentum is smaller than the inverse size of the zero-mode. The energy of the zero-mode is $\omega=k$.

For free-particle decays $k \to q_1 + q_2 + \ldots$, the matrix element
\begin{align}
\braket{q_1,\ldots q_n | k}
\label{a.2}
\end{align}
involves a space-time integral
\begin{align}
\int {\rm d}^4 x\ e^{i (q_1 + \ldots q_n) \cdot x} e^{-i k \cdot x}\,.
\label{a.3}
\end{align}
The time integral gives an energy-conserving $\delta$-function $\delta(q_1^0 + \ldots + q_n^0 -k^0)$
and the space integral gives a momentum-conserving $\delta$-function $\delta(\mathbf{q}_1 + \ldots + \mathbf{q}_n -\mathbf{k})$. We consider a matrix element $\braket{q_1,\ldots q_n | \psi_0} $ where the final states are plane-waves, but the intial state is the zero-mode trapped on a loop. The time integral still gives an energy-conserving $\delta$-function $\delta(q_1^0 + \ldots + q_n^0 - \omega)$. However, the space integral
\begin{align}
I &=\int {\rm d}^3 \mathbf{x}\ e^{-i \mathbf{q} \cdot \mathbf{x}}\ \psi_0(\mathbf{x},0) \,,
\label{a.4}
\end{align}
with $\mathbf{q}=\mathbf{q}_1+\ldots +\mathbf{q}_n$
no longer gives a $\delta$-function. The integral Eq.~\eqref{a.4} is now
\begin{align}
I & =  \sqrt{\frac{1}{\Delta^2 L}} \frac{2}{q_z} \sin \frac{q_z \Delta}{2}
\int r\,\rd r\, \rd \phi \sum_n (-i)^n  J_n( \abs{\mathbf{q}_\parallel} r) e^{i n \phi} e^{- i k R \phi} \ \theta(\abs{r-R} \le \Delta/2)
\label{a.5}
\end{align}
expanding the exponential in Eq.~\eqref{a.4} in terms of Bessel functions. Here $\mathbf{q}_\parallel$ is the component of $\mathbf{q}$ in the plane of the loop. The $\phi$ integral picks out the term with $n=kR$,
\begin{align} 
I & =    \sqrt{\frac{1}{\Delta^2 L}}  \frac{4\pi}{q_z} \sin \frac{q_z \Delta}{2}
\int_{R-\Delta/2}^{R+\Delta/2} r\,\rd r\,  (-i)^{kR}  J_{kR}( \abs{\mathbf{q}_\parallel} r)
\label{a.6}
\end{align}
In the regime of interest, $\abs{\mathbf{q}} \Delta \ll 1$ so the Bessel function is approximately constant, over the integration domain, and (using $L=2\pi R$)
\begin{align} 
I 
\approx \sqrt{2 \pi   R}\, \Delta
  (-i)^{kR}  J_{kR}( \abs{\mathbf{q}_\parallel} R)\,.
\label{int}
\end{align}
The square of the matrix element has the factor
\begin{align}
\abs{I}^2 &=2 \pi R \Delta^2 \abs{ J_{kR}( \abs{\mathbf{q}_\parallel} R)}^2\,.
\label{a.8}
\end{align}
which is the analog of
\begin{align}
\abs{(2\pi)^3 \delta^{(3)}(\mathbf{q}- \mathbf{k})}^2 &= V \, (2\pi)^3 \delta^{(3)}(\mathbf{q}- \mathbf{k})
\end{align}
where $V$ is the volume of space.
We want the limit of Eq.~\eqref{a.8} for $ \abs{\mathbf{q}_\parallel} R \gg 1$, i.e.\ for a large loop. From Ref.~\cite[\S8.4]{Watson:1944},
\begin{align}
J_\nu(\nu \operatorname{sech} \alpha) & \sim_{\nu \to \infty}  \frac{ e^{\nu (\tanh \alpha - \alpha)} }{\sqrt{2 \pi \nu \tanh \alpha}}
\end{align}
Using this asymptotic expansion in Eq.~\eqref{a.8} with
\begin{align}
\nu &= k R, &
\operatorname{sech} \alpha &= \frac{\abs{\mathbf{q}_\parallel}}{k}, &
\tanh \alpha &= \frac{\sqrt{k^2-\abs{\mathbf{q}_\parallel}^2}}{k},
\end{align}
gives
\begin{align}
\abs{\mathcal{I}}^2 &\sim  \Delta^2 \frac{1}{ \sqrt{k^2-\abs{\mathbf{q}_\parallel}^2} }\ e^{+2 R k \left[  \frac{ \sqrt{k^2-\abs{\mathbf{q}_\parallel}^2} }{k} - \tanh^{-1} \frac{\sqrt{k^2-\abs{\mathbf{q}_\parallel}^2}}{k}  \right]} \nn
&\sim  \Delta^2 \frac{1}{ \sqrt{k^2-\abs{\mathbf{q}_\parallel}^2} }\ e^{-\frac23 R k^{-2} \left(k^2-\abs{\mathbf{q}_\parallel}^2 \right)^{3/2} } \,.
\label{a.12}
\end{align}
For the two-body decay $\psi_0 \to q_1 + q_2$, the decay rate using the asymptotic expansion Eq.~\eqref{a.12} is
\begin{align}
\Gamma  
&\sim \int \frac{1}{2k} \frac{\rd^3 q_1}{(2\pi)^3 2E_1} \frac{\rd^3 q_2}{(2\pi)^3 2E_2} \abs{\mathcal{M}}^2 2\pi \ \delta(k-E_1-E_2)   \Delta^2 \frac{1}{ \sqrt{k^2-\abs{\mathbf{q}_\parallel}^2} } e^{-\frac23 R k^{-2}  \left(k^2-\abs{\mathbf{q}_\parallel}^2 \right)^{3/2} }
\label{A.13}
\end{align}
We can find the decay rate for large $R$ by an asymptotic expansion of Eq.~\eqref{A.13}. If the final state particles are massless, then $\mathbf{q}_\parallel$, with $\mathbf{q}=\mathbf{q}_1+\mathbf{q}_2$ can be as large as $\abs{\mathbf{q}_\parallel}=k$, if the two massless particles are emitted in the same direction. The asymptotic form of the integral for large $R$ is dominated by this region, and is
\begin{align}
\frac{\Gamma}{\Gamma_0} \sim  \frac{( k \Delta)^2}{2 \pi^2}  \frac{\Gamma(\frac23)}{2^{2/3} 3^{1/3}(k R)^{2/3}}
\label{A.16}
\end{align}
where $\Gamma_0$ is the free-particle decay rate (i.e.\ for a free particle not trapped on the string).\footnote{The $\Gamma(\frac23)=1.354$ on the r.h.s.\ is the Gamma function.} Eq.~\eqref{A.16} has $(k \Delta )^2$ and $1/(k R )^{2/3}$ suppression factors relative to the free particle decay rate. The final result vanishes as $R \to \infty$, as it must by the boost invariance argument for a straight string.

If the final particles are massive, with masses $m_1$ and $m_2$, then the maximum value of $\abs{\mathbf{q}_\parallel}$ is when both particles are emitted in the same direction with the same velocity, with boost factor $\gamma$ given by
\begin{align}
    \gamma &= \frac{E}{m_1+m_2}
\end{align}
and the smallest value of $k^2-\abs{\mathbf{q}_\parallel}^2$ is $(m_1+m_2)^2$. The asymptotic form of the integral is dominated by this point, and is
\begin{align}
\frac{\Gamma}{\Gamma_0} \sim  \frac{(k\Delta)^2}{8 \pi} 
 \frac{1}{(k R)^2}\frac{k^4\sqrt{m_1 m_2}}{(m_1+m_2)^5} e^{-\frac23\frac{ R  (m_1+m_2)^3}{k^2} }
\label{a.14}
\end{align}
provided the exponent is much larger than unity.

\section{Vorton Abundance for Small $y$}\label{sec:vortonapp}

In this appendix we estimate the vorton relic density in the absence of leakage processes, and \textit{without} a PMF. While in the case of KSVZ fermions with a Yukawa coupling we have seen that vorton are generically unstable, here we provide the complementary analysis for the case that the leakage processes are suppressed. Though we do not have a particular UV model in mind that allows for the decay of KSVZ fermions while preventing current leakage, we are interested in the projected vorton abundance in this case.

In order to estimate the vorton energy density, we first estimate the number of vortons formed at a given Hubble time. From the evolution of the string network, we can calculate the total length of strings disappering from the network at that time. The network reduces its length by radiating axions or forming loops. Analytic estimates show the string shrinkage by axion emissions is at most the same order contribution as one by loop formations\,\cite{Martins:2000cs}. For simplicity, we assume all the extra length forms loops and eventually becomes vortons.

The comoving number density of strings (per unit area) is
\begin{eqnarray}
n_\text{str} = \xi \, R^2(t)\, H^2,
\end{eqnarray}
where $R(t)$ is the scale factor of the universe.  
$n_\text{str}$ would have been conserved if not for the string reconnections.
Thus, the total length of the string which forms loops, $L_\text{tot}(t)$, is proportional to the time derivative of $n_\text{str}$;
\begin{eqnarray}
\frac{d L_\text{loop, tot}}{dt} = -\left(\frac{1}{RH}\right)^3 \frac{dn_\text{str}}{dt}.
\end{eqnarray}
Within a Hubble time,
\begin{eqnarray}
L_\text{loop, tot} \simeq \frac{2\xi}{H} - \frac{\dot{\xi}}{H^2},
\end{eqnarray}
where we take $R(t) = 1$ without any loss of generality.
The loops of string eventually ends up as vortons. Thus, the number of the vortons at a given Hubble time is given by the ratio between $L_\text{loop, tot}$ and the typical size of each loop, $L_\text{loop}$. Here, for simplicity, we assume that the sizes of all loops at a Hubble time are the same. 
For $L_\text{loop} = L_0 = \xi^{-1/2} H^{-1}$, the number of the loops is
\begin{eqnarray}
\label{eq:VortonNumber}
N_\text{loop} = \frac{L_\text{loop, tot}}{L_\text{loop}} = 2\xi^{3/2} - \xi^{1/2} \frac{\dot{\xi}}{H}
\equiv k \, \xi^{3/2},
\end{eqnarray}
where 
\begin{align}
\label{eq:RecEff}
k \equiv 2 - \frac{\dot{\xi}}{\xi H}
\end{align}
denotes the efficiency of string reconnections.

The vorton energy density at a given Hubble time is $\rho_v \simeq E_v N_\text{loop} H^3$.
In order to compare it with the dark matter number density, it is convenient to normalize it by the entropy density $s$:
\begin{eqnarray}
\label{eq:VortonAbundance}
\frac{\rho_v}{s} \simeq 2.1 \left(\frac{g_\star}{106.75}\right)^\frac14 \sqrt{\mathcal{C} / \pi} \, k\, \xi^\frac{5}{4} \left(\frac{T}{M_p}\right)^\frac52 f_a,
\end{eqnarray}
which is a comoving quantity. In the above estimate, assumed that the vorton does not annihilates after it forms. This may be a reasonable assumption for particles, but since the size of the loop is macroscopic at the beginning, there may be some annihilation with other vortons. The opposite limit from no annihilation is to assume that the vorton annihilates efficiently, and only the net current asymmetry of the initial vorton distribution survives. In this case, we can simply replace $N_\text{loop} \to \sqrt{N_\text{loop}}$, obtaining the same equation
as Eq.\,\eqref{eq:VortonAbundance} with  $k \, \xi^{5/4} \to \sqrt{k}\, \xi^{1/2}$.
Note that we have focused on the electromagnetic current, but the same dynamics holds for the color current on the axion string as well. A similar analysis of the color current gives the same order-of-magnitude estimate of the vorton abundance.

So far, we have assumed that all loops formed at a Hubble time are initially the same size, $L_0 = \xi^{-1/2}H^{-1}$. In reality, the sizes of loops distributes over the wide range of scales\,\cite{BlancoPillado:2011dq,Blanco-Pillado:2013qja}. If we take smaller loop sizes $\tilde{L}$, the vorton charge and mass decreases by $\sqrt{\tilde{L} / L_0}$ whereas the number density increases by $L_0 / \tilde{L}$ for the fixed energy of loops. On the other hand, the estimation of the total charge asymmetry inside a horizon does not change. Thus, if we included more detailed spectrum of string loops, the na\"ive estimation would increase but the conservative one would not change.

We now estimate the value of $\xi$, the number of strings per Hubble. In the early universe, the friction force due to scattering with plasma particle decelerates the string and $\xi > 1$ in general\,\cite{Vachaspati:1984dz,Vilenkin:1991zk}. The scattering cross section between the string and thermal particles with a non-zero Peccei-Quinn charge can be as large as allowed by unitarity, $\lambda \sim 1 / T$\,\cite{Nagasawa:1997zn}. In the KSVZ scenario, such PQ charged particles are absent in the thermal bath. The axion may exist in the universe, but the scattering cross section between the axion and the string is expected to be geometrical. This is an analogue of the scattering between the magnetic monopole and the photon as the magnetic dual of the axion couples to the string. However, since the string supports the (color) current, a (color) magnetic fields winding the string play a role of an effective thickness of the string. This results a scattering rate as large as the unitarity bound. Ref.\,\cite{Chudnovsky:1986hc,Dimopoulos:1997xa,Carter:2000fv} estimate the cross section as
\begin{eqnarray}
\label{eq:MagRad}
\lambda \sim \frac{I}{\sqrt{\rho}},
\end{eqnarray}
where $\rho$ is the energy density of the thermal plasma. This roughly corresponds to the radius where the pressure between the thermal plasma and the magnetic field induced by the current balances.
According to the standard picture\,\cite{Vilenkin:2000jqa}, $\xi$ and the string velocity $v_s$ are determined in the following way.
First, the frictional force per unit length of string is
\begin{eqnarray}
\label{eq:FricForce}
F_\text{fric} \simeq \langle \lambda v_\text{rel} n q \rangle \sim T^2 I v_s,
\end{eqnarray}
where $\langle\rangle$ is the thermal average for the scattering particles, $v_\text{rel}$ is the relative velocity between the particles and the string, $n$ is the number density of the scattering particles and $q$ is the momentum transfer. The typical time scale due to friction is the relaxation time of the string momentum,
\begin{eqnarray}
\label{eq:FricTime}
\tau_\text{fric} = \frac{\mu v_s}{F_\text{fric}}\simeq \frac{\mu}{T^2I}.
\end{eqnarray}
For a typical string distance scale $L \sim \xi^{-1/2} H^{-1}$,
the force on the string per unit length is from the string curvature
\begin{eqnarray}
\label{eq:StrForce}
F_\text{str} \simeq \frac{\mu}{L}.
\end{eqnarray}
The string accelerates for a time $\sim\tau_{\text{fric}}$ and the string velocity is thus limited to
\begin{eqnarray}
\label{eq:StrVelFric}
v_s \simeq \frac{F_\text{str}}{\mu} \tau_\text{fric} \simeq \frac{\mu}{L T^2I}.
\end{eqnarray}
On the other hand, $\xi$ is also given as $L \simeq v_\text{s} H^{-1}$ if we assume the reconnection process occurs, so that $\dot{n}_\text{str} \ne 0$. Then,
\begin{align}
v_s &\simeq \xi^{-\frac12}, &
\xi &\simeq \dfrac{T^2I}{\mu H}, \label{eq:xiThermFric}
\end{align}
which justify the assumption. Note that since $v_s < 1$, $\xi$ stops decreasing when it reaches a value of order one. For simplicity, we thus assume the number of the string per Hubble horizon is $\max\,(\xi, 1)$. In the above estimation, we assume the number of the string just after the PQ phase transition at $T_\text{PQ}$, where we assume $T_\text{PQ} = f_a$, is more than $\xi(T_\text{PQ})$ so that the number of the string is monotonically decreasing after the phase transition, which is a reasonable assumption\,\cite{Murayama:2009nj}.

Using Eq.\,\eqref{eq:xiThermFric} with Eq.\,\eqref{eq:VortonAbundance}, we can calculate $\rho_v / s$ for a given Hubble time and obtain $\rho_v / s \sim T^{15/4}$. Thus, the temperature $T_v$ when most vortons form is $T_v=T_\text{leak}$ and the relic abundance of vortons today is $\Omega_v = \frac{\rho_v(T_\text{leak})}{s(T_\text{leak})}\frac{s_0}{\rho_0}$, where $s_0(\rho_0)$ is the entropy (energy) density of the universe today.

\begin{figure}
\begin{center}
\includegraphics{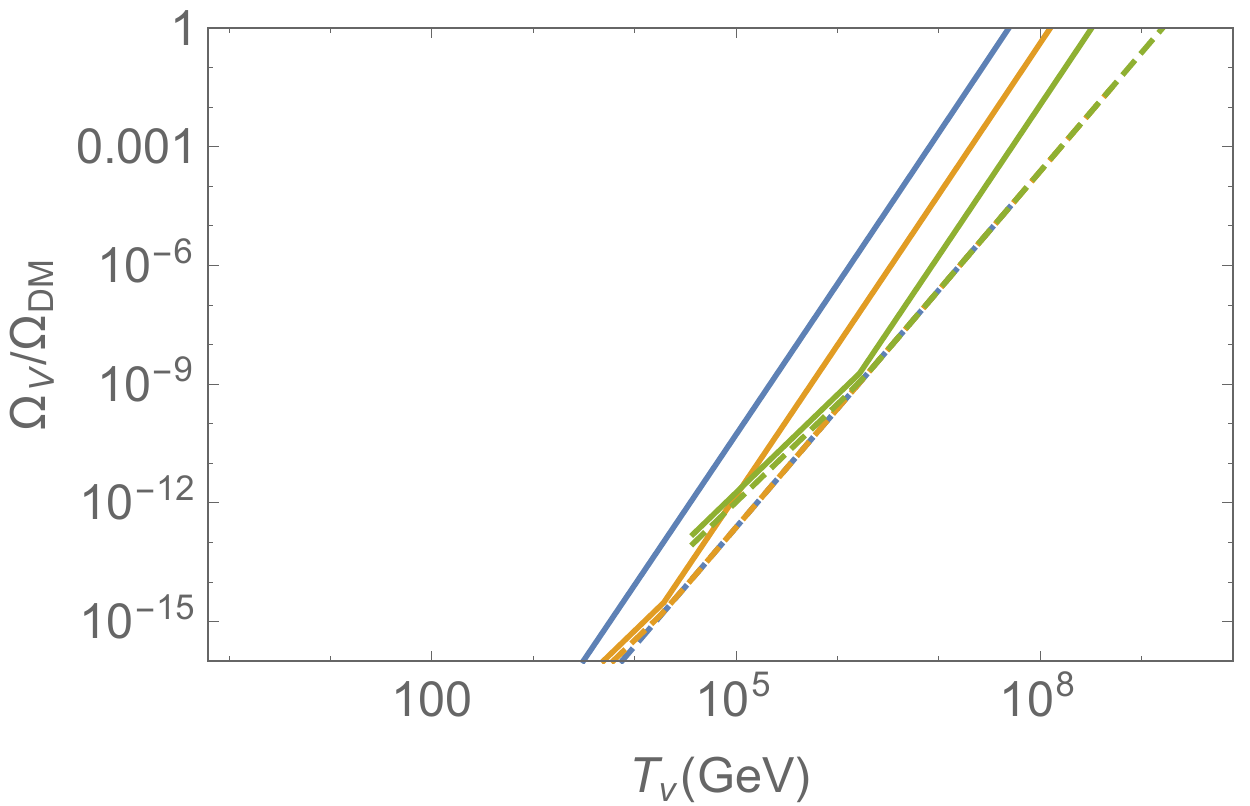}
\end{center}
\caption{The relic abundance of the vorton, $\Omega_v/\Omega_\text{DM}$, in terms of the vorton formation temperature $T_v$. The blue, orange and green lines corresponds to $f_a = 10^9, 10^{10} \text{, and\ }10^{11}\,\text{GeV}$. The solid line assumes vortons never annihilate after being formed. The dashed line assumes vortons efficiently annihilate and only the asymmetric component remains. The blue dashed line is overlapped by the orange dashed line. See the text.
}
\label{fig:noPMFVorton}
\end{figure}

We plot the vorton abundance normalized by the dark matter density in terms of $T_v$ in Fig.\,\ref{fig:noPMFVorton}. As we have discussed, we show two estimations, assuming no annihilation or efficient annihilation of vortons, as a solid and dashed line, respectively. As the universe gets cold and $\xi$ decreases to $\xi \sim \mathcal{O}(1)$, the two estimates are expected to coincide since the number vortons in the Hubble horizon is $\xi \sim \mathcal{O}(1)$. Indeed, one can see that after the kink in the plot, at which temperature $\xi$ reaches $\mathcal{O}(1)$, the solid and dashed line go together.

Note that our analysis of vortons is different from previous work\,\cite{Carter:1993wu,Brandenberger:1996zp,Martins:1998gb,Martins:1998th,Carter:1999an} in several respects. As far as we can tell, our work is the first to show that vortons are inevitably generated by axion strings.  Secondly, our estimate for the overall charge per vorton differs from previous analyses \cite{Carter:1999an}.  The underlying reason for our different estimate is that the Goldstone-Wilczek in the bulk serves to average the current and the charge over the string loop, leading to an overall charge of $Q\propto \sqrt{TL}$ instead of $Q\propto TL$. Consequently, our analysis results in smaller vorton abundances. Finally, though previous analyses concluded that the chiral current on the vortons is completely conserved, we explore several scattering processes that lead to current leakage from the string. When current leakage is efficient enough, the vorton abundance is greatly suppressed, and there is no risk of vortons overclosing the universe. In particular, since $T_\text{leak} < f_a / b < f_a$, the vorton abundance is insensitive to physics at the PQ scale.

After the QCD phase transition, each string loop becomes the boundary of a domain wall. As the whole system shrinks, the edge current on the string increases. When the current reaches Eq.\,\eqref{eq:CurrentUpperBound}, where the energy of the string is at the minimum in terms of the length, the shrinking of the loop tends to be halted since it is no more energetically favored. On the other hand, the wall is energetically likely to continue to shrink. In order that the wall shrinks with the perimeter constant, it is expected that the domain wall-string system is twisted and eventually breaks into many vortons whose size is around the domain wall thickness. Vortons produced by this process are negligible since the energy is only a tiny fraction of the total energy of the string-domain wall network giving the axion dark matter. Note that if the size of the vorton is larger than the axion compton wavelength (i.e.\ the domain wall thickness), the same process happens and the vorton breaks into many small vortons. However, the total energy of the vorton is conserved, since it is proportional to the total charge, which is conserved during this breaking process.

It is worth asking whether the superconductivity alters the domain wall dynamics. A domain wall supports some topological field theory to compensate the anomaly on the string around the wall\,\cite{Gaiotto:2017tne}. For example, the electromagnetic anomaly on the string infers that the Chern-Simons theory for the electromagnetism lives on the domain wall, just like the (anomalous) Hall effect\,\cite{Hidaka:2020iaz}. The wall thickness is of order of the axion mass and much thicker than $T^{-1}$, the typical scale of the thermal bath particles. Thus, we expect such IR structure of the wall does not qualitatively change the picture of the time evolution of the wall, although detailed studies may be needed to determine the precise abundance of axions from domain walls.

Finally, let us comment on the potential detection of vortons from the early universe. The charge of the vorton is gigantic $\gtrsim 10^3$ as Eq.\,\eqref{eq:vortonCond} requires. The situation is similar to an atomic nucleus with a very large atomic number. Once the charge exceeds $1 / \alpha \sim 137$, Schwinger pair production\,\cite{Schwinger:1951nm} occurs and the charge is screened to be $Q \lesssim 137$, although the magnetic flux cannot be screened. Similarly, the quark and gluon cloud neutralize the color charge of the vorton after QCD confinement. The color magnetic flux is Higgsed due to the dual Meissner effect\,\cite{tHooft:1975krp,Mandelstam:1974pi}. Thus, the vorton looks like a charged exotic hadron with a large magnetic dipole moment. The detection of such an object is not easy\,\cite{Agrawal:2016quu,DeLuca:2018mzn}. The vorton is screened by the Earth's crust and underground direct detection experiments cannot detect it. In addition, the vorton is much heavier than $f_a$ and the number density is low. Thus, searches for exotic hadrons do not place any constraints. Consequently, we require $\Omega_v < \Omega_\text{DM}$ based on Ref.\,\cite{Carter:1999an} although a more detailed discussion of the constraints on vorton abundance remains to be studied.

\end{appendix}

\bibliographystyle{JHEP}
\bibliography{papers}

\providecommand{\href}[2]{#2}\begingroup\raggedright\begin{thebibliography}{10}

\bibitem{Ubaldi:2008nf}
L.~Ubaldi, {\it {Effects of theta on the deuteron binding energy and the
  triple-alpha process}},  {\em Phys. Rev. D} {\bf 81} (2010) 025011,
  [\href{http://arxiv.org/abs/0811.1599}{{\tt arXiv:0811.1599}}].

\bibitem{Peccei:1977hh}
R.~Peccei and H.~R. Quinn, {\it {CP Conservation in the Presence of
  Instantons}},  {\em Phys. Rev. Lett.} {\bf 38} (1977) 1440--1443.

\bibitem{Peccei:1977ur}
R.~Peccei and H.~R. Quinn, {\it {Constraints Imposed by CP Conservation in the
  Presence of Instantons}},  {\em Phys. Rev. D} {\bf 16} (1977) 1791--1797.

\bibitem{Vafa:1984xg}
C.~Vafa and E.~Witten, {\it {Parity Conservation in QCD}},  {\em Phys. Rev.
  Lett.} {\bf 53} (1984) 535.

\bibitem{Weinberg:1977ma}
S.~Weinberg, {\it {A New Light Boson?}},  {\em Phys. Rev. Lett.} {\bf 40}
  (1978) 223--226.

\bibitem{Wilczek:1977pj}
F.~Wilczek, {\it {Problem of Strong {$P$} and {$T$} Invariance in the Presence
  of Instantons}},  {\em Phys. Rev. Lett.} {\bf 40} (1978) 279--282.

\bibitem{Preskill:1982cy}
J.~Preskill, M.~B. Wise, and F.~Wilczek, {\it {Cosmology of the Invisible
  Axion}},  {\em Phys. Lett. B} {\bf 120} (1983) 127--132.

\bibitem{Abbott:1982af}
L.~Abbott and P.~Sikivie, {\it {A Cosmological Bound on the Invisible Axion}},
  {\em Phys. Lett. B} {\bf 120} (1983) 133--136.

\bibitem{Dine:1982ah}
M.~Dine and W.~Fischler, {\it {The Not So Harmless Axion}},  {\em Phys. Lett.
  B} {\bf 120} (1983) 137--141.

\bibitem{Aghanim:2018eyx}
{\bf Planck} Collaboration, N.~Aghanim et~al., {\it {Planck 2018 results. VI.
  Cosmological parameters}},  \href{http://arxiv.org/abs/1807.06209}{{\tt
  arXiv:1807.06209}}.

\bibitem{Hikage:2012tf}
C.~Hikage, M.~Kawasaki, T.~Sekiguchi, and T.~Takahashi, {\it {Extended analysis
  of CMB constraints on non-Gaussianity in isocurvature perturbations}},  {\em
  JCAP} {\bf 03} (2013) 020, [\href{http://arxiv.org/abs/1212.6001}{{\tt
  arXiv:1212.6001}}].

\bibitem{Kobayashi:2013nva}
T.~Kobayashi, R.~Kurematsu, and F.~Takahashi, {\it {Isocurvature Constraints
  and Anharmonic Effects on QCD Axion Dark Matter}},  {\em JCAP} {\bf 09}
  (2013) 032, [\href{http://arxiv.org/abs/1304.0922}{{\tt arXiv:1304.0922}}].

\bibitem{Kibble:1980mv}
T.~Kibble, {\it {Some Implications of a Cosmological Phase Transition}},  {\em
  Phys. Rept.} {\bf 67} (1980) 183.

\bibitem{Kibble:1976sj}
T.~Kibble, {\it {Topology of Cosmic Domains and Strings}},  {\em J. Phys. A}
  {\bf 9} (1976) 1387--1398.

\bibitem{Zurek:1985qw}
W.~Zurek, {\it {Cosmological Experiments in Superfluid Helium?}},  {\em Nature}
  {\bf 317} (1985) 505--508.

\bibitem{Zurek:1996sj}
W.~Zurek, {\it {Cosmological experiments in condensed matter systems}},  {\em
  Phys. Rept.} {\bf 276} (1996) 177--221,
  [\href{http://arxiv.org/abs/cond-mat/9607135}{{\tt cond-mat/9607135}}].

\bibitem{Murayama:2009nj}
H.~Murayama and J.~Shu, {\it {Topological Dark Matter}},  {\em Phys. Lett. B}
  {\bf 686} (2010) 162--165, [\href{http://arxiv.org/abs/0905.1720}{{\tt
  arXiv:0905.1720}}].

\bibitem{Nagasawa:1997zn}
M.~Nagasawa, {\it {Scaling distribution of axionic strings and estimation of
  axion density from axionic domain walls}},  {\em Prog. Theor. Phys.} {\bf 98}
  (1997) 851, [\href{http://arxiv.org/abs/hep-ph/9712341}{{\tt
  hep-ph/9712341}}].

\bibitem{Bennett:1989yp}
D.~P. Bennett and F.~R. Bouchet, {\it {High resolution simulations of cosmic
  string evolution. 1. Network evolution}},  {\em Phys. Rev. D} {\bf 41} (1990)
  2408.

\bibitem{Allen:1990tv}
B.~Allen and E.~Shellard, {\it {Cosmic string evolution: a numerical
  simulation}},  {\em Phys. Rev. Lett.} {\bf 64} (1990) 119--122.

\bibitem{Vanchurin:2005yb}
V.~Vanchurin, K.~Olum, and A.~Vilenkin, {\it {Cosmic string scaling in flat
  space}},  {\em Phys. Rev. D} {\bf 72} (2005) 063514,
  [\href{http://arxiv.org/abs/gr-qc/0501040}{{\tt gr-qc/0501040}}].

\bibitem{Olum:2006ix}
K.~D. Olum and V.~Vanchurin, {\it {Cosmic string loops in the expanding
  Universe}},  {\em Phys. Rev. D} {\bf 75} (2007) 063521,
  [\href{http://arxiv.org/abs/astro-ph/0610419}{{\tt astro-ph/0610419}}].

\bibitem{Hiramatsu:2010yu}
T.~Hiramatsu, M.~Kawasaki, T.~Sekiguchi, M.~Yamaguchi, and J.~Yokoyama, {\it
  {Improved estimation of radiated axions from cosmological axionic strings}},
  {\em Phys. Rev. D} {\bf 83} (2011) 123531,
  [\href{http://arxiv.org/abs/1012.5502}{{\tt arXiv:1012.5502}}].

\bibitem{BlancoPillado:2011dq}
J.~J. Blanco-Pillado, K.~D. Olum, and B.~Shlaer, {\it {Large parallel cosmic
  string simulations: New results on loop production}},  {\em Phys. Rev. D}
  {\bf 83} (2011) 083514, [\href{http://arxiv.org/abs/1101.5173}{{\tt
  arXiv:1101.5173}}].

\bibitem{Hiramatsu:2012gg}
T.~Hiramatsu, M.~Kawasaki, K.~Saikawa, and T.~Sekiguchi, {\it {Production of
  dark matter axions from collapse of string-wall systems}},  {\em Phys. Rev.
  D} {\bf 85} (2012) 105020, [\href{http://arxiv.org/abs/1202.5851}{{\tt
  arXiv:1202.5851}}]. [Erratum: Phys.Rev.D 86, 089902 (2012)].

\bibitem{Kawasaki:2014sqa}
M.~Kawasaki, K.~Saikawa, and T.~Sekiguchi, {\it {Axion dark matter from
  topological defects}},  {\em Phys. Rev. D} {\bf 91} (2015), no.~6 065014,
  [\href{http://arxiv.org/abs/1412.0789}{{\tt arXiv:1412.0789}}].

\bibitem{Fleury:2015aca}
L.~Fleury and G.~D. Moore, {\it {Axion dark matter: strings and their cores}},
  {\em JCAP} {\bf 01} (2016) 004, [\href{http://arxiv.org/abs/1509.00026}{{\tt
  arXiv:1509.00026}}].

\bibitem{Klaer:2017ond}
V.~B. Klaer and G.~D. Moore, {\it {The dark-matter axion mass}},  {\em JCAP}
  {\bf 11} (2017) 049, [\href{http://arxiv.org/abs/1708.07521}{{\tt
  arXiv:1708.07521}}].

\bibitem{Klaer:2017qhr}
V.~B. Klaer and G.~D. Moore, {\it {How to simulate global cosmic strings with
  large string tension}},  {\em JCAP} {\bf 10} (2017) 043,
  [\href{http://arxiv.org/abs/1707.05566}{{\tt arXiv:1707.05566}}].

\bibitem{Gorghetto:2018myk}
M.~Gorghetto, E.~Hardy, and G.~Villadoro, {\it {Axions from Strings: the
  Attractive Solution}},  {\em JHEP} {\bf 07} (2018) 151,
  [\href{http://arxiv.org/abs/1806.04677}{{\tt arXiv:1806.04677}}].

\bibitem{Vaquero:2018tib}
A.~Vaquero, J.~Redondo, and J.~Stadler, {\it {Early seeds of axion
  miniclusters}},  {\em JCAP} {\bf 04} (2019) 012,
  [\href{http://arxiv.org/abs/1809.09241}{{\tt arXiv:1809.09241}}].

\bibitem{Buschmann:2019icd}
M.~Buschmann, J.~W. Foster, and B.~R. Safdi, {\it {Early-Universe Simulations
  of the Cosmological Axion}},  \href{http://arxiv.org/abs/1906.00967}{{\tt
  arXiv:1906.00967}}.

\bibitem{Hindmarsh:2019csc}
M.~Hindmarsh, J.~Lizarraga, A.~Lopez-Eiguren, and J.~Urrestilla, {\it {Scaling
  Density of Axion Strings}},  {\em Phys. Rev. Lett.} {\bf 124} (2020), no.~2
  021301, [\href{http://arxiv.org/abs/1908.03522}{{\tt arXiv:1908.03522}}].

\bibitem{Klaer:2019fxc}
V.~B. Klaer and G.~D. Moore, {\it {Global cosmic string networks as a function
  of tension}},  \href{http://arxiv.org/abs/1912.08058}{{\tt
  arXiv:1912.08058}}.

\bibitem{Saikawa_IPMU_slide}
K.~Saikawa, ``{Axion dark matter mass: Towards a reliable estimate}.''
  \url{http://research.ipmu.jp/seminar/sysimg/seminar/2360.pdf}.

\bibitem{Gorghetto:2020qws}
M.~Gorghetto, E.~Hardy, and G.~Villadoro, {\it {More Axions from Strings}},
  \href{http://arxiv.org/abs/2007.04990}{{\tt arXiv:2007.04990}}.

\bibitem{DiLuzio:2020wdo}
L.~Di~Luzio, M.~Giannotti, E.~Nardi, and L.~Visinelli, {\it {The landscape of
  QCD axion models}},  {\em Phys. Rept.} {\bf 870} (2020) 1--117,
  [\href{http://arxiv.org/abs/2003.01100}{{\tt arXiv:2003.01100}}].

\bibitem{Kim:1979if}
J.~E. Kim, {\it {Weak Interaction Singlet and Strong CP Invariance}},  {\em
  Phys. Rev. Lett.} {\bf 43} (1979) 103.

\bibitem{Shifman:1979if}
M.~A. Shifman, A.~Vainshtein, and V.~I. Zakharov, {\it {Can Confinement Ensure
  Natural CP Invariance of Strong Interactions?}},  {\em Nucl. Phys. B} {\bf
  166} (1980) 493--506.

\bibitem{Callan:1984sa}
C.~G. Callan, Jr. and J.~A. Harvey, {\it {Anomalies and Fermion Zero Modes on
  Strings and Domain Walls}},  {\em Nucl. Phys.} {\bf B250} (1985) 427--436.

\bibitem{Davis:1988jq}
R.~Davis and E.~Shellard, {\it {The Physics of Vortex Superconductivity. 2}},
  {\em Phys. Lett. B} {\bf 209} (1988) 485--490.

\bibitem{Carter:1993wu}
B.~Carter and X.~Martin, {\it {Dynamic instability criterion for circular
  (Vorton) string loops}},  {\em Annals Phys.} {\bf 227} (1993) 151--171,
  [\href{http://arxiv.org/abs/hep-th/0306111}{{\tt hep-th/0306111}}].

\bibitem{Brandenberger:1996zp}
R.~H. Brandenberger, B.~Carter, A.-C. Davis, and M.~Trodden, {\it {Cosmic
  vortons and particle physics constraints}},  {\em Phys. Rev. D} {\bf 54}
  (1996) 6059--6071, [\href{http://arxiv.org/abs/hep-ph/9605382}{{\tt
  hep-ph/9605382}}].

\bibitem{Martins:1998gb}
C.~Martins and E.~Shellard, {\it {Vorton formation}},  {\em Phys. Rev. D} {\bf
  57} (1998) 7155--7176, [\href{http://arxiv.org/abs/hep-ph/9804378}{{\tt
  hep-ph/9804378}}].

\bibitem{Martins:1998th}
C.~Martins and E.~Shellard, {\it {Limits on cosmic chiral vortons}},  {\em
  Phys. Lett. B} {\bf 445} (1998) 43--51,
  [\href{http://arxiv.org/abs/hep-ph/9806480}{{\tt hep-ph/9806480}}].

\bibitem{Carter:1999an}
B.~Carter and A.-C. Davis, {\it {Chiral vortons and cosmological constraints on
  particle physics}},  {\em Phys. Rev. D} {\bf 61} (2000) 123501,
  [\href{http://arxiv.org/abs/hep-ph/9910560}{{\tt hep-ph/9910560}}].

\bibitem{Witten:1984eb}
E.~Witten, {\it {Superconducting Strings}},  {\em Nucl. Phys.} {\bf B249}
  (1985) 557--592.

\bibitem{Weinberg:1981eu}
E.~J. Weinberg, {\it {Index Calculations for the Fermion-Vortex System}},  {\em
  Phys. Rev.} {\bf D24} (1981) 2669.

\bibitem{Jackiw:1981ee}
R.~Jackiw and P.~Rossi, {\it {Zero Modes of the Vortex - Fermion System}},
  {\em Nucl. Phys.} {\bf B190} (1981) 681--691.

\bibitem{Goldstone:1981kk}
J.~Goldstone and F.~Wilczek, {\it {Fractional Quantum Numbers on Solitons}},
  {\em Phys. Rev. Lett.} {\bf 47} (1981) 986--989. [,365(1981); ,365(1981)].

\bibitem{Kaplan:1987kh}
D.~B. Kaplan and A.~Manohar, {\it {Anomalous Vortices and Electromagnetism}},
  {\em Nucl. Phys.} {\bf B302} (1988) 280--290.

\bibitem{Naculich:1987ci}
S.~G. Naculich, {\it {Axionic Strings: Covariant Anomalies and Bosonization of
  Chiral Zero Modes}},  {\em Nucl. Phys.} {\bf B296} (1988) 837--867.

\bibitem{Fukuda_Yonekura}
H.~Fukuda and K.~Yonekura, {\it {Witten effect, anomaly inflow, and charge
  teleportation}},  \href{http://arxiv.org/abs/2010.02221}{{\tt
  arXiv:2010.02221}}.

\bibitem{Peskin:1995ev}
M.~E. Peskin and D.~V. Schroeder, {\em {An Introduction to quantum field
  theory}}.
\newblock Addison-Wesley, Reading, USA, 1995.

\bibitem{Manohar:1988gv}
A.~Manohar, {\it {Anomalous Vortices and Electromagnetism. 2.}},  {\em Phys.
  Lett.} {\bf B206} (1988) 276. [Erratum: Phys. Lett.209,543(1988)].

\bibitem{Harari:1992ea}
D.~Harari and P.~Sikivie, {\it {Effects of a Nambu-Goldstone boson on the
  polarization of radio galaxies and the cosmic microwave background}},  {\em
  Phys. Lett.} {\bf B289} (1992) 67--72.

\bibitem{Fedderke:2019ajk}
M.~A. Fedderke, P.~W. Graham, and S.~Rajendran, {\it {Axion Dark Matter
  Detection with CMB Polarization}},  {\em Phys. Rev.} {\bf D100} (2019), no.~1
  015040, [\href{http://arxiv.org/abs/1903.02666}{{\tt arXiv:1903.02666}}].

\bibitem{Barr:1987xm}
S.~M. Barr and A.~Matheson, {\it {Weak Resistance in Superconducting Cosmic
  Strings}},  {\em Phys. Rev. D} {\bf 36} (1987) 2905.

\bibitem{Sen:1992yt}
A.~Sen, {\it {Macroscopic charged heterotic string}},  {\em Nucl. Phys. B} {\bf
  388} (1992) 457--473, [\href{http://arxiv.org/abs/hep-th/9206016}{{\tt
  hep-th/9206016}}].

\bibitem{Schiff:1955vw}
L.~I. Schiff, {\em Quantum Mechanics}.
\newblock McGraw-Hill, second edition~ed., 1955.

\bibitem{Vilenkin:1991zk}
A.~Vilenkin, {\it {Cosmic string dynamics with friction}},  {\em Phys. Rev. D}
  {\bf 43} (1991) 1060--1062.

\bibitem{Martins:2000cs}
C.~Martins and E.~Shellard, {\it {Extending the velocity dependent one scale
  string evolution model}},  {\em Phys. Rev. D} {\bf 65} (2002) 043514,
  [\href{http://arxiv.org/abs/hep-ph/0003298}{{\tt hep-ph/0003298}}].

\bibitem{Babul:1987me}
A.~Babul, T.~Piran, and D.~N. Spergel, {\it {BOSONIC SUPERCONDUCTING COSMIC
  STRINGS. 1. CLASSICAL FIELD THEORY SOLUTIONS}},  {\em Phys. Lett. B} {\bf
  202} (1988) 307--314.

\bibitem{BlancoPillado:2000ep}
J.~J. Blanco-Pillado, K.~D. Olum, and A.~Vilenkin, {\it {Dynamics of
  superconducting strings with chiral currents}},  {\em Phys. Rev. D} {\bf 63}
  (2001) 103513, [\href{http://arxiv.org/abs/astro-ph/0004410}{{\tt
  astro-ph/0004410}}].

\bibitem{Davis:2000cx}
A.~Davis, T.~Kibble, M.~Pickles, and D.~A. Steer, {\it {Dynamics and properties
  of chiral cosmic strings in Minkowski space}},  {\em Phys. Rev. D} {\bf 62}
  (2000) 083516, [\href{http://arxiv.org/abs/astro-ph/0005514}{{\tt
  astro-ph/0005514}}].

\bibitem{Kawasaki:2004qu}
M.~Kawasaki, K.~Kohri, and T.~Moroi, {\it {Big-Bang nucleosynthesis and
  hadronic decay of long-lived massive particles}},  {\em Phys. Rev. D} {\bf
  71} (2005) 083502, [\href{http://arxiv.org/abs/astro-ph/0408426}{{\tt
  astro-ph/0408426}}].

\bibitem{Parker:1955zz}
E.~N. Parker, {\it {Hydromagnetic Dynamo Models}},  {\em Astrophys. J.} {\bf
  122} (1955) 293.

\bibitem{Neronov:1900zz}
A.~Neronov and I.~Vovk, {\it {Evidence for strong extragalactic magnetic fields
  from Fermi observations of TeV blazars}},  {\em Science} {\bf 328} (2010)
  73--75, [\href{http://arxiv.org/abs/1006.3504}{{\tt arXiv:1006.3504}}].

\bibitem{Durrer:2013pga}
R.~Durrer and A.~Neronov, {\it {Cosmological Magnetic Fields: Their Generation,
  Evolution and Observation}},  {\em Astron. Astrophys. Rev.} {\bf 21} (2013)
  62, [\href{http://arxiv.org/abs/1303.7121}{{\tt arXiv:1303.7121}}].

\bibitem{Subramanian:2015lua}
K.~Subramanian, {\it {The origin, evolution and signatures of primordial
  magnetic fields}},  {\em Rept. Prog. Phys.} {\bf 79} (2016), no.~7 076901,
  [\href{http://arxiv.org/abs/1504.02311}{{\tt arXiv:1504.02311}}].

\bibitem{Jedamzik:2018itu}
K.~Jedamzik and A.~Saveliev, {\it {Stringent Limit on Primordial Magnetic
  Fields from the Cosmic Microwave Background Radiation}},  {\em Phys. Rev.
  Lett.} {\bf 123} (2019), no.~2 021301,
  [\href{http://arxiv.org/abs/1804.06115}{{\tt arXiv:1804.06115}}].

\bibitem{Martins:1996jp}
C.~Martins and E.~Shellard, {\it {Quantitative string evolution}},  {\em Phys.
  Rev. D} {\bf 54} (1996) 2535--2556,
  [\href{http://arxiv.org/abs/hep-ph/9602271}{{\tt hep-ph/9602271}}].

\bibitem{Banerjee:2004df}
R.~Banerjee and K.~Jedamzik, {\it {The Evolution of cosmic magnetic fields:
  From the very early universe, to recombination, to the present}},  {\em Phys.
  Rev. D} {\bf 70} (2004) 123003,
  [\href{http://arxiv.org/abs/astro-ph/0410032}{{\tt astro-ph/0410032}}].

\bibitem{Demozzi:2009fu}
V.~Demozzi, V.~Mukhanov, and H.~Rubinstein, {\it {Magnetic fields from
  inflation?}},  {\em JCAP} {\bf 08} (2009) 025,
  [\href{http://arxiv.org/abs/0907.1030}{{\tt arXiv:0907.1030}}].

\bibitem{Durrer:2003ja}
R.~Durrer and C.~Caprini, {\it {Primordial magnetic fields and causality}},
  {\em JCAP} {\bf 11} (2003) 010,
  [\href{http://arxiv.org/abs/astro-ph/0305059}{{\tt astro-ph/0305059}}].

\bibitem{Kamada:2018kyi}
K.~Kamada, Y.~Tsai, and T.~Vachaspati, {\it {Magnetic Field Transfer From A
  Hidden Sector}},  {\em Phys. Rev. D} {\bf 98} (2018) 043501,
  [\href{http://arxiv.org/abs/1803.08051}{{\tt arXiv:1803.08051}}].

\bibitem{Thompson:1988jn}
A.~C. Thompson, {\it {Cosmological Effects of Superconducting Strings}},
  thesis, 10, 1988.

\bibitem{Mijic:1988ag}
M.~B. Mijic, {\it {Turning on a superconducting cosmic string}},  {\em Phys.
  Rev. D} {\bf 39} (1989) 2864.

\bibitem{Carter:2000fv}
B.~Carter, R.~H. Brandenberger, A.~Davis, and G.~Sigl, {\it {Prolongation of
  friction dominated evolution for superconducting cosmic strings}},  {\em
  JHEP} {\bf 11} (2000) 019, [\href{http://arxiv.org/abs/hep-ph/0009278}{{\tt
  hep-ph/0009278}}].

\bibitem{Dimopoulos:1997xa}
K.~Dimopoulos and A.-C. Davis, {\it {Friction domination with superconducting
  strings}},  {\em Phys. Rev. D} {\bf 57} (1998) 692--701,
  [\href{http://arxiv.org/abs/hep-ph/9705302}{{\tt hep-ph/9705302}}].

\bibitem{Martins:2018dqg}
C.~Martins, {\it {Scaling properties of cosmological axion strings}},  {\em
  Phys. Lett. B} {\bf 788} (2019) 147--151,
  [\href{http://arxiv.org/abs/1811.12678}{{\tt arXiv:1811.12678}}].

\bibitem{Vilenkin:2000jqa}
A.~Vilenkin and E.~S. Shellard, {\em {Cosmic Strings and Other Topological
  Defects}}.
\newblock Cambridge University Press, 7, 2000.

\bibitem{Yamaguchi:2005gp}
M.~Yamaguchi, {\it {Cosmological evolution of cosmic strings with time
  dependent tension}},  {\em Phys. Rev. D} {\bf 72} (2005) 043533,
  [\href{http://arxiv.org/abs/hep-ph/0503227}{{\tt hep-ph/0503227}}].

\bibitem{Zyla:2020zbs}
{\bf Particle Data Group} Collaboration, P.~Zyla et~al., {\it {Review of
  Particle Physics}},  {\em PTEP} {\bf 2020} (2020), no.~8 083C01.

\bibitem{Ibe:2021ctf}
M.~Ibe, S.~Kobayashi, Y.~Nakayama, and S.~Shirai, {\it {On Stability of
  Fermionic Superconducting Current in Cosmic String}},  {\em JHEP} {\bf 05}
  (2021) 217, [\href{http://arxiv.org/abs/2102.05412}{{\tt arXiv:2102.05412}}].

\bibitem{Watson:1944}
G.~Watson, {\em {A Treatise on the Theory of Bessel Functions}}.
\newblock Cambridge University Press, 1944.

\bibitem{Blanco-Pillado:2013qja}
J.~J. Blanco-Pillado, K.~D. Olum, and B.~Shlaer, {\it {The number of cosmic
  string loops}},  {\em Phys. Rev. D} {\bf 89} (2014), no.~2 023512,
  [\href{http://arxiv.org/abs/1309.6637}{{\tt arXiv:1309.6637}}].

\bibitem{Vachaspati:1984dz}
T.~Vachaspati and A.~Vilenkin, {\it {Formation and Evolution of Cosmic
  Strings}},  {\em Phys. Rev. D} {\bf 30} (1984) 2036.

\bibitem{Chudnovsky:1986hc}
E.~Chudnovsky, G.~Field, D.~Spergel, and A.~Vilenkin, {\it {Superconducting
  cosmic strings}},  {\em Phys. Rev. D} {\bf 34} (1986) 944--950.

\bibitem{Gaiotto:2017tne}
D.~Gaiotto, Z.~Komargodski, and N.~Seiberg, {\it {Time-reversal breaking in
  QCD$_{4}$, walls, and dualities in 2 + 1 dimensions}},  {\em JHEP} {\bf 01}
  (2018) 110, [\href{http://arxiv.org/abs/1708.06806}{{\tt arXiv:1708.06806}}].

\bibitem{Hidaka:2020iaz}
Y.~Hidaka, M.~Nitta, and R.~Yokokura, {\it {Higher-form symmetries and 3-group
  in axion electrodynamics}},  {\em Phys. Lett. B} {\bf 808} (2020) 135672,
  [\href{http://arxiv.org/abs/2006.12532}{{\tt arXiv:2006.12532}}].

\bibitem{Schwinger:1951nm}
J.~S. Schwinger, {\it {On gauge invariance and vacuum polarization}},  {\em
  Phys. Rev.} {\bf 82} (1951) 664--679. [,116(1951)].

\bibitem{tHooft:1975krp}
G.~'t~Hooft, {\it {Gauge Fields with Unified Weak, Electromagnetic, and Strong
  Interactions}},  in {\em {1975 High-Energy Particle Physics Divisional
  Conference of EPS (includes 8th biennial conf on Elem. Particles)}}, p.~1225,
  9, 1975.

\bibitem{Mandelstam:1974pi}
S.~Mandelstam, {\it {Vortices and Quark Confinement in Nonabelian Gauge
  Theories}},  {\em Phys. Rept.} {\bf 23} (1976) 245--249.

\bibitem{Agrawal:2016quu}
P.~Agrawal, F.-Y. Cyr-Racine, L.~Randall, and J.~Scholtz, {\it {Make Dark
  Matter Charged Again}},  {\em JCAP} {\bf 05} (2017) 022,
  [\href{http://arxiv.org/abs/1610.04611}{{\tt arXiv:1610.04611}}].

\bibitem{DeLuca:2018mzn}
V.~De~Luca, A.~Mitridate, M.~Redi, J.~Smirnov, and A.~Strumia, {\it {Colored
  Dark Matter}},  {\em Phys. Rev. D} {\bf 97} (2018), no.~11 115024,
  [\href{http://arxiv.org/abs/1801.01135}{{\tt arXiv:1801.01135}}].

\end{thebibliography}\endgroup




\end{document}